\documentclass[aps,prd,twocolumn,preprintnumbers,superscriptaddress,
               nofootinbib,10pt,floatfix]{revtex4-2}

\usepackage{amsmath,amssymb,amsfonts,bm}
\usepackage{graphicx}
\usepackage{xcolor}
\usepackage[colorlinks=true,linkcolor=magenta,citecolor=blue,
            urlcolor=blue]{hyperref}
\usepackage[nameinlink]{cleveref}
\usepackage{enumitem}
\usepackage{ragged2e}
\usepackage[labelfont=bf,justification=justified,singlelinecheck=false]{caption}
\usepackage{subcaption}
\usepackage{orcidlink}
\usepackage{newunicodechar}
\usepackage{times}
\usepackage{natbib}
\usepackage{soul}
\usepackage[outdir=./]{epstopdf}
\usepackage[normalem]{ulem}
\DeclareCaptionJustification{justified}{\justifying}
\newunicodechar{⊙}{\odot}

\newcommand{\Msun}{M_\odot}
\newcommand{\fR}{f_{R0}}
\newcommand{\Geff}{G_{\rm eff}}
\newcommand{\Edd}{\lambda_{\rm Edd}}

\newcommand{\dd}{\mathrm{d}}

\graphicspath{{./}{figures/}}

\begin{document}

\title{Primordial Black Hole Seeds for Little Red Dots in $f(R)$ Gravity}

\author{Saeed Fakhry\orcidlink{0000-0002-6349-8489}}
\email{saeed.fakhry@uv.es}
\affiliation{Departamento de Astronom\'{\i}a y Astrof\'{\i}sica,
Universitat de Val\`encia, Avenida  Vicent Andrés Estellés 19,
46100 Burjassot (Val\`encia), Spain}


\begin{abstract}
The little red dots (LRDs) discovered by the James Webb Space Telescope host black holes of up to $10^{9}\,\Msun$ within the first billion years, posing a challenge to conventional seeding scenarios. We investigate whether such seeds can form in dense clusters of primordial black holes (PBHs), in which the central object grows by gravitational capture of the surrounding PBHs while accreting gas, within Hu-Sawicki $f(R)$ gravity, where the scalar field acts on the matter surrounding the PBHs. For the adopted cluster configuration, our results show that accretion and mergers reinforce one another: for time-averaged Eddington ratios $\lambda_{\rm Edd}\gtrsim0.3$, the seed consumes its cluster by $z\simeq5.5$-$7.3$ and subsequently grows into the LRD mass range. Calibrated on the observed black-hole mass function, the model reproduces the abundance of LRDs near the peak of their population without further adjustment, while it underpredicts the abundance at $z\simeq7.5$ and overpredicts it at $z\simeq2$. Our findings indicate that the earliest seeds may accrete about twice as efficiently as later ones, and that the finite cluster reservoir can imprint a lower edge on the mass function near the cluster mass. Within the adopted screening and halo-growth prescriptions, the scalar field enhances the abundance of active seeds by up to $\sim 13\%$ at lower redshifts and by $\sim 7\%$ at $z \simeq 7.5$ for the strongest model considered, modestly moving the high-redshift prediction towards the data at fixed normalization. Also, the remnant spins can further discriminate between feeding modes.
\end{abstract}

\maketitle

\section{\label{sec:intro}Introduction}
Deep near-infrared spectroscopy from the James Webb Space Telescope (JWST) has uncovered an unexpectedly abundant population of active black holes at high redshifts. Broad-line active galactic nuclei (AGN) and compact red sources, known as little red dots (LRDs), are observed in abundance at $z \sim 4\text{-}9$, with virial black-hole masses inferred in the range $M_{\rm BH} \sim 10^6\text{-}10^9\,\Msun$~\citep{Labbe2023,Harikane2023,Maiolino2024,Matthee2024,Kokorev2024,Greene2024,Akins2024}. The presence of such massive objects within the first gigayear of cosmic history further highlights a long-standing challenge for models of supermassive black-hole (SMBH) assembly~\citep{Volonteri2010,Inayoshi2020,Fan2023, 2025PhRvD.112l3503F, 2026ApJ...998..178F}. Even with continuous Eddington-limited accretion, a $\sim100\,\Msun$ Population~III remnant~\citep{Madau2001,Bromm2004} cannot plausibly grow to $\sim10^9\,\Msun$ by $z\sim7$ without sustained super-Eddington accretion or highly idealized duty cycles. Heavy-seed scenarios alleviate this timing problem but face their own limitations: direct-collapse black holes require pristine atomic-cooling halos exposed to an intense Lyman-Werner radiation field to suppress fragmentation~\citep{BrommLoeb2003,Lodato2006,Regan2009,Latif2013}, while runaway stellar collisions in dense nuclear clusters require finely tuned physical conditions~\citep{Devecchi2009,Katz2015}. Both channels remain observationally unconfirmed, which motivates seed scenarios decoupled from the star-formation history.

Primordial black holes (PBHs), formed through the gravitational collapse of large-amplitude density perturbations during the radiation-dominated era, provide a nonstellar channel for black-hole seeding~\citep{Zeldovich1967,Hawking1971,Carr1974,CarrHawking1974,Khlopov2010}. Recent studies have explored PBHs as a possible origin of massive black-hole seeds at high redshift~\citep{2025JCAP...04..040Z} and, more specifically, as a pathway for producing the massive black-hole seeds associated with the LRDs observed by JWST~\citep{2026PhRvL.136w1402D}. Alternative primordial formation scenarios have also been proposed in which strongly non-Gaussian curvature perturbations generated by a curvaton can produce supermassive PBHs, providing a direct primordial explanation for the black holes powering LRDs~\citep{2026arXiv260628296A}. Interest in PBHs was renewed by the LIGO-Virgo-KAGRA (LVK) detections of tens-of-solar-mass black holes~\citep{Abbott2016,Bird2016,Sasaki2016,Clesse2017}, and PBHs remain a viable dark-matter component over several mass ranges~\citep{Carr2016,CarrKuhnel2020,CarrKuhnel2021,Green2021,Escriva2022b}. PBHs need not form a homogeneous Poisson distribution: primordial non-Gaussianity, spectator-field dynamics, and domain-wall or phase-transition mechanisms can generate significant spatial clustering at formation~\citep{Chisholm2006,Tada2015,Young2015,Suyama2019,Atal2020,DeLuca2021}. A fraction of the PBH population may then form within, or rapidly assemble into, dense virialized substructures, where the focus shifts from isolated binary formation~\citep{Nakamura1997,Ioka1998,AliHaimoud2017} to the collective dynamical evolution of a self-gravitating PBH population, where close encounters, two-body relaxation, and gravitational-wave (GW) capture act concurrently.

The kinetics of PBH binary formation and mergers in virialized halos have been studied extensively over the past decade, both for the smooth cosmological PBH population and for populations embedded in dark-matter halos~\citep{Bird2016,Sasaki2016,Raidal2017,Ballesteros2018,Vaskonen2020,Jedamzik2020,Hutsi2021,Franciolini2022}. The predicted merger rate depends sensitively on the structure and collapse geometry of the host halos: ellipsoidal rather than spherical collapse can substantially enhance the merger rate~\citep{Fakhry2021ellip}, while self-interacting dark-matter profiles modify the probability of close encounters~\citep{Fakhry2021sidm}. Furthermore, high-precision semianalytic mass functions can alter the abundance of potential host halos~\citep{Fakhry2022}, whereas dense environments, such as dark-matter spikes around SMBHs and the interiors of cosmic voids, provide distinct dynamical settings for PBH capture~\citep{Fakhry2023voids,Fakhry2023spikes}. The PBH merger rate thus depends on the local environment as well as on the PBH abundance.

In scalar-tensor theories of gravity, the environment also determines the strength of a screened fifth force. In metric $f(R)$ gravity, the Hu-Sawicki model provides a canonical and observationally viable realization~\citep{HuSawicki2007,Sotiriou2010,DeFelice2010,Nojiri2011, 2017PhR...692....1N}. In this framework, the scalar field mediates a fifth force that enhances the effective gravitational coupling $\Geff$ in low-density environments. At the same time, the chameleon mechanism suppresses this force in deep Newtonian potentials, allowing the theory to remain consistent with solar-system and laboratory constraints~\citep{Khoury2004,Brax2008,Lombriser2014,Burrage2018, 2024Univ...10..297F}. The fifth force, however, does not act on all bodies equally. Black holes carry no scalar charge in scalar-tensor theories~\citep{Hawking1972,Sotiriou2012}: they neither source the scalar field nor respond to an external scalar gradient, a manifestation of the violation of the strong equivalence principle~\citep{HuiNicolisStubbs2009}. The two-body interaction between PBHs, and therefore their GW radiation capture, is consequently that of general relativity (GR) in every $f(R)$ model. The influence of the scalaron on PBHs embedded in dark-matter halos is instead environmental, acting on the particle dark matter and gas that make up the host. In line with this picture, the PBH merger rate in $f(R)$ gravity has been shown to depend on the background field amplitude $|\fR|$~\citep{Fakhry2024fR}. Furthermore, the merger dynamics of compact binaries has been investigated in both the Hu-Sawicki and normal-branch Dvali-Gabadadze-Porrati (DGP) models. In these models, modifications to the halo mass function, concentration, and velocity dispersion combine to produce a redshift-dependent enhancement of the merger rate~\citep{Fakhry2024pbhns, Fakhry:2024kjj}. Current cluster-abundance analyses bound $|\fR|\lesssim1.6\times10^{-5}$~\citep{Cataneo2015}, and galaxy-scale tests are considerably tighter~\citep{DesmondFerreira2020}; the largest value considered below, $|\fR|=10^{-4}$, is therefore included as an illustrative upper case.

Despite this progress, several aspects of PBH growth through repeated mergers remain to be explored. Many existing treatments focus on the merger rate for a prescribed PBH mass function rather than evolving the mass spectrum through successive merger generations~\citep{Liu2019,Wu2020,Liu2023}. Although the importance of multiple mergers depends on the initial mass function and can be subdominant in some scenarios~\citep{Liu2023}, successive mergers can redistribute PBHs toward higher masses and therefore provide a natural channel for building a high-mass tail. Capturing this evolution requires following the full mass spectrum rather than a representative binary, for which the nonlinear Smoluchowski coagulation equation provides a natural kinetic framework~\citep{Smoluchowski1916,Silk1993}. Previous work has not combined such a coagulation solver with the screening of the host environment, evaluated with the redshift-dependent background field, nor connected merger-driven PBH growth to the active black holes observed at $z\sim4\text{-}7$. Moreover, the internal dynamical viability of dense PBH clusters must be established rather than assumed. Core collapse, two-body relaxation, mass segregation, and the retention or ejection of GW-recoiled merger remnants can all affect the efficiency of hierarchical growth~\citep{SigurdssonHernquist1993,NunoSilesGarciaBellido2025}. 

In this work, we ask whether PBH clusters can seed the massive black holes powering LRDs, and how $f(R)$ gravity affects their growth. In Section~\ref{sec:framework}, we introduce the cosmological background, Hu-Sawicki $f(R)$ gravity, the screening and scale-dependent growth of the host matter, and the dense PBH cluster configuration. In Section~\ref{sec:assembly}, we develop the hybrid mass-assembly model: the GW radiation-capture kernel, the Smoluchowski coagulation formalism and its numerical implementation, and the coupled merger and accretion growth of the central seed with its finite reservoir. In Section~\ref{sec:diagnostics}, we define the black-hole mass function and the abundance comparison with JWST data. Section~\ref{sec:results} presents the growth of the seed, its calibration on the mass function, the comparison with the LRD abundance and the environmental effect of the scalar field, and Section~\ref{sec:conclusions} summarizes our conclusions. Three supplementary diagnostics, which rest on additional assumptions, are developed in the appendices: the stochastic GW background (Appendix~\ref{sec:app_sgwb}), the spatial clustering of PBHs (Appendix~\ref{sec:app_clustering}) and the remnant spins (Appendix~\ref{sec:app_spin}); Appendix~\ref{sec:app_numerics} documents the reservoir normalization and a self-consistent test of the field depletion.

\section{\label{sec:framework}Theoretical and Physical Framework}

\subsection{\label{ssec:cosmo}Cosmological Background}

We consider a spatially flat $\Lambda$CDM cosmology to determine the
cosmic time available for PBH mergers and subsequent black-hole growth.
The cosmological background enters the calculation primarily through
the relation between redshift and cosmic time. This relation sets the
duration of the merger-coagulation phase and provides the time interval
over which the resulting black-hole seeds can subsequently grow through
gas accretion.

The Hubble expansion rate is given by
\begin{equation}
H(z)
=
H_0
\sqrt{\Omega_m(1+z)^3+\Omega_\Lambda},
\label{eq:Hz}
\end{equation}
where $H_0$ is the present-day Hubble constant, while $\Omega_m$ and
$\Omega_\Lambda$ denote the present-day matter and dark-energy density
parameters, respectively. We adopt the Planck 2018 best-fit values,
$H_0=67.4~{\rm km\,s^{-1}\,Mpc^{-1}}$, $\Omega_m=0.3134$, and
$\Omega_\Lambda=1-\Omega_m$~\citep{Planck2018}, since these determine the
absolute cosmic-time budget available for PBH coagulation and set the
overall normalization of the growth calculation that follows. The
cosmic time corresponding to a given redshift is then obtained from
\begin{equation}
t(z)
=
\int_z^\infty
\dfrac{\dd z'}{(1+z')H(z')},
\label{eq:t_of_z}
\end{equation}
where $z'$ is an integration variable.

The calculation starts at $z_{\rm ini}=11.5$, within the epoch commonly associated with Population
III star formation and the assembly of the first dense stellar
systems~\citep{Madau2001,Bromm2004}, and the seed growth is followed down to $z=3.5$, the lowest redshift of the JWST data used for the black-hole mass function (BHMF) comparison (Sec.~\ref{ssec:bhmf_framework}). The coagulation of the field population is integrated to $z=5$; below this redshift the field population changes slowly and is held at its $z=5$ state. No redshift is imposed for the end of the merger phase. Mergers act for as long as the PBH reservoir of the cluster is not exhausted (Sec.~\ref{ssec:accretion_model}). For reference, the interval from $z_{\rm ini}$ to $z=6$ corresponds to a cosmic-time budget of $\simeq0.54$~Gyr. Gas accretion is
\emph{not} postponed until after the merger phase: the merger and accretion channels act concurrently from $z_{\rm ini}$ and are solved together for the central seed of each cluster. The initial PBH mass is fixed at
$m_{\rm PBH}^{(0)}=30\,\Msun$, matching the characteristic component
mass of the stellar-mass binary black holes detected by LVK,
which is also the mass scale for which PBHs have been most actively
discussed as a dark-matter and seeding candidate~\citep{Abbott2016,Bird2016,Sasaki2016}.

We adopt a lognormal initial mass distribution
\begin{equation}
\dfrac{\dd N}{\dd\ln m}
\propto
\exp\!\left[
-\dfrac{
(\ln m-\ln m_{\rm PBH}^{(0)})^2
}{2\sigma_{\ln m}^2}\right],
\label{eq:lognormal}
\end{equation}
where $N$ denotes the number of PBHs and $m$ is the PBH mass. The
logarithmic width is set to $\sigma_{\ln m}=0.40$, a value typical of
the extended mass functions predicted by broad formation mechanisms
such as curvaton-field or near-critical gravitational collapse
scenarios, and is used here to avoid the artificial sharpness of a
monochromatic spectrum in the coagulation calculation. The
distribution is normalized to $N_{\rm cl}=10^5$ PBHs per cluster. This number is an adopted input, not a derived quantity. Together with $m_{\rm PBH}^{(0)}$ and the number density of Sec.~\ref{ssec:cluster_dynamics}, it fixes the cluster mass, radius and velocity dispersion, and therefore enters the physics of the merger channel as well as the statistical sampling of the mass spectrum. This initial population provides the starting
point for the subsequent merger-coagulation evolution.

\subsection{\label{ssec:screening} Hu-Sawicki $f(R)$ Gravity}
In $f(R)$ gravity, the Einstein-Hilbert Lagrangian is generalized to a function of the Ricci scalar. The corresponding action is given by
\begin{equation}
S = \dfrac{1}{16\pi G} \int \dd^4x\,\sqrt{-g}\,f(R) + S_{\rm m}[g_{\mu\nu},\Psi], 
\label{eq:fR_action}
\end{equation}
where $g_{\mu\nu}$ is the spacetime metric, $g$ is its determinant,
$R$ is the Ricci scalar, and $S_{\rm m}$ denotes the matter action with
$\Psi$ representing the matter fields. The additional gravitational
degree of freedom, commonly referred to as the scalaron, is
characterized by $f_R\equiv \dd f/\dd R$. In this work, we adopt the
Hu-Sawicki form~\citep{HuSawicki2007},
\begin{equation}
f(R)=R-m^2\dfrac{c_1(R/m^2)^n}{c_2(R/m^2)^n+1},
\label{eq:HS_model}
\end{equation}
where $m^2$ sets the characteristic curvature scale, $c_1$ and $c_2$
are dimensionless model parameters, and $n$ controls the curvature
dependence of the modification. In the high-curvature regime, the
Hu-Sawicki model approaches the GR limit while retaining a nontrivial
scalar degree of freedom. We characterize the strength of the
modification by the present-day background scalaron value,
$f_{R0}\equiv f_R(z=0)$, and use $|f_{R0}|$ to label the different
modified-gravity cases considered below.

The scalaron can mediate an additional attractive force when it is
sufficiently unscreened. In the Hu-Sawicki model, this fifth force is
suppressed by the chameleon mechanism in sufficiently dense
environments or in regions with deep gravitational potentials, so that
the effective gravitational interaction depends on the surrounding
environment.

Black holes, however, do not take part in this interaction. For a stationary black hole the no-hair property of scalar-tensor theories implies a vanishing scalar charge, $q_s=0$~\citep{Hawking1972,Sotiriou2012}. This statement is not restricted to isolated black holes in vacuum. A black hole placed in an external scalar gradient, such as that produced by a surrounding matter distribution with $\nabla f_R\neq0$, neither sources the scalar nor responds to the gradient, and therefore falls at a different rate from non-relativistic matter, which feels the fifth force~\citep{HuiNicolisStubbs2009}. Scalar hair can be induced on a black hole only by a time-dependent background~\citep{Jacobson1999,HorbatschBurgess2012} or supported by a spatial gradient~\citep{Berti2013}, and the resulting charge is suppressed by the ratio of the horizon scale to the scale of the background variation. For a $30\,\Msun$ PBH at $z\sim10$ this ratio is $GM H/c^3\sim10^{-20}$, further multiplied by $|f_R(z)|\ll1$. The two-body interaction between PBHs, and hence the GW radiation-capture kernel of Sec.~\ref{ssec:kernel}, is therefore that of GR in all models considered here. The scalaron instead acts on the non-relativistic matter of the environment: the particle dark matter and the gas of the halo that hosts the PBH cluster.

The strength of the modification at redshift $z$ is set by the background scalaron field. For the Hu-Sawicki model in the high-curvature regime $R\gg m^2$,
\begin{equation}
|f_R(z)|
=
|f_{R0}|
\left[
\dfrac{\Omega_m+4\Omega_\Lambda}{\Omega_m(1+z)^3+4\Omega_\Lambda}
\right]^{n+1},
\label{eq:fRz}
\end{equation}
which decreases steeply toward high redshift. We adopt $n=1$ throughout, for which $|f_R(z)|/|f_{R0}|\simeq7.7\times10^{-4}$ at $z=6$ and $2.5\times10^{-5}$ at $z=11.5$. The range of the fifth force is set by the physical Compton wavelength of the scalaron,
\begin{eqnarray}
&\lambda_{C}(z)
=
\sqrt{\dfrac{3(n+1)\,|f_R(z)|}{R(z)}},\nonumber\\
&R(z)=\dfrac{3H_0^2}{c^2}\left[\Omega_m(1+z)^3+4\Omega_\Lambda\right],
\label{eq:lambdaC}
\end{eqnarray}
which gives $\lambda_C\simeq3.6$, $11$ and $36$~Mpc at $z=0$, but only $\simeq17$, $53$ and $166$~kpc at $z=6$ for $|f_{R0}|=10^{-6}$, $10^{-5}$ and $10^{-4}$, respectively.

The environment in which the scalaron acts is the host halo of the PBH cluster. We model a host of mass $M_h$ at redshift $z$ as a virialized halo with radius and circular velocity
\begin{equation}
R_{\rm vir}=\left[\dfrac{3M_h}{4\pi\Delta_{\rm vir}(z)\rho_c(z)}\right]^{1/3},
\qquad
v_c^2=\dfrac{GM_h}{R_{\rm vir}},
\label{eq:host_halo}
\end{equation}
where $\rho_c(z)$ is the critical density and $\Delta_{\rm vir}(z)$ is the virial overdensity of~\citet{BryanNorman1998}. The depth of the host potential is $|\Phi_h|=v_c^2/c^2$. For $M_h=10^{9}$ and $10^{10}\,\Msun$, this gives $|\Phi_h|\simeq1.0\times10^{-8}$ and $4.8\times10^{-8}$ at $z=6$, and $1.9\times10^{-8}$ and $8.6\times10^{-8}$ at $z=11.5$.

The degree of chameleon screening of the host is related to its potential depth through the thin-shell condition. In the
quasi-static limit, we define the thin-shell ratio
\begin{equation}
x_{\rm scr}(M_h,z)
\equiv
\dfrac{1.5\,|f_R(z)|}
{|\Phi_h(M_h,z)|},
\label{eq:xscr}
\end{equation}
where the numerical prefactor of $1.5$ is fixed by the Hu-Sawicki
thin-shell relation, and the background field is evaluated at the redshift of interest rather than at $z=0$. We adopt a smooth interpolation prescription between the screened and unscreened regimes,
\begin{equation}
f_{\rm scr}
=
\dfrac{x_{\rm scr}}{1+x_{\rm scr}},
\label{eq:fscr}
\end{equation}
which reduces to $f_{\rm scr}\simeq x_{\rm scr}$ in the strongly screened regime $x_{\rm scr}\ll1$ and approaches unity asymptotically as $x_{\rm scr}$ increases. Thus, $f_{\rm scr}\rightarrow1$ corresponds to the unscreened limit, whereas $f_{\rm scr}\rightarrow0$ describes the strongly screened, GR-like regime. This prescription is an effective description of the environmental suppression of the fifth force; it is not derived from an explicit solution for the nonlinear scalaron profile, and the resulting values should be regarded as indicative. It also does not account for the finite range of the force. At $z\simeq11.5$, the Compton wavelength for $|f_{R0}|=10^{-6}$ ($\simeq1.3$~kpc) is smaller than the virial radius of a $10^{9}\,\Msun$ host ($\simeq2.6$~kpc), so that Eq.~\eqref{eq:fscr} overestimates the modification for small field amplitudes at the highest redshifts.

In the absence of chameleon screening, the scalar contribution
enhances the Newtonian gravitational coupling felt by non-relativistic matter by a factor of $1/3$,
such that $G_{\rm eff}=4G/3$ on scales below $\lambda_C$. Environmental screening suppresses this
contribution according to the screening factor $f_{\rm scr}$, so that
\begin{equation}
G_{\rm eff}
=
G\left(
1+\dfrac{1}{3}f_{\rm scr}
\right),
\qquad
G\leq G_{\rm eff}\leq\dfrac{4}{3}G.
\label{eq:Geff}
\end{equation}
Equation~\eqref{eq:Geff} describes the coupling felt by the particle dark matter and gas of the host. It does not apply to the PBHs, which carry no scalar charge.

For concreteness, we consider Hu-Sawicki models with $|f_{R0}|=10^{-6}$, $10^{-5}$, and $10^{-4}$, corresponding to progressively stronger departures from the GR limit~\citep{HuSawicki2007}. As noted in Sec.~\ref{sec:intro}, $|f_{R0}|=10^{-4}$ is excluded by current cluster-abundance constraints~\citep{Cataneo2015} and is retained as an illustrative upper case. For orientation, Table~\ref{tab:geff} lists the present-day screening and effective coupling for non-relativistic matter in an environment of potential depth $\Phi_{\rm tot}\simeq2.85\times10^{-6}$, the value for the reference cluster in a Milky-Way-like host. These $z=0$ values are not used in the seed calculation. At the redshifts of interest, Eq.~\eqref{eq:fRz} reduces the background field by three to five orders of magnitude, and the screening of high-redshift hosts is computed from Eqs.~\eqref{eq:fRz}-\eqref{eq:fscr}; it is shown in Fig.~\ref{fig:growth}(b).

\begin{table}[h!]
\centering
\caption{\label{tab:geff}Present-day ($z=0$) thin-shell screening and the corresponding effective coupling for non-relativistic matter in an environment of potential depth $\Phi_{\rm tot}\simeq2.85\times10^{-6}$, corresponding to the reference cluster ($N_{\rm cl}=10^5$, $m_{\rm PBH}^{(0)}=30\,\Msun$, $n_{\rm cl}=10^8\,{\rm pc}^{-3}$) in a Milky-Way-like host with $v_{\rm host}=220\,{\rm km\,s^{-1}}$. The values are computed with $|f_{R0}|$ and are shown for orientation only. They are not used for the high-redshift calculation, which employs $|f_R(z)|$ of Eq.~\eqref{eq:fRz}, and they do not apply to PBH-PBH interactions.}
\begin{ruledtabular}
\begin{tabular}{lccc}
Model & $x_{\rm scr}$ & $f_{\rm scr}$ & $\Geff/G$ \\
\hline
GR & $0$ & $0$ & $1.000$  \\
$|f_{R0}|=10^{-6}$ & $0.526$ & $0.345$ & $1.115$ \\
$|f_{R0}|=10^{-5}$ & $5.26$ & $0.840$ & $1.280$  \\
$|f_{R0}|=10^{-4}$ & $52.6$ & $0.981$ & $1.327$  \\
\end{tabular}
\end{ruledtabular}
\end{table}

\subsection{\label{ssec:growth_mg}Scale-Dependent Growth of the Host Matter}

In scalar-tensor theories such as the Hu-Sawicki $f(R)$ model~\citep{HuSawicki2007}, the additional scalar degree of freedom modifies the gravitational coupling on scales below its Compton wavelength. In the quasi-static linear regime, this scale dependence can be expressed through the effective gravitational coupling,
\begin{equation}
\mu(k,a) \equiv \dfrac{G_{\rm eff}(k,a)}{G} = \dfrac{1 + \frac{4}{3}\left(\frac{k}{a,m_{f_R}(a)}\right)^2}{1 + \left(\frac{k}{a,m_{f_R}(a)}\right)^2},
\label{eq:mu_hs}
\end{equation}
where $a=1/(1+z)$ is the scale factor, $m_{f_R}(a)$ is the inverse physical Compton wavelength of the scalaron, $\lambda_C(a)$. For the Hu-Sawicki model, $\lambda_C(a)$ is given by Eq.~\eqref{eq:lambdaC} with the background field of Eq.~\eqref{eq:fRz}; it shrinks from tens of Mpc at $z=0$ to $\lesssim0.2$~Mpc at $z\gtrsim6$ for the models considered.
Thus, on scales well below the Compton wavelength, $k\gg a\,m_{f_R}$, the effective coupling approaches $\mu\to4/3$, corresponding to the enhancement of gravity by $4/3$. On scales much larger than the Compton wavelength, $k\ll a\,m_{f_R}$, the scalar-mediated contribution is suppressed and the GR limit is recovered.

The coupling $\mu(k,a)$ applies to non-relativistic matter. PBHs carry no scalar charge and do not feel the fifth force (Sec.~\ref{ssec:screening}); they respond to the gravitational potential of the matter perturbations with Newton's constant. On linear scales, the growth factors of the particle dark matter, $D_m(k,a)$, and of the PBHs, $D_{\rm P}(k,a)$, therefore obey
\begin{align}
D_m''+\left(2-\tfrac{3}{2}\Omega_m(a)\right)D_m'&=\tfrac{3}{2}\Omega_m(a)\,\mu(k,a)\,D_m,
\label{eq:growth_m}\\
D_{\rm P}''+\left(2-\tfrac{3}{2}\Omega_m(a)\right)D_{\rm P}'&=\tfrac{3}{2}\Omega_m(a)\,D_m,
\label{eq:growth_P}
\end{align}
where primes denote derivatives with respect to $\ln a$, and both reduce to the GR growth factor $D_{\rm GR}(a)$ for $\mu=1$. We characterize the modification by the ratios $g_m(k)\equiv D_m/D_{\rm GR}$ and $g_{\rm P}(k)\equiv D_{\rm P}/D_{\rm GR}$, and include the screening of collapsed structures at the mass scale $M(k)=(4\pi/3)\rho_m(\pi/k)^3$ of each mode through the host screening factor of Eq.~\eqref{eq:fscr}, $g\to1+f_{\rm scr}[M(k),z]\,(g-1)$. The PBH clustering quantities of Appendix~\ref{sec:app_clustering} are evaluated at $z=6$.

\subsection{\label{ssec:cluster_dynamics}Cluster Dynamics and Retention}

The cluster model is specified by the PBH number density $n_{\rm cl}$ and
the total number of PBHs $N_{\rm cl}$. For a spherical distribution with
uniform number density, the cluster volume and radius are
\begin{equation}
V_{\rm cl}
=
\dfrac{N_{\rm cl}}{n_{\rm cl}},
\qquad
R_{\rm cl}
=
\left(
\frac{3N_{\rm cl}}
{4\pi n_{\rm cl}}
\right)^{1/3},
\label{eq:Rcl}
\end{equation}
where $R_{\rm cl}$ denotes the characteristic cluster radius. For the
parameters considered throughout this work, this relation gives
$R_{\rm cl}\approx0.062~{\rm pc}$. The adopted density, $n_{\rm cl}=10^{8}\,{\rm pc}^{-3}$, is comparable to the cores of the densest known stellar systems, such as young massive
and nuclear star clusters, and is chosen as an upper-end
environment in which a clustered PBH sub-population could reside. As shown below, it is also a density for which the
internal relaxation time is short enough for the cluster to remain
dynamically active over the available cosmic-time budget. We stress that $N_{\rm cl}$ and $n_{\rm cl}$ are adopted inputs, not derived quantities. For comparison, if the cluster is identified with the PBH content of a virialized host halo, a $3\sigma$ halo collapsing at $z\simeq10$ with $f_{\rm PBH}=10^{-3}$ contains $M_{\rm cl}\simeq2\times10^{6}\,\Msun$, close to the adopted cluster mass, but the halo-averaged PBH density is only $\sim3\times10^{2}\,{\rm pc}^{-3}$. Reaching $n_{\rm cl}\sim10^{8}\,{\rm pc}^{-3}$ therefore requires a mechanism that concentrates the PBHs far beyond the halo average, such as mass segregation and gas dynamical friction toward the halo center. Such a sinking of initially clustered PBHs into a dense core, followed by runaway mergers, has been proposed as a channel for massive seeds~\citep{2025JCAP...04..040Z,NunoSilesGarciaBellido2025}, although the efficiency of dynamical friction in clumpy high-redshift galaxies is debated~\citep{Ma2021}. At the halo-averaged density, the merger channel described below becomes negligible. The merger-driven results of this work should therefore be read as conditional on the adopted cluster density.

For a virialized cluster we adopt the characteristic speed
$v_{\rm vir}\equiv\sqrt{G M_{\rm cl}/R_{\rm cl}}\approx475~{\rm km\,s^{-1}}$
for the reference parameters above. The escape speed from the center of the uniform cluster, where the central seed resides, is $v_{\rm esc}=\sqrt{3}\,v_{\rm vir}\simeq820~{\rm km\,s^{-1}}$ ($\sqrt{2}\,v_{\rm vir}\simeq670~{\rm km\,s^{-1}}$ at its edge). The associated crossing time is
\begin{equation}
t_{\rm cross}
=
\dfrac{R_{\rm cl}}{v_{\rm vir}}
\approx130~{\rm yr},
\label{eq:tcross}
\end{equation}
while the two-body relaxation time is
\begin{equation}
t_{\rm relax}
\approx
\dfrac{0.1N_{\rm cl}}
{\ln N_{\rm cl}}
\,t_{\rm cross}
\approx0.11~{\rm Myr}.
\label{eq:trelax}
\end{equation}
As evident from this relation, the relaxation time is much shorter than the available cosmic interval, $\Delta t_{\rm merge}\simeq0.54$~Gyr. The PBH cluster can therefore undergo significant internal dynamical evolution during the merger epoch. In particular, two-body relaxation can redistribute the PBHs and drive the formation of a denser central region with $n_{\rm core}>n_{\rm cl}$, thereby enhancing the local encounter rate.

GW recoil can eject merger remnants: comparable-mass binaries with modest spins receive kicks of tens to a few hundred ${\rm km\,s^{-1}}$, and tuned spin configurations up to $\sim10^3\,{\rm km\,s^{-1}}$~\citep{Campanelli2007,Gonzalez2007,Baker2006}. The central escape speed of the reference cluster is $v_{\rm esc}=\sqrt{3}\,v_{\rm vir}\simeq820\,{\rm km\,s^{-1}}$, and the recoil computed in our spin Monte Carlo ejects the central seed in $\lesssim0.4\%$ of the realizations (Appendix~\ref{sec:app_spin}). We therefore assume efficient retention and include no recoil-loss term; this may need revision for clusters with shallower potentials.

We formulate the coagulation dynamics using the characteristic cluster
density $n_{\rm cl}$ and the virial speed specified by the initial configuration, thereby providing a controlled description of the merger
dynamics at fixed cluster conditions. The cluster is treated as a closed reservoir: the central seed can capture at most the PBH mass the cluster contains, $M_{\rm cl}$, and no PBHs are supplied from outside. This fixed-environment description ceases to apply once the seed dominates the cluster potential. The gravitational sphere of influence of a seed of mass $M$, $r_h=GM/v_{\rm vir}^2$, reaches $R_{\rm cl}$ when $M\simeq M_{\rm cl}$, at which point the independent-pair capture kernel is no longer appropriate. We therefore do not follow merger-driven growth beyond the exhaustion of the reservoir.

\section{\label{sec:assembly}Hybrid Mass Assembly Model}

The assembly of massive black holes is modeled as a hybrid of two
concurrent channels. Within a sufficiently dense cluster, PBHs undergo
repeated binary formation and coalescence. The resulting evolution of
the PBH mass distribution is described by the Smoluchowski coagulation
equation, with the merger kernel determined by GW radiation capture.
The same kernel also provides the rate at which the central seed of the cluster captures the surrounding PBHs. In parallel, radiatively efficient gas accretion
can operate once seeds inhabit gas-rich galactic environments. The two channels are solved together for the central seed (Sec.~\ref{ssec:accretion_model}): mergers dominate the growth until the PBH reservoir of the cluster is exhausted, after which the seed grows by accretion alone. Since the PBHs carry no scalar charge (Sec.~\ref{ssec:screening}), every element of this assembly model is evaluated with Newton's constant $G$.

\subsection{\label{ssec:kernel}GW Radiation-Capture Merger Kernel}

In a sufficiently dense PBH cluster, initially unbound PBHs can form bound 
binaries through GW radiation capture during close hyperbolic encounters. 
For two PBHs with masses $m_i$ and $m_j$, the total and reduced masses are 
defined as
\begin{equation}
M_{ij}=m_i+m_j,
\qquad
\mu_{ij}
=
\dfrac{m_i m_j}{m_i+m_j}.
\end{equation}
Since black holes carry no scalar charge, the orbital dynamics, gravitational focusing and GW emission of a PBH pair are those of GR in every model considered here (Sec.~\ref{ssec:screening}). The energy radiated in GWs during an
encounter with relative velocity $v$ and pericenter distance $r_p$
has the dependence
\begin{equation}
\Delta E_{\rm GW}
\propto
\dfrac{
G^{7/2}
\mu_{ij}^{\,2}
M_{ij}^{\,5/2}
}{
c^5 r_p^{7/2}
}.
\label{eq:EGW}
\end{equation}
GW capture occurs when the radiated energy is sufficient to overcome the 
initial kinetic energy of the relative orbit,
\begin{equation}
\Delta E_{\rm GW}
\geq
\dfrac{1}{2}\mu_{ij}v^2.
\label{eq:capture_condition}
\end{equation}
The equality in Eq.~\eqref{eq:capture_condition} determines the largest 
pericenter distance for which capture is possible. Combining 
Eqs.~\eqref{eq:EGW} and \eqref{eq:capture_condition} gives
\begin{equation}
r_{p,\rm max}
\propto
G
\mu_{ij}^{\,2/7}
M_{ij}^{\,5/7}
c^{-10/7}
v^{-4/7}.
\label{eq:rpmax}
\end{equation}

For the nonrelativistic encounters considered here, $v\ll c$, gravitational 
focusing dominates the relation between the impact parameter and the 
pericenter distance. In this regime,
\begin{equation}
b_{\rm max}^2
\simeq
\dfrac{2GM_{ij}r_{p,\rm max}}{v^2},
\label{eq:bmax}
\end{equation}
and the corresponding capture cross section is 
$\sigma_{\rm cap}=\pi b_{\rm max}^2$. Substitution of 
Eq.~\eqref{eq:rpmax} then yields
\begin{equation}
\sigma_{\rm cap}
\propto
\dfrac{
G^{\,2}
M_{ij}^{\,12/7}
\mu_{ij}^{\,2/7}
}{
c^{10/7}v^{18/7}
}.
\label{eq:sigmacap}
\end{equation}

The PBH-PBH kernel is therefore independent of $|f_{R0}|$. The modified-gravity coupling of Eq.~\eqref{eq:Geff} applies to the particle dark matter and gas of the host halo, and enters this work only through the host environment.

The merger kernel entering the coagulation equation is the velocity-averaged capture rate, $K_{ij}=\left\langle\sigma_{\rm cap}v\right\rangle$. For an isotropic three-dimensional Maxwellian distribution of relative 
velocities, this average gives
\begin{equation}
K_{ij}
=
C_{\rm GW}
\dfrac{
G^{\,2}
(m_i+m_j)^{10/7}
m_i^{\,2/7}
m_j^{\,2/7}
}{
c^{10/7}
v_{\rm vir}^{11/7}
},
\label{eq:Kij}
\end{equation}
where
\begin{equation}
C_{\rm GW}
=
2\pi
\left(
\dfrac{85\pi}{6\sqrt{2}}
\right)^{2/7}
\dfrac{2}{\sqrt{\pi}}
\,
\Gamma\!\left(\dfrac{3}{14}\right)
\left(\dfrac{4}{3}\right)^{-11/14}.
\label{eq:Cgw}
\end{equation}

\subsection{\label{ssec:smolu}Smoluchowski Coagulation Formalism}

The nonlinear evolution of the PBH mass distribution through repeated 
binary coalescences is described by the continuous Smoluchowski coagulation 
equation \citep{Smoluchowski1916}. Let $n(m,t)$ denote the comoving number 
density of PBHs per unit mass at mass $m$ and cosmic time $t$, and let 
$K(m_1,m_2)$ denote the merger kernel for a pair of PBHs with masses 
$m_1$ and $m_2$. The mass distribution evolves according to
\begin{align}
\dfrac{\partial n(m,t)}{\partial t}
={}&
\dfrac{1}{2}
\int_0^m
K(m',m-m')
n(m',t)
n(m-m',t)
\,\dd m'
\nonumber\\
&-
n(m,t)
\int_0^\infty
K(m,m')
n(m',t)
\,\dd m' .
\label{eq:smolu_cont}
\end{align}
The first term describes the formation of PBHs of mass $m$ through mergers 
of lighter objects, whereas the second term accounts for the removal of 
PBHs of mass $m$ through mergers with the rest of the population.

For the numerical evolution, the continuous mass distribution is 
discretized on $N_{\rm bins}$ logarithmically spaced mass nodes 
$\{m_k\}$. The number of PBHs contained in mass bin $k$ is denoted by $N_k$. For a cluster of volume $V_{\rm cl}$, the merger-induced depletion rate of this bin is then
\begin{equation}
\left.
\dfrac{\dd N_k}{\dd t}
\right|_{\rm loss}
=
-\dfrac{N_k}{V_{\rm cl}}
\sum_j K_{kj}N_j,
\label{eq:loss}
\end{equation}
where $K_{kj}=K(m_k,m_j)$. The corresponding formation rate of merger 
pairs involving PBHs in bins $i$ and $j$ is
\begin{equation}
\dot{N}_{ij}^{\rm pair}
=
\dfrac{1}{V_{\rm cl}}
\begin{cases}
K_{ij}N_iN_j,
& i\neq j,\\[3pt]
\dfrac{1}{2}K_{ii}N_i^2,
& i=j,
\end{cases}
\label{eq:pair_rate}
\end{equation}
where the overdot denotes differentiation with respect to time, and the normalization of the interaction rates is fixed by Eq.~\eqref{eq:Rcl}.

A merger remnant has mass $m_{\rm new}=m_i+m_j$, which in general does not coincide with a mass-grid node. For 
$m_k\leq m_{\rm new}\leq m_{k+1}$, the remnant is therefore distributed 
between the two neighboring nodes according to the interpolation weights
\begin{equation}
w_k
=
\dfrac{
m_{k+1}-m_{\rm new}
}{
m_{k+1}-m_k
},
\qquad
w_{k+1}=1-w_k.
\label{eq:weights}
\end{equation}
These weights preserve the remnant mass exactly, since
\begin{equation}
w_km_k+w_{k+1}m_{k+1}
=
m_{\rm new}.
\end{equation}
The discretized evolution therefore conserves the PBH mass within the 
resolved mass range. Consequently, merger remnants with
$m_{\rm new}>m_{N_{\rm bins}}$ are recorded as flux through the upper mass 
boundary and excluded from the resolved mass distribution. We integrate Eq.~\eqref{eq:smolu_cont} with the GR kernel of Eq.~\eqref{eq:Kij} from $z_{\rm ini}=11.5$ to $z=5$. The resulting population $n(m,t)$ supplies the field PBHs captured by the central seed (Sec.~\ref{ssec:accretion_model}).

\subsection{\label{ssec:merger_growth}Merger-Driven Mass-Growth Rate}

The coagulation equation describes the evolution of the entire PBH mass 
distribution, whereas the seed-growth model requires the mass 
growth rate of an individual object, the central seed of the cluster. This quantity can be obtained directly 
from the merger kernel. Consider a PBH of mass $M$ embedded in the cluster. 
The rate at which it merges with PBHs in the mass interval 
$[m,m+\dd m]$ is
\begin{equation}
\dd\Gamma_{\rm merger}(M,m,t)
=
K(M,m)\,n(m,t)\,\dd m.
\label{eq:dGamma_merger}
\end{equation}
Under the assumption adopted in the coagulation model that a merger
between a PBH of mass $M$ and a companion of mass $m$ produces a remnant
of mass $M+m$, the merger-driven mass-growth rate of a PBH of mass $M$
is
\begin{equation}
\dot{M}_{\rm merger}(M,t)
=
\int_0^\infty
m\,K(M,m)\,n(m,t)\,\dd m.
\label{eq:Mdot_merger_cont}
\end{equation}
The quantity $\dot{M}_{\rm merger}$ denotes the
per-object mass-growth rate, while the total cluster mass remains
conserved under dry coalescences.

For the discretized mass distribution, Eq.~\eqref{eq:Mdot_merger_cont} 
becomes
\begin{equation}
\dot{M}_{{\rm merger},i}
=
\sum_j
K_{ij}
\dfrac{N_j}{V_{\rm cl}}
m_j,
\label{eq:Mdot_merger_disc}
\end{equation}
where $\dot{M}_{{\rm merger},i}$ is the merger-driven growth rate of a PBH 
with mass $m_i$. Equivalently, the corresponding merger timescale can be 
defined as
\begin{equation}
t_{{\rm merger},i}
=
\dfrac{m_i}
{\dot{M}_{{\rm merger},i}}.
\label{eq:tmerger}
\end{equation}
Consequently, the merger-driven growth rate is determined self-consistently
from the same interaction kernel that governs the Smoluchowski evolution,
maintaining consistency between the mass-growth and coagulation dynamics. For $M\gg m$, the kernel of Eq.~\eqref{eq:Kij} scales as $K(M,m)\propto M^{12/7}$, so that $\dot M_{\rm merger}$ grows faster than linearly with the seed mass. A seed that becomes appreciably heavier than its companions therefore captures them at an accelerating rate, a runaway that is limited only by the finite PBH content of the cluster (Sec.~\ref{ssec:accretion_model}).

\subsection{\label{ssec:numerics}Numerical Implementation and Stability}

Because the capture kernel has a superlinear mass dependence,
$K_{ij}\sim M_{ij}^{10/7}$, the coagulation equations can become
numerically stiff as the high-mass tail develops. We therefore solve
the discretized Smoluchowski equation with an implicit,
variable-order backward differentiation formula (BDF) scheme on a
logarithmically spaced mass grid. We use 300 mass bins spanning
\textcolor{blue}{$0.1\,m_0$} to $10^6\,m_0$, with relative and absolute tolerances of
${\rm rtol}=2\times10^{-7}$ and ${\rm atol}=10^{-10}$, respectively. The seed-growth equation of Sec.~\ref{ssec:accretion_model} is integrated with an adaptive LSODA scheme at the same relative tolerance.
The upper mass boundary is chosen sufficiently far from the masses
reached during the evolution for the cluster parameters considered.

We monitor the numerical completeness of the mass distribution through
the unresolved mass fraction
\begin{equation}
f_{\rm unres}(t)
=
1-\dfrac{M_{\rm tot}(t)}{M_{\rm tot}(t_{\rm ini})},
\qquad
M_{\rm tot}(t)=\sum_k m_kN_k(t),
\label{eq:funres}
\end{equation}
which measures the fraction of the initial cluster mass that crosses
the upper mass boundary. For the reference cluster, $f_{\rm unres}<10^{-5}$ over the full integration, so boundary losses are negligible. We also
test the numerical convergence by increasing the number of mass bins from 300 to 400 and the upper mass boundary from $10^6\,m_0$ to $10^7\,m_0$. These changes alter $M_{\rm max}$ at $z=6$ and the seed masses at $z=6$ and $z=5$ by less than $1\%$, and the redshift of reservoir exhaustion by less than $0.01$, demonstrating that the
reported results are insensitive to the adopted numerical resolution.

\subsection{\label{ssec:accretion_model}Coupled Merger and Accretion Growth of the Central Seed}

The central seed of each cluster grows through two concurrent channels: the capture of field PBHs, and radiatively efficient gas accretion once it inhabits a gas-rich environment. Its total mass-growth rate is written as the sum of the two contributions,
\begin{equation}
\dfrac{\dd M}{\dd t}
=
\dot{M}_{\rm merger}
+
\dot{M}_{\rm acc}.
\label{eq:hybrid}
\end{equation}
The accretion contribution is parameterized by the Eddington ratio 
$\Edd\equiv L/L_{\rm Edd}$ and the radiative efficiency $\epsilon$:
\begin{equation}
\dot{M}_{\rm acc}
=
\dfrac{1-\epsilon}{\epsilon}
\dfrac{\Edd M}{t_{\rm Edd}},
\qquad
t_{\rm Edd}\simeq0.45\,{\rm Gyr},
\label{eq:Mdot_acc}
\end{equation}
where $t_{\rm Edd}$ is the Salpeter time and the factor 
$(1-\epsilon)/\epsilon$ accounts for the fraction of the accreted 
rest-mass energy that contributes to black-hole growth after radiative 
losses. We take $\epsilon=0.10$ and treat $\Edd$ as a time-averaged parameter. We apply efficient accretion only to the central seed, which sits at the bottom of the host potential where gas accumulates. The bulk of the PBHs, distributed through the cluster and surrounded by lower-density gas, accrete far below the Eddington rate, consistent with the strong suppression of PBH accretion in the high-redshift intergalactic medium~\citep{AliHaimoud2017}.

The merger term is the capture rate of Eq.~\eqref{eq:Mdot_merger_cont}, evaluated with the field population $n(m,t)$ of the Smoluchowski solution and weighted by the fraction of the cluster that has not yet been captured. Combining Eqs.~\eqref{eq:Mdot_merger_cont} and \eqref{eq:Mdot_acc}, Eq.~\eqref{eq:hybrid} becomes
\begin{eqnarray}
\dfrac{\dd M}{\dd t}
=
\left(1-\dfrac{M_{\rm cap}}{M_{\rm cl}}\right)
\int_0^\infty
m\,K(M,m)\,n(m,t)\,\dd m\nonumber\\
+
\dfrac{1-\epsilon}{\epsilon}
\dfrac{\Edd M}{t_{\rm Edd}},
\label{eq:full_growth}
\end{eqnarray}
where
\begin{equation}
M_{\rm cap}(t)=\int_{t_{\rm ini}}^{t}\dot M_{\rm merger}\,\dd t'
\label{eq:Mcap}
\end{equation}
is the PBH mass already captured by the seed. The reservoir factor $(1-M_{\rm cap}/M_{\rm cl})$ enforces the closed-cluster assumption of Sec.~\ref{ssec:cluster_dynamics}: the seed captures at most $M_{\rm cl}=N_{\rm cl}\langle m\rangle=N_{\rm cl}\,m_c\,e^{\sigma_{\ln m}^{2}/2}=3.25\times10^{6}\,\Msun$, the total mass of the initial distribution, and the merger term vanishes once the reservoir is exhausted. The field population entering $\dot M_{\rm merger}$ is taken from the coagulation of Sec.~\ref{ssec:smolu}, solved independently of the seed, and the reservoir factor represents its depletion by the seed. A self-consistent calculation in which each captured PBH is removed from the field reproduces the exhaustion redshifts to within $0.01$ and the seed masses to within $0.2\%$ (Appendix~\ref{sec:app_numerics}). The field population itself, however, is not depleted in the independent coagulation, so that field-field mergers continue after the seed has consumed its cluster; this affects the field-field merger rate used for the SGWB (Appendix~\ref{sec:app_sgwb}). No redshift cutoff is imposed on the merger channel. The seed starts at $z_{\rm ini}$ as the most massive PBH of the initial distribution, $N(>M)=1$ per cluster, which gives $M(t_{\rm ini})\simeq170\,\Msun$ for the lognormal of Eq.~\eqref{eq:lognormal}. Equation~\eqref{eq:full_growth} is integrated jointly for both channels; the merger and accretion contributions are not combined multiplicatively.

Once the reservoir is exhausted at time $t_{\rm exh}$, the seed grows by accretion alone. For a constant Eddington ratio, integration from $M(t_{\rm exh})\simeq M_{\rm cl}$ to a target mass $M_{\rm target}$ over the interval $\Delta t=t_{\rm obs}-t_{\rm exh}$ gives
\begin{equation}
M_{\rm target}
=
M(t_{\rm exh})
\exp\left[
\dfrac{1-\epsilon}{\epsilon}
\dfrac{\Edd\Delta t}{t_{\rm Edd}}
\right],
\label{eq:accretion_solution}
\end{equation}
and the corresponding Eddington ratio required to reach $M_{\rm target}$ by $z_{\rm obs}$ is
\begin{equation}
\lambda_{\rm Edd,\; req}
=
\dfrac{\epsilon}{1-\epsilon}
\dfrac{t_{\rm Edd}}{\Delta t}
\ln\left(
\dfrac{M_{\rm target}}
{M(t_{\rm exh})}
\right).
\label{eq:lambda_req}
\end{equation}
Because $t_{\rm exh}$ itself depends on $\Edd$ through the coupled growth, Eq.~\eqref{eq:lambda_req} must be solved together with Eq.~\eqref{eq:full_growth}; it applies only after the reservoir is exhausted. Where the reservoir is not exhausted by $z_{\rm obs}$, the seed mass is obtained directly from Eq.~\eqref{eq:full_growth}.

\section{\label{sec:diagnostics}Observables and Diagnostics}

The primary observable of this work is the abundance of active seeds, which we compare with the measured black-hole mass function and the abundance of LRDs. This section defines that comparison. Three secondary diagnostics, the spins of the seeds, the stochastic GW background from mergers within the clusters, and the spatial clustering of PBHs, rest on additional assumptions and are developed in Appendices~\ref{sec:app_spin}, \ref{sec:app_sgwb} and~\ref{sec:app_clustering}, respectively.

\subsection{\label{ssec:bhmf_framework}Black-Hole Mass Function and High-Redshift Data}

The coagulation solver yields the PBH mass distribution within each cluster. Its mass-weighted form is
\begin{equation}
\dfrac{\dd F}{\dd\ln m}(m,z)
=
\dfrac{m^2\,N(m,z)}{\int_0^\infty m'^2\,N(m',z)\,\dd\ln m'},
\label{eq:dfdlnm}
\end{equation}
where $N(m,z)$ is the number of PBHs per unit mass within the cluster, so that $\int(\dd F/\dd\ln m)\,\dd\ln m=1$. The maximum black-hole mass of the field population, $M_{\rm max}(z)$, follows from the cumulative cluster population
\begin{equation}
N(>M_{\rm max},z) = \int_{M_{\rm max}}^\infty N(m,z)\,\dd m = 1,
\label{eq:mmax_def}
\end{equation}
solved for $M_{\rm max}(z)$ by log-linear interpolation.

To convert the cluster-level result into a cosmological active BHMF, we count clusters rather than individual PBHs. Each cluster hosts one efficiently accreting central seed (Sec.~\ref{ssec:accretion_model}). The comoving number density of clusters is
\begin{eqnarray}
n_{\rm cl}^{\rm cos}
=
\dfrac{f_{\rm dense}\,f_{\rm PBH}\,\rho_{\rm DM}}{M_{\rm cl}}
\simeq
10\,f_{\rm dense}\left(\dfrac{f_{\rm PBH}}{10^{-3}}\right)\nonumber\\
\times\left(\dfrac{3.25\times10^6\,\Msun}{M_{\rm cl}}\right){\rm Mpc}^{-3},
\label{eq:nPBH}
\end{eqnarray}
where $\rho_{\rm DM} = \Omega_{\rm DM}\rho_{\rm crit,0} \approx 3.3 \times 10^{10}\,M_\odot\,{\rm Mpc}^{-3}$ is the present-day comoving dark-matter density~\citep{Planck2018}, $f_{\rm PBH}$ is the PBH dark-matter fraction, and $f_{\rm dense}$ is the fraction of PBHs residing in dense clusters. Equation~\eqref{eq:nPBH} is fixed by the initial conditions and is not affected by accretion: accretion increases the mass of each seed but not their number, while mergers within a cluster do not change the number of clusters. This normalization conserves the number of seeds, and the mass accreted onto active seeds, $\simeq3\times10^{4}\,\Msun\,{\rm Mpc^{-3}}$, is far below the baryon density. Applying the accretion growth to all PBHs instead would add a mass density $f_{\rm PBH}\rho_{\rm DM}(f_{\rm acc}-1)$, which exceeds $\Omega_b\rho_c$ once the mean growth factor reaches $f_{\rm acc}\simeq190$ at $f_{\rm PBH}=10^{-3}$. Hosts in the upper tail of the Eddington-ratio distribution grow by more than this factor, so applying the accretion growth to every PBH would exceed the baryon budget; the number density must therefore be normalized per cluster.

A seed can accrete efficiently only if its cluster resides in a host halo able to retain it and to supply gas. We take this to require a host mass $M_h\geq M_{\rm min}=10^{8}\,\Msun$, comparable to the atomic-cooling threshold at these redshifts, and write the fraction of clusters in such hosts as the collapsed mass fraction
\begin{equation}
F_{\rm coll}(>M_{\rm min},z)
=
\dfrac{1}{\rho_{m}}\int_{M_{\rm min}}^{\infty}M\,\dfrac{\dd n}{\dd M}\,\dd M,
\label{eq:Fcoll}
\end{equation}
computed from the~\citet{ShethTormen1999} mass function. This is the only place where the scalaron enters the BHMF. In the unscreened regime, the linear growth of particle-dark-matter perturbations is scale dependent, with $\mu(k,a)$ of Eq.~\eqref{eq:mu_hs} and the Compton wavelength of Eq.~\eqref{eq:lambdaC}; we integrate the linear growth equation for each wavenumber and evaluate $\sigma(M,z)$ with the resulting power spectrum. Screening of the hosts is included by interpolating the mass function between the GR and unscreened limits with the host screening factor of Eq.~\eqref{eq:fscr}, $\dd n/\dd M=\dd n_{\rm GR}/\dd M+f_{\rm scr}^{\rm host}(\dd n_{\rm unscr}/\dd M-\dd n_{\rm GR}/\dd M)$. The GR mass function uses the collapse threshold $\delta_c^{\rm GR}(z)$ of~\citet{NakamuraSuto1997}, and the unscreened one the threshold for collapse with $G_{\rm eff}=4G/3$, $\delta_{c,\rm unscr}(z)=1.7063\,[1-0.0136\log_{10}(1+(\Omega_m^{-1}-1)(1+z)^{-3})]$~\citep{Schmidt2009,Hagstotz2019}, which is slightly higher than in GR ($1.706$ versus $1.686$ at $z=5.5$): the enhanced abundance in $f(R)$ arises from the faster linear growth, not from a lower threshold. Calibrated $f(R)$ thresholds with explicit mass dependence~\citep{Kopp2013,Hagstotz2019} are fitted to cluster-scale halos at $z\lesssim1$ and do not extend to the $10^{8}$-$10^{10}\,\Msun$ hosts at $z\gtrsim5$ considered here.

The masses of the seeds follow from Eq.~\eqref{eq:full_growth}. Throughout this work, $\lambda_{\rm Edd}$ denotes the time-averaged Eddington ratio of an individual seed, as in Eq.~\eqref{eq:Mdot_acc} and Fig.~\ref{fig:growth}. Since the accretion rate differs between hosts, $\lambda_{\rm Edd}$ is distributed across the population; we take this distribution to be log-normal, with median $\bar\lambda_{\rm Edd}$ and width $\sigma$ in dex, and map each $\lambda_{\rm Edd}$ to a seed mass $M_{\rm seed}(z;\lambda_{\rm Edd})$. We adopt $\sigma=0.2$~dex. The active BHMF per decade in mass is then
\begin{equation}
\Phi_{\rm active}(M,z)
=
n_0\,F_{\rm coll}(>M_{\rm min},z)\,\dfrac{\dd P_{\rm seed}}{\dd\log_{10}M}(M,z).
\label{eq:phi_theory}
\end{equation}
In the above relation, $n_0\equiv\delta\,n_{\rm cl}^{\rm cos}$, $\delta\leq1$ is the AGN duty cycle and $\dd P_{\rm seed}/\dd\log_{10}M=p(\log_{10}\lambda_{\rm Edd})\,|\dd\log_{10}M_{\rm seed}/\dd\log_{10}\lambda_{\rm Edd}|^{-1}$ is the normalized distribution of seed masses. Since each measurement averages over a redshift range, we compare the data with the comoving-volume-weighted average of Eq.~\eqref{eq:phi_theory} over the same range,
\begin{equation}
\bar\Phi(M)=\dfrac{\int_{z_1}^{z_2}\Phi_{\rm active}(M,z)\,(\dd V_c/\dd z)\,\dd z}{\int_{z_1}^{z_2}(\dd V_c/\dd z)\,\dd z},
\label{eq:phi_avg}
\end{equation}
with $3.5<z<6$ for the measurements of~\cite{Taylor2025} and~\cite{Matthee2024}, and $6<z<7$ for the lower limits from the highest-redshift objects of~\cite{Taylor2025}. The normalization $n_0$ and the median $\bar\lambda_{\rm Edd}$ for $3.5<z<6$ are fitted to the measurements in GR and then held fixed for the $f(R)$ models. The black holes at $z>6$ require a higher median, and we allow the median to evolve: for $6<z<7$ we adopt the smallest value consistent with the lower limits.

JWST spectroscopy of broad-line AGN and LRDs, with inferred black-hole masses of $10^6$-$10^9\,\Msun$ at $z\gtrsim4$~\citep{Labbe2023,Matthee2024,Taylor2025,Jones2025}, provides a direct test of Eq.~\eqref{eq:phi_theory}. We use two independent, completeness-corrected BHMF measurements. The first is the broad-line H$\alpha$ active BHMF at $z=4.2$-$5.5$ from~\cite{Matthee2024}, based on 20 emitters from the EIGER and FRESCO surveys ($V_c=5.7\times10^5\,{\rm cMpc}^3$). The second is the broad-line AGN mass function at $z=3.5$-$6.0$ from~\cite{Taylor2025}, based on 62 CEERS and RUBIES NIRSpec detections ($V_c=1.60\times10^6\,{\rm cMpc}^3$); both lie within the range $3.5<z<6$ used in Eq.~\eqref{eq:phi_avg}. At $z=6$-$7$, where a complete BHMF is not yet available, we instead use the individual detections of~\cite{Taylor2025} to derive empirical lower limits on the number density ($V_c\approx4.4\times10^5\,{\rm cMpc}^3$), and compare the high-mass tail of our prediction with the LRD mass estimates compiled by~\cite{Jones2025}.

\section{\label{sec:results}Results}

We first present the growth of the central seed and the conditions for runaway growth (Secs.~\ref{ssec:growth_res} and~\ref{ssec:full_accretion}). We then calibrate the model to the measured black-hole mass function (Sec.~\ref{ssec:bhmf_results}), compare the calibrated model with the abundance of LRDs and quantify the environmental effect of the scalar field (Sec.~\ref{ssec:lowz}), and summarize the remnant spins (Sec.~\ref{ssec:spin_summary}).

\subsection{\label{ssec:growth_res}Central-Seed Growth and Host Screening}

\begin{figure}[!ht]
\centering
\includegraphics[width=\columnwidth]{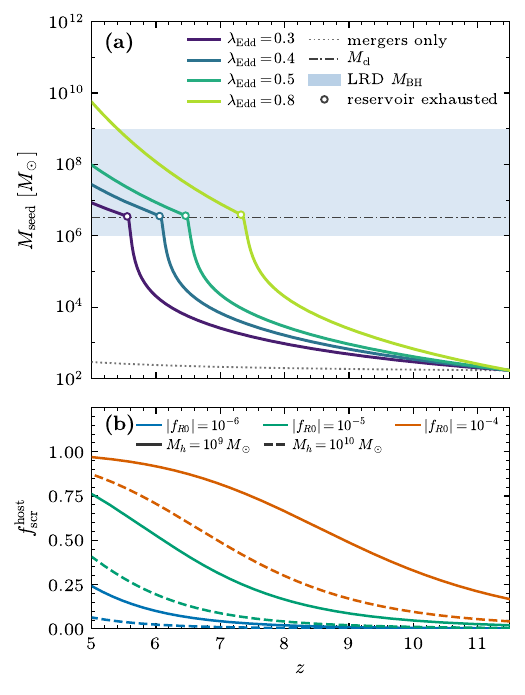}
\caption{\label{fig:growth}\textbf{(a)}~Central-seed mass $M_{\rm seed}(z)$ from the coupled merger and accretion growth of Eq.~\eqref{eq:full_growth}, for $N_{\rm cl}=10^{5}$ PBHs of initial mass $30\,\Msun$ at $n_{\rm cl}=10^{8}\,{\rm pc}^{-3}$ (adopted; see Sec.~\ref{ssec:cluster_dynamics}), for four time-averaged Eddington ratios $\lambda_{\rm Edd}$ (colour) with $\epsilon=0.10$. The dotted line is the mergers-only track. Open markers show where the cluster reservoir is exhausted ($99\%$ of $M_{\rm cl}=3.25\times10^{6}\,\Msun$ captured; dash-dotted line). In the $f(R)$ models considered, the scalar field acts on the medium surrounding the PBHs rather than on the PBH-PBH interaction, so these tracks hold for every $|f_{R0}|$; its effect enters through the host halo, panel (b). The shaded band marks the LRD black-hole mass range, $10^{6}\text{-}10^{9}\,\Msun$. \textbf{(b)}~Screening factor $f_{\rm scr}$ of Eq.~\eqref{eq:fscr} for particle-dark-matter host halos of mass $10^{9}$ (solid) and $10^{10}\,\Msun$ (dashed), for $|f_{R0}|=10^{-6}$, $10^{-5}$ and $10^{-4}$, evaluated with the background field $|f_R(z)|$ of Eq.~\eqref{eq:fRz}. Unscreened hosts have $f_{\rm scr}\to1$.}
\end{figure}

Figure~\ref{fig:growth}(a) shows the mass of the central seed obtained from Eq.~\eqref{eq:full_growth} for the reference cluster and four time-averaged Eddington ratios. Since the scalar field acts on the medium surrounding the PBHs rather than on their mutual capture, these tracks hold for all three Hu-Sawicki models. They are conditional on the adopted cluster parameters of Sec.~\ref{ssec:cluster_dynamics}.

Without accretion (dotted line), the central seed grows only slowly, from $\simeq170\,\Msun$ at $z_{\rm ini}$ to $\simeq240\,\Msun$ at $z=6$ and $\simeq290\,\Msun$ at $z=5$. For comparison, the most massive object of the coagulating field population reaches $M_{\rm max}\simeq490\,\Msun$ at $z=6$ [Eq.~\eqref{eq:mmax_def}]; the two differ because $M_{\rm max}$ is a statistic of the whole population, whereas the dotted line follows a single object. Dry mergers alone therefore remain far below the LRD mass range.

Concurrent accretion changes this picture qualitatively. As the seed grows, its capture rate rises approximately as $M^{12/7}$ (Sec.~\ref{ssec:merger_growth}), and once the seed is appreciably heavier than the surrounding PBHs, the capture of the cluster runs away. This produces the steep rise in each track. The tracks then turn over where the reservoir is exhausted (open markers), when $M_{\rm cap}\to M_{\rm cl}$ and the merger term of Eq.~\eqref{eq:full_growth} vanishes. Beyond this point, which is also where the seed comes to dominate the cluster potential (Sec.~\ref{ssec:cluster_dynamics}), the growth is purely accretion-driven and follows Eq.~\eqref{eq:accretion_solution}. Both features are consequences of the model: the runaway follows from the mass dependence of the kernel, and the turnover from the closed-reservoir assumption.

For $\lambda_{\rm Edd}=0.3$, $0.4$, $0.5$ and $0.8$, the reservoir is exhausted at $z\simeq5.55$, $6.06$, $6.46$ and $7.32$, respectively, when the seed has reached $M\simeq M_{\rm cl}=3.25\times10^{6}\,\Msun$. The merger channel thus assembles a seed of order the cluster mass, and no further PBH mergers are possible once the cluster is consumed. Growth above $M_{\rm cl}$ is accretion-driven, so the massive end of the population is set by the accretion history rather than by repeated mergers of comparably massive black holes. By $z=5$, the seed reaches $\simeq8\times10^{6}$, $3\times10^{7}$ and $1\times10^{8}\,\Msun$ for $\lambda_{\rm Edd}=0.3$, $0.4$ and $0.5$, respectively, within the LRD range of $10^{6}\text{-}10^{9}\,\Msun$. For $\lambda_{\rm Edd}=0.8$, the seed already reaches $\simeq10^{8}\,\Msun$ at $z=6$ and exceeds the LRD range by $z=5$. LRD masses at $z\simeq5\text{-}6.5$ are therefore reached with sub-Eddington time-averaged accretion. The Eddington ratios used in the BHMF comparison of Sec.~\ref{ssec:bhmf_results} must be chosen consistently with these tracks.

Figure~\ref{fig:growth}(b) shows where the scalaron enters the calculation: the screening factor of the host halo, Eq.~\eqref{eq:fscr}, evaluated with the background field $|f_R(z)|$ of Eq.~\eqref{eq:fRz} and the host potential of Eq.~\eqref{eq:host_halo}. Because $|f_R(z)|$ grows toward low redshift, hosts become progressively less screened as $z$ decreases. For a $10^{9}\,\Msun$ host, $f_{\rm scr}\simeq0.002$, $0.02$ and $0.17$ at $z=11.5$ for $|f_{R0}|=10^{-6}$, $10^{-5}$ and $10^{-4}$, rising to $\simeq0.24$, $0.76$ and $0.97$ at $z=5$. A $10^{10}\,\Msun$ host, with a deeper potential, remains more strongly screened at all redshifts. The modification of the host environment therefore grows with time and is strongest for the largest field amplitude. Its consequences for the PBH population enter through the host-halo abundance and the BHMF (Sec.~\ref{ssec:bhmf_results}), not through the merger kernel.

\subsection{\label{ssec:full_accretion}Accretion Parameter Space}

\begin{figure*}[!ht]
\centering
\includegraphics[width=\textwidth]{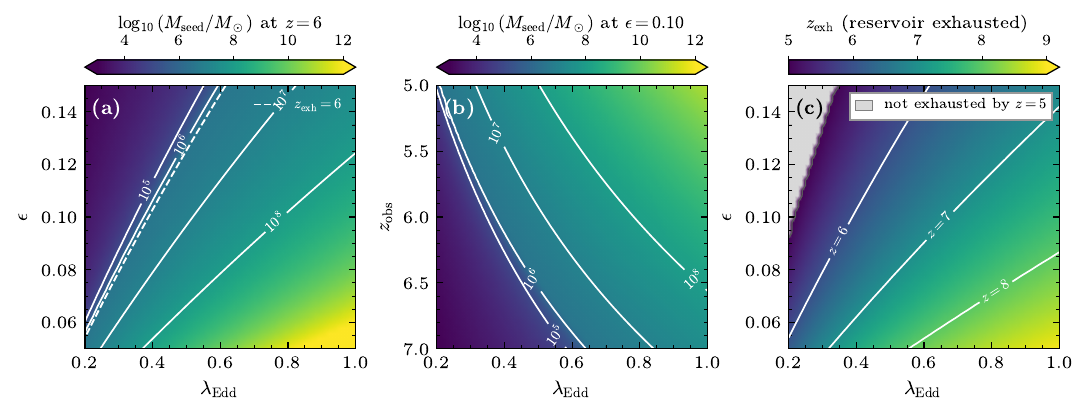}
\caption{Accretion parameter space of the central seed, from the coupled merger and accretion growth of Eq.~\eqref{eq:full_growth} for the reference cluster. The scalar field acts on the medium surrounding the PBHs, so these maps hold for every $|f_{R0}|$ considered. White contours mark $M_{\rm seed}=10^{5}$, $10^{6}$, $10^{7}$ and $10^{8}\,\Msun$.
  \textbf{(a)}~$\log_{10}(M_{\rm seed}/\Msun)$ at $z=6$ as a function of $\lambda_{\rm Edd}$ and $\epsilon$. The dashed line marks reservoir exhaustion at $z=6$.
  \textbf{(b)}~$\log_{10}(M_{\rm seed}/\Msun)$ as a function of $\lambda_{\rm Edd}$ and the observation redshift $z_{\rm obs}$, at $\epsilon=0.10$.
  \textbf{(c)}~Redshift $z_{\rm exh}$ at which the cluster reservoir is exhausted ($99\%$ of $M_{\rm cl}$ captured), as a function of $\lambda_{\rm Edd}$ and $\epsilon$. In the grey region the reservoir is not exhausted by $z=5$.}
\label{fig:reqlamb}
\end{figure*}

In Fig.~\ref{fig:reqlamb}, we have mapped the seed mass over the accretion parameters entering Eq.~\eqref{eq:Mdot_acc}. Every point is an independent solution of the coupled growth equation, Eq.~\eqref{eq:full_growth}, for the reference cluster. These maps hold for all three Hu-Sawicki models, since the scalar field acts on the surrounding medium; its effect on the population enters through the host abundance (Secs.~\ref{ssec:bhmf_results} and~\ref{ssec:lowz}). White contours mark $M_{\rm seed}=10^{5}$, $10^{6}$, $10^{7}$ and $10^{8}\,\Msun$.

In Fig.~\ref{fig:reqlamb}(a), the seed mass at $z=6$ is shown over $\lambda_{\rm Edd}\in[0.2,1]$ and $\epsilon\in[0.05,0.15]$. The accretion term depends on the two parameters only through the combination $\lambda_{\rm Edd}(1-\epsilon)/\epsilon$, so the contours run diagonally: a lower radiative efficiency is equivalent to a higher Eddington ratio. The $10^{5}$ and $10^{6}\,\Msun$ contours lie close together, and close to the dashed line marking reservoir exhaustion at $z=6$. This crowding is the runaway of Sec.~\ref{ssec:growth_res}: just above this threshold, the seed captures the whole cluster within a short interval and its mass jumps to $\sim M_{\rm cl}$. For $\epsilon=0.10$, $M_{\rm seed}(z=6)$ reaches $10^{6}$, $10^{7}$ and $10^{8}\,\Msun$ for $\lambda_{\rm Edd}\simeq0.38$, $0.52$ and $0.79$, respectively. At fixed $\lambda_{\rm Edd}=0.5$, lowering $\epsilon$ from $0.15$ to $0.05$ raises the seed mass at $z=6$ from $\simeq3\times10^{4}$ to $\simeq10^{9}\,\Msun$. The radiative efficiency is therefore as important a parameter as the Eddington ratio.

In Fig.~\ref{fig:reqlamb}(b), the seed mass is shown as a function of the observation redshift $z_{\rm obs}\in[5,7]$ and $\lambda_{\rm Edd}$, at $\epsilon=0.10$. A lower $z_{\rm obs}$ allows a longer growth time, so the Eddington ratio required for a given mass decreases toward low redshift. A seed of $10^{7}\,\Msun$ requires $\lambda_{\rm Edd}\simeq0.32$, $0.52$ and $0.85$ at $z_{\rm obs}=5$, $6$ and $7$, respectively, and a seed of $10^{8}\,\Msun$ requires $\lambda_{\rm Edd}\simeq0.51$ at $z_{\rm obs}=5$, rising to $\simeq0.98$ at $z_{\rm obs}=6.5$. LRD masses at $z\simeq5\text{-}6$ are therefore reached with sub-Eddington time-averaged accretion. At $z\gtrsim6.5$, near-Eddington rates are needed, and $10^{8}\,\Msun$ is not reached by $z=7$ for $\lambda_{\rm Edd}\le1$.

Figure~\ref{fig:reqlamb}(c) shows the redshift $z_{\rm exh}$ at which the cluster reservoir is exhausted. For $\epsilon=0.10$, the reservoir is exhausted before $z=5$ for $\lambda_{\rm Edd}\gtrsim0.22$, and the exhaustion moves to higher redshift as accretion becomes more efficient, reaching $z_{\rm exh}\simeq9$ in the lower-right corner of the panel. In the grey region, the seed does not consume its cluster by $z=5$, and growth through mergers remains incomplete. The merger channel thus delivers a seed of order $M_{\rm cl}$ over most of the parameter space considered, and the redshift at which it does so is set by the accretion parameters.

\subsection{\label{ssec:bhmf_results}Active Black-Hole Mass Function}

\begin{figure*}[!ht]
\centering
\includegraphics[width=0.83\textwidth]{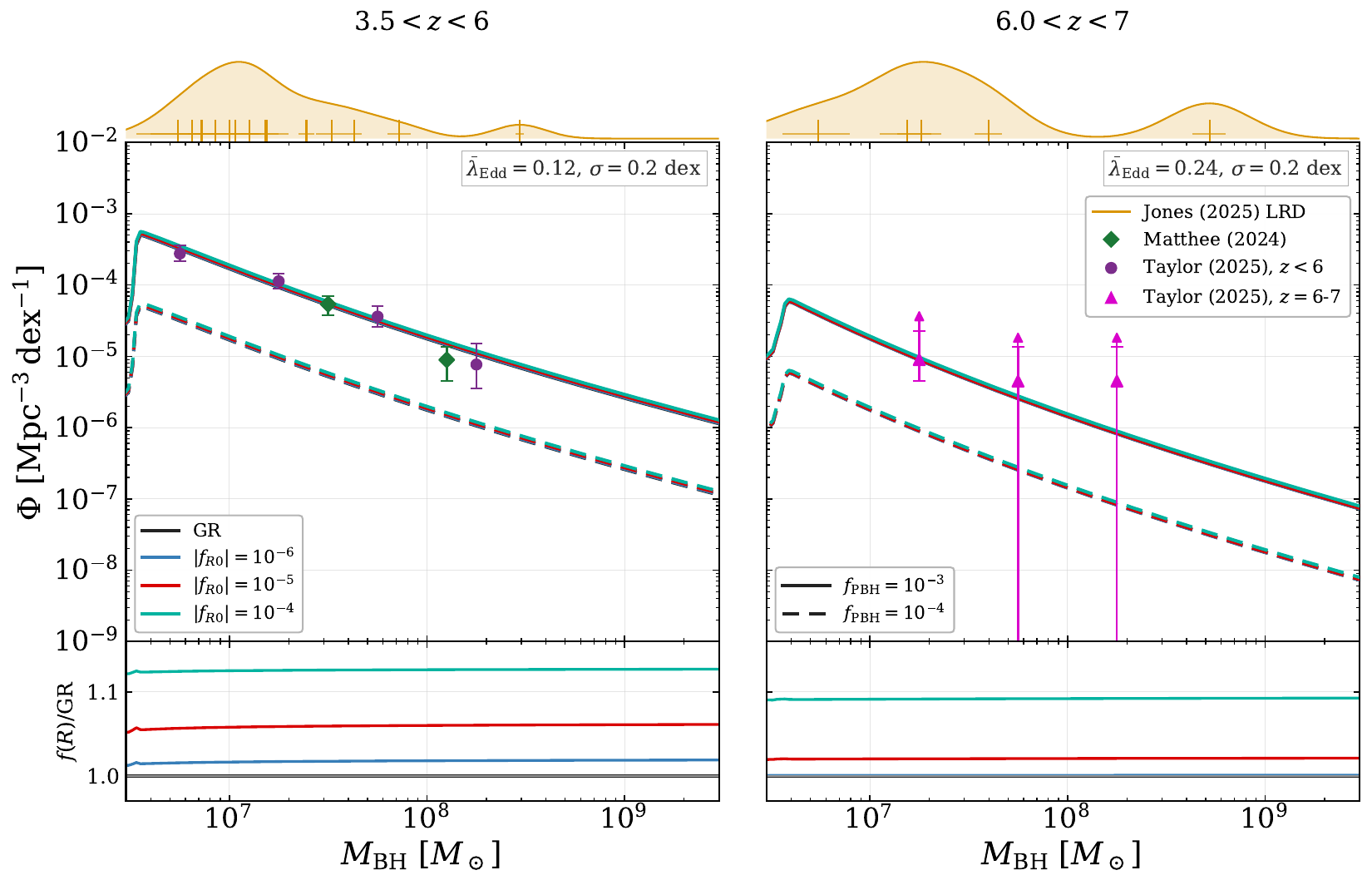}
\caption{Active black-hole mass function of PBH-cluster seeds, volume-averaged over $3.5<z<6$ (\textbf{left}) and $6<z<7$ (\textbf{right}), Eq.~\eqref{eq:phi_avg}, for GR and Hu-Sawicki models with $|f_{R0}|=10^{-6}$, $10^{-5}$ and $10^{-4}$. Seed masses follow from the coupled growth of Eq.~\eqref{eq:full_growth} for a log-normal distribution of time-averaged Eddington ratios $\lambda_{\rm Edd}$ across hosts (width $0.2$~dex), with median $\bar\lambda_{\rm Edd}=0.12$ for $3.5<z<6$ (calibrated) and \textcolor{blue}{$0.245$} for $6<z<7$ (the minimum allowed by the lower limits). The number density of active seeds scales with the collapsed fraction of host halos above $10^{8}\,\Msun$, Eq.~\eqref{eq:Fcoll}. The normalization is fitted to the $3.5<z<6$ data in GR for $f_{\rm PBH}=10^{-3}$ (solid) and held fixed for the $f(R)$ models; dashed curves show $f_{\rm PBH}=10^{-4}$ at the same $\delta f_{\rm dense}$. Data: completeness-corrected measurements of~\cite{Matthee2024} and~\cite{Taylor2025}, lower limits from the $z=6$-$7$ objects of~\cite{Taylor2025}, and the LRD masses of~\cite{Jones2025} (top strips). Lower panels: ratio of the $f(R)$ to the GR mass function.}
\label{fig:bhmf}
\end{figure*}

In Fig.~\ref{fig:bhmf}, we have compared the volume-averaged active BHMF of Eq.~\eqref{eq:phi_avg} with JWST measurements for $3.5<z<6$ and $6<z<7$. Completeness-corrected measurements from~\cite{Matthee2024} and~\cite{Taylor2025} are shown in the first range, and lower limits from the $z=6$-$7$ objects of~\cite{Taylor2025} in the second; the kernel-density strips along the top show the LRD masses of~\cite{Jones2025}. All quantities are shown per decade in mass.

We calibrate the model on the mass function. For $3.5<z<6$, the fit to the six measurements fixes $\bar\lambda_{\rm Edd}\simeq0.12$ and $n_0\simeq3\times10^{-3}\,{\rm Mpc}^{-3}$, with $\chi^2\simeq2.8$; these parameters are then held fixed in all subsequent comparisons. The model reproduces both the amplitude and the slope of the observed BHMF between $10^{6.75}$ and $10^{8.25}\,\Msun$. The slope follows from the distribution of Eddington ratios: since $M_{\rm seed}$ grows exponentially with $\lambda_{\rm Edd}$ once the reservoir is exhausted, the log-normal tail of $\lambda_{\rm Edd}$ maps onto a power-law-like decline of $\Phi$ with mass, in line with the featureless shape reported by~\cite{Taylor2025}. The median and width are not separately constrained by these data: widths of $0.15$, $0.2$, $0.3$ and $0.4$~dex are fitted with $\bar\lambda_{\rm Edd}\simeq0.15$, $0.12$, $0.07$ and $0.03$ ($\chi^2\simeq1.1$, $2.8$, $5.1$ and $6.3$). The robust quantity is the Eddington ratio of the hosts that produce the observed masses, which is fixed by the growth tracks: at the effective redshift $z\simeq4.75$, black holes of $10^{6.75}$-$10^{8.25}\,\Msun$ require $\lambda_{\rm Edd}\simeq0.23$-$0.49$, within the range of the tracks in Fig.~\ref{fig:growth}(a), independently of the width. For the adopted distribution, these hosts lie $1.3$-$3.0\sigma$ above the median. Hosts with $\lambda_{\rm Edd}\lesssim0.18$, including the median, remain below the runaway threshold and contain seeds of $\lesssim10^{5}\,\Msun$. The seeds that consumed their cluster produce a sharp lower edge of the BHMF at $M\simeq M_{\rm cl}$, near $3\times10^{6}\,\Msun$, below the range currently probed.

The fitted normalization implies $\delta f_{\rm dense}f_{\rm PBH}\simeq3\times10^{-7}$, i.e., $\delta f_{\rm dense}\simeq3\times10^{-4}$ for $f_{\rm PBH}=10^{-3}$. Since $\Phi_{\rm active}$ depends on the PBH abundance only through this product, the data constrain $\delta f_{\rm dense}f_{\rm PBH}$ rather than $f_{\rm PBH}$ alone. At the same $\delta f_{\rm dense}$, a fraction $f_{\rm PBH}=10^{-4}$ (dashed curves) falls an order of magnitude below the measurements; reproducing them would require $\delta f_{\rm dense}\simeq3\times10^{-3}$, which remains plausible. The BHMF therefore requires $f_{\rm PBH}\gtrsim3\times10^{-7}$ for $\delta f_{\rm dense}\leq1$, and the viable range extends to the abundance limit $f_{\rm PBH}\lesssim10^{-3}$ for $\sim30\,\Msun$ PBHs~\citep{Hutsi2021}. The mass density in active seeds is $\simeq3\times10^{4}\,\Msun\,{\rm Mpc}^{-3}$ for $3.5<z<6$, more than five orders of magnitude below the baryon density.

For $6<z<7$ [right panel of Fig.~\ref{fig:bhmf}], a distribution with the same median would fall below the lower limits by more than an order of magnitude: the shorter growth time leaves too few seeds above $10^{7}\,\Msun$. No width between $0.2$ and $0.5$~dex removes this deficit with a non-evolving median. The lower limits are met at the $84\%$ Poisson level for a median $\bar\lambda_{\rm Edd}\geq0.25$ at $6<z<7$, about twice the value for $3.5<z<6$; with this minimum value, the prediction matches the limit at $\sim2\times10^{7}\,\Msun$ and lies within the single-object Poisson uncertainties of the limits at $\sim6\times10^{7}$ and $\sim2\times10^{8}\,\Msun$. The BHMF thus requires more efficient accretion onto the seeds observed at $z>6$, consistent with the higher gas fractions and inflow rates expected at earlier times. A complete sample at $z\gtrsim6$ would turn this lower bound into a measurement of the evolution.

The modified-gravity imprint enters through $F_{\rm coll}$ alone and is therefore a mass-independent rescaling of the BHMF, shown in the lower panels of Fig.~\ref{fig:bhmf}. For $3.5<z<6$, the volume-averaged BHMF increases by $1.7\%$, $5.9\%$ and $12.6\%$ for $|f_{R0}|=10^{-6}$, $10^{-5}$ and $10^{-4}$, respectively; for $6<z<7$ the increase is $0.1\%$, $2.1\%$ and $9.2\%$. The enhancement is larger in the lower-redshift range because the background field grows with time. Using the GR threshold for the unscreened mass function would raise these values only slightly; the enhancement is limited by the background field rather than by the collapse model. These enhancements are small because the background field is suppressed at high redshift [Eq.~\eqref{eq:fRz}] and the scale-dependent growth has had little time to act, consistent with the conclusion that viable $f(R)$ models cannot substantially raise the abundance of high-redshift structures~\citep{WangLiu2023}. Because the enhancement is a mass-independent rescaling, it leaves the shape of the mass function and the Eddington-ratio evolution unchanged. With $n_0$ fixed by the GR calibration, as adopted throughout, it raises the abundance of active seeds; if $n_0$ were instead recalibrated separately in each Hu-Sawicki model, the enhancement in the calibration range would be absorbed into $n_0$, and only its redshift dependence would remain.

\subsection{\label{ssec:lowz}Seed Abundance Across Cosmic Time and the Scalaron Signature}

The mass-function analysis of Sec.~\ref{ssec:bhmf_results} fixes the normalization $n_0$ and the Eddington-ratio distribution of the hosts by a fit to broad-line AGN at $3.5<z<6$, with the lower limits at $6<z<7$ setting the minimum median at higher redshift. Here we confront the same model, without further adjustment, with the abundance of LRDs and broad-line AGN over $2\lesssim z\lesssim8$. Since each survey reaches a different depth, we apply the selection of each survey to the model. A seed of mass $M_{\rm seed}$ accreting at $\lambda_{\rm Edd}$ enters a sample with optical limit $M_{\rm opt}$ if $\lambda_{\rm Edd}L_{\rm Edd}(M_{\rm seed})$ exceeds $L_{\rm bol}=\kappa\,\lambda L_{5100}(M_{\rm opt})$, with bolometric corrections $\kappa=5.4\text{-}9$~\citep{Kapoor2026}. The predicted number density above the limit is
\begin{eqnarray}
n(z;M_{\rm opt})=n_0\,F_{\rm coll}(>M_{\rm min},z)\int p(\log_{10}\lambda_{\rm Edd})\nonumber\\
\times\Theta\!\left[\lambda_{\rm Edd}L_{\rm Edd}(M_{\rm seed})-L_{\rm bol}\right]\dd\log_{10}\lambda_{\rm Edd},
\label{eq:n_lowz}
\end{eqnarray}
where the median of $p$ evolves by $0.17$~dex per unit redshift, as determined by the two mass-function bins, and the parameter set ($\sigma$, $\bar\lambda_{\rm Edd}$, $n_0$) spans all models allowed by the mass-function fit ($\Delta\chi^2\leq6.18$). We use three limits: $M_{\rm opt}<-19$ for the deep JWST samples, $M_{5100}<-20$ for the LRDs of~\citet{Kapoor2026}, and $M_{5500}<-22.5$ for the ground-based sample of~\citet{Ma2026lowz}. The data comprise the LRD luminosity function of~\citet{Kocevski2025} integrated at $4.5<z<6.5$ and $6.5<z<8.5$, the LRDs and the full broad-line AGN population of~\citet{Kapoor2026} at $z=1.9\text{-}2.5$, and the number densities of~\citet{Ma2026lowz} at $1.7<z<6$.

\begin{figure*}[!ht]
\centering
\includegraphics[width=\textwidth]{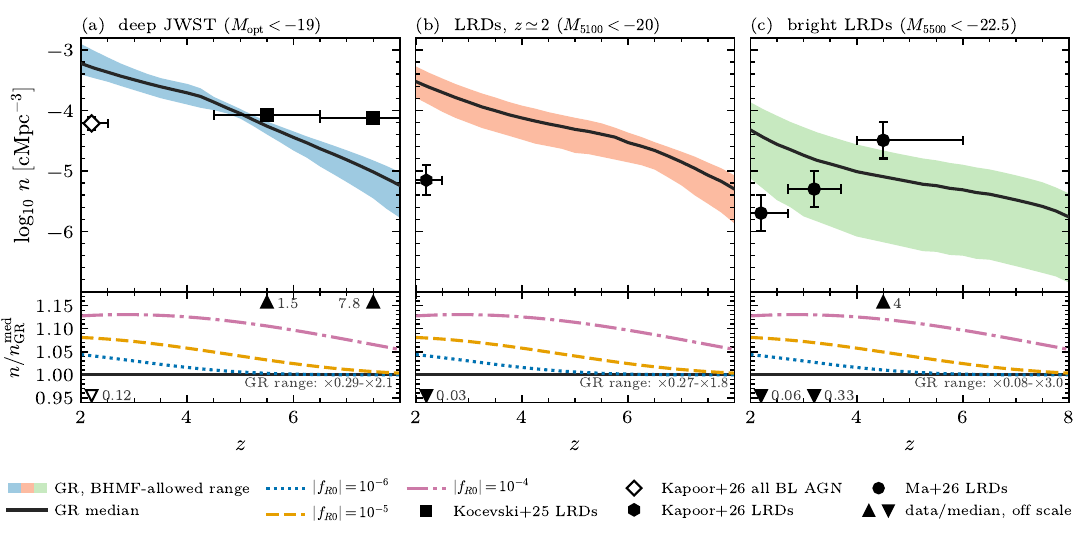}
\caption{Seed abundance across cosmic time in GR and Hu-Sawicki $f(R)$ gravity. Each panel shows the number density of active PBH-cluster seeds, Eq.~\eqref{eq:n_lowz}, above the optical limit of one sample: \textbf{(a)}~deep JWST surveys ($M_{\rm opt}<-19$), \textbf{(b)}~LRDs at $z\simeq2$ ($M_{5100}<-20$) and \textbf{(c)}~luminous LRDs ($M_{5500}<-22.5$). Upper sub-panels: shaded regions span the GR predictions of all models allowed by the mass-function fit ($\Delta\chi^2\leq6.18$) with bolometric corrections $5.4\text{-}9$, and solid lines are the GR medians. Lower sub-panels: predictions relative to the GR median on a linear scale, for $|f_{R0}|=10^{-6}$, $10^{-5}$ and $10^{-4}$ with $n_0$ fixed by the GR fit; since black holes are not coupled to the scalaron, the enhancement arises entirely from the host abundance $F_{\rm coll}(>10^{8}\,\Msun,z)$, with the scale-dependent growth of Sec.~\ref{ssec:growth_mg} and the host screening of Sec.~\ref{ssec:screening}. The full GR range is quoted in each strip, and triangles at the edges give the ratio of each measurement to the GR median. Data: LRDs of~\citet{Kocevski2025} (squares), LRDs (hexagon) and all broad-line AGN (open diamond) of~\citet{Kapoor2026}, and LRDs of~\citet{Ma2026lowz} (circles). No parameter is fitted to these data.}
\label{fig:lowz}
\end{figure*}

Figures~\ref{fig:lowz}(a)-(c) show the result, one panel per selection; the lower strip of each panel shows the predictions relative to the GR median, where the Hu-Sawicki models can be distinguished. Since $n_0$ and the Eddington-ratio distribution are calibrated on the mass function (Sec.~\ref{ssec:bhmf_results}), this comparison involves no further adjustment to the LRD data. The calibrated model reproduces the LRD abundance near the peak of the population: the predicted density lies within the measurements at $z\simeq3.2$ and $4.5$ for the luminous sample, and slightly below the deep sample at $z\simeq5.5$, where the ratio of the predicted to the measured density is $0.5\text{-}0.8$. Two departures delimit the regime of validity. At $z\simeq7.5$ the model underpredicts the deep sample by a factor of $5\text{-}20$, which indicates that the median Eddington ratio rises more steeply at $z>6$ than the linear trend inferred from the two mass-function bins, consistent with the lower bound $\bar\lambda_{\rm Edd}\gtrsim0.25$ at $6<z<7$. At $z\simeq2.2$ the model overpredicts the LRD counts by factors of $2\text{-}60$, and the full broad-line AGN population of~\citet{Kapoor2026} by a factor of $6\text{-}16$. Since LRDs constitute $\lesssim3\%$ of AGN at $z\simeq2$~\citep{Kapoor2026}, this excess implies that accretion onto cluster seeds must decline faster below $z\simeq3$ than the extrapolated trend, through a falling duty cycle or Eddington ratio. In the calibrated model, the luminous phase of the seeds is therefore confined to $z\gtrsim3$, in line with the observed decline of the LRD population.

The lower strips of Fig.~\ref{fig:lowz} show the environmental
effect of the scalar field. In this model, the scalar field acts
on the dark matter and gas surrounding the PBHs, rather than
directly on PBH-PBH capture, so its signature enters through
the abundance of host halos. For the adopted threshold
$M_{\min}=10^8\,M_\odot$, the host abundance is enhanced by
up to $\sim 13\%$ for $|f_{R0}|=10^{-4}$ at
$z\simeq 3$--$3.5$, decreasing to $\sim 8\%$ by $z=7$.
For $|f_{R0}|=10^{-5}$, the enhancement is $\sim 6.5\%$
at $z=3.5$ and $\sim 2.4\%$ at $z=6$. Since the modification
acts on the host halo as a whole, the predictions for all three
samples are rescaled according to the redshift-dependent
enhancement of the host abundance, with $n_0$ fixed by the
GR fit as in Fig.\,\ref{fig:bhmf}.

For $|f_{R0}|=10^{-4}$, the seed abundance rises by $13\%$
at $z\simeq 2.2$, $12\%$ at $z\simeq 4.5$, $11\%$ at
$z\simeq 5.5$, and $7\%$ at $z\simeq 7.5$. For
$|f_{R0}|=10^{-5}$, the corresponding enhancements are
$8\%$, $5\%$, $3\%$, and $0.7\%$, respectively; for
$|f_{R0}|=10^{-6}$, they are $4\%$, $1\%$, $0.3\%$,
and below $0.1\%$. The modified-gravity correction therefore
moves the prediction towards the data at $z\simeq 7.5$,
where the model underpredicts the abundance, and further
away from the data at $z\simeq 2.2$, where it overpredicts
the counts. In both cases, however, the correction is smaller
than the spread allowed by the mass-function fit. These
enhancements follow from the adopted host-screening
prescription, Eq.~\eqref{eq:fscr}, and the interpolation of the
halo mass function between the GR and unscreened limits
(Sec.~\ref{ssec:bhmf_framework}). They quantify the response
of this environmental model; the enhancement for
$|f_{R0}|=10^{-4}$, the largest value considered, should not
be interpreted as a generic prediction of $f(R)$ gravity.
The agreement near the peak of the population, as well as
the two departures at its edges, persists across all three
Hu--Sawicki models.

\subsection{\label{ssec:spin_summary}Remnant Spins}

The spin of the seed records its feeding history; the framework and the full results are given in Appendix~\ref{sec:app_spin}. Along the growth track with $\lambda_{\rm Edd}=0.5$, coherent accretion drives the seed to the Thorne limit, $\chi=0.998$, aligned with the disk, and anisotropic fuelling with a persistent net orientation gives $\chi\simeq0.94$. Chaotic accretion through self-gravity-limited episodes gives $\chi\simeq0.28$ with isotropic orientations, consistent with the equilibrium of~\citet{King2008}, and the few unequal-mass captures of a non-accreting seed give $\chi\simeq0.37$, below the equal-mass value of $0.686$. During the runaway, isotropic minor mergers temporarily spin every accreting seed down to $\chi\simeq0.1$ at $z\simeq6.5$. These results assume instantaneous alignment with each accretion episode, non-spinning PBHs and isotropic capture orbits, and the spin can identify the growth channel only together with the epoch and mass of the black hole.

\section{\label{sec:conclusions}Conclusions}

We have investigated whether dense clusters of PBHs can produce the massive black-hole seeds of LRDs through the coupled action of GW captures and gas accretion, and how the surrounding environment in Hu-Sawicki $f(R)$ gravity affects the resulting population. Our findings are conditional on the adopted cluster configuration, $N_{\rm cl}=10^{5}$ PBHs of median mass $30\,\Msun$ at $n_{\rm cl}=10^{8}\,{\rm pc^{-3}}$, which corresponds to $M_{\rm cl}=3.25\times10^{6}\,\Msun$. They show what happens if such clusters exist; whether they form and survive is not established here.

\textit{Seed growth.} Mergers and accretion reinforce one another: the accreted mass raises the capture rate, which scales approximately as $M^{12/7}$, and for time-averaged Eddington ratios $\lambda_{\rm Edd}\gtrsim0.3$ the capture of the cluster runs away. The seed consumes its cluster at $z\simeq5.5$-$7.3$ for $\lambda_{\rm Edd}=0.3$-$0.8$, after which growth is purely accretion-driven and carries the seed into the LRD mass range by $z\simeq5$. Dry mergers alone build seeds of only a few hundred solar masses by $z\simeq6$.

\textit{Observational comparison.} Calibrated on the measured black-hole mass function at $3.5<z<6$, which fixes the normalization and a median Eddington ratio $\bar\lambda_{\rm Edd}\simeq0.12$, the model requires the median to rise to $\gtrsim0.25$ at $6<z<7$. Compared without further adjustment with the LRD abundance at $2\lesssim z\lesssim8$, using the selection of each survey, the calibrated model reproduces the measurements near the peak of the population, $3\lesssim z\lesssim5.5$, underpredicts the abundance at $z\simeq7.5$, and overpredicts it at $z\simeq2$. These departures imply that accretion onto the seeds rises more steeply above $z\simeq6$ and declines below $z\simeq3$. The finite reservoir also predicts a lower edge of the mass function near $M_{\rm cl}$, just below the masses probed so far.

\textit{Environmental modified gravity and secondary diagnostics.} The scalar field acts on the dark matter and gas surrounding the PBHs rather than on their mutual capture. Within the adopted host-screening prescription and halo mass-function interpolation, it enhances the abundance of active seeds by up to $\sim13\%$ for $|f_{R0}|\leq10^{-4}$, most strongly in the least massive hosts and at the lowest redshifts, and, for a normalization fixed by the GR calibration, moves the prediction at $z\simeq7.5$ towards the data. The remnant spins, the stochastic GW background and the clustering of PBHs provide complementary diagnostics (Appendices~\ref{sec:app_spin}, \ref{sec:app_sgwb} and~\ref{sec:app_clustering}): coherent, anisotropic and chaotic feeding produce distinct spin distributions, the background from mergers within the clusters peaks near $6$~Hz with an amplitude set by an illustrative merger rate, and the PBH correlation function is enhanced by up to $\simeq10\%$ for a phenomenological power spectrum.

The main limitations follow directly. The results depend on the dense cluster configuration, whose formation and survival remain to be demonstrated, and the density dependence of the coupled solution has been tested only at one additional density. The reservoir is normalized to the mean mass of the initial distribution; a self-consistent treatment of the field depletion reproduces the growth of the seed, but the field population is not depleted after exhaustion, which affects the field-field contribution to the GW background. The $f(R)$ enhancement depends on the screening and halo mass-function prescriptions. Mass segregation, which would accelerate the runaway, is not included, and the two-body capture kernel becomes approximate once the seed dominates its cluster. The principal contribution of this work is a coupled treatment of merger and accretion growth in PBH clusters, calibrated on the mass function and tested against the LRD abundance; the open questions are the formation of the clusters, the accretion history that would reconcile the departures at the highest and lowest redshifts, and complete black-hole samples at $z\gtrsim6$ that could test the predicted evolution of the Eddington ratio.

\section*{Acknowledgements}
This work is supported by the Conselleria de Innovación, Universidades, Ciencia y Sociedad Digital of the Generalitat Valenciana and the European Social Fund through the postdoctoral fellowship APOSTD 2025 (CIAPOS/2024/461), by the Spanish Agencia Estatal de Investigación (grant PID2024-159689NB-C21), funded by MICIU/AEI/10.13039/501100011033 and FEDER/EU, by the Generalitat Valenciana (Prometeo grant CIPROM/2022/49), and by the European Horizon Europe Marie Skłodowska-Curie Staff Exchanges programme (HORIZON-MSCA-2021-SE-01, grant agreement No. 101086251, NewFunFiCO).

\bibliographystyle{apsrev4-2}
\bibliography{draft_ml}

@ARTICLE{Labbe2023,
       author = {{Labb{\'e}}, I. and {van Dokkum}, P. and {Nelson}, E. and {Bezanson}, R. and {Suess}, K.~A. and {Leja}, J. and {Brammer}, G. and {Whitaker}, K. and {Mathews}, E. and {Stefanon}, M. and {Wang}, B.},
        title = "{A population of red candidate massive galaxies 600 Myr after the Big Bang}",
      journal = {Nature},
         year = 2023,
        month = apr,
       volume = {616},
        pages = {266-269},
          doi = {10.1038/s41586-023-05786-2},
archivePrefix = {arXiv},
       eprint = {2207.12446},
 primaryClass = {astro-ph.GA},
       adsurl = {https://ui.adsabs.harvard.edu/abs/2023Natur.616..266L}
}

@ARTICLE{Harikane2023,
       author = {{Harikane}, Y. and {Zhang}, Y. and {Nakajima}, K. and {Ouchi}, M. and {Isobe}, Y. and {Ono}, Y. and {Umeda}, H. and {Xu}, Y. and {Yoshida}, N.},
        title = "{A JWST/NIRSpec First Census of Broad-line AGNs at z = 4-7: Detection of 10 Faint AGNs with $M_{\rm BH}\sim10^{6}$-$10^{8}$ $M_{\odot}$ and Their Host Galaxy Properties}",
      journal = {Astrophys. J.},
         year = 2023,
        month = dec,
       volume = {959},
       number = {1},
          eid = {39},
        pages = {39},
          doi = {10.3847/1538-4357/ad029b},
archivePrefix = {arXiv},
       eprint = {2303.11946},
 primaryClass = {astro-ph.GA},
       adsurl = {https://ui.adsabs.harvard.edu/abs/2023ApJ...959...39H}
}

@ARTICLE{Maiolino2024,
       author = {{Maiolino}, R. and {Scholtz}, J. and {Witstok}, J. and {Carniani}, S. and {et al.}},
        title = "{A small and vigorous black hole in the early Universe}",
      journal = {Nature},
         year = 2024,
        month = mar,
       volume = {627},
        pages = {59-63},
          doi = {10.1038/s41586-024-07052-5},
archivePrefix = {arXiv},
       eprint = {2305.12492},
 primaryClass = {astro-ph.GA},
       adsurl = {https://ui.adsabs.harvard.edu/abs/2024Natur.627...59M}
}

@ARTICLE{Matthee2024,
       author = {{Matthee}, J. and {Naidu}, R.~P. and {Brammer}, G. and {Matthee}, A. and {Oesch}, P. and {Stefanon}, M. and {Maseda}, M.~V. and {Labbe}, I. and {Bezanson}, R. and {Dayal}, P. and {de Graaff}, A. and {Illingworth}, G. and {Lin}, X. and {Riechers}, D. and {van Dokkum}, P. and {Whitaker}, K. and {Wuyts}, S.},
        title = "{Little Red Dots: An Abundant Population of Faint Active Galactic Nuclei at z $\sim$ 5 Revealed by the EIGER and FRESCO JWST Surveys}",
      journal = {Astrophys. J.},
         year = 2024,
        month = mar,
       volume = {963},
       number = {2},
          eid = {129},
        pages = {129},
          doi = {10.3847/1538-4357/ad2345},
archivePrefix = {arXiv},
       eprint = {2306.05448},
 primaryClass = {astro-ph.GA},
       adsurl = {https://ui.adsabs.harvard.edu/abs/2024ApJ...963..129M}
}

@ARTICLE{Kokorev2024,
       author = {{Kokorev}, V. and {Caputi}, K.~I. and {Greene}, J.~E. and {Iani}, E. and {Rinaldi}, P. and {Akhlaghi}, M. and {Illingworth}, G. and {Stefanon}, M. and {Wilkins}, S.~M.},
        title = "{A Census of Photometrically Selected Little Red Dots at 4 $<$ z $<$ 9 in JWST Blank Fields}",
      journal = {Astrophys. J.},
         year = 2024,
        month = jun,
       volume = {968},
       number = {1},
          eid = {38},
        pages = {38},
          doi = {10.3847/1538-4357/ad4265},
archivePrefix = {arXiv},
       eprint = {2401.09981},
 primaryClass = {astro-ph.GA},
       adsurl = {https://ui.adsabs.harvard.edu/abs/2024ApJ...968...38K}
}

@ARTICLE{Greene2024,
       author = {{Greene}, Jenny E. and {Labbe}, Ivo and {Goulding}, Andy D. and {et al.}},
        title = "{UNCOVER Spectroscopy Confirms the Surprising Ubiquity of Active Galactic Nuclei in Red Sources at z > 5}",
      journal = {Astrophys. J.},
         year = 2024,
        month = mar,
       volume = {964},
       number = {1},
          eid = {39},
        pages = {39},
          doi = {10.3847/1538-4357/ad1e5f},
archivePrefix = {arXiv},
       eprint = {2309.05714},
 primaryClass = {astro-ph.GA},
       adsurl = {https://ui.adsabs.harvard.edu/abs/2024ApJ...964...39G}

    }

@ARTICLE{Akins2024,
        author = {{Akins}, Hollis B. and {Casey}, Caitlin M. and {Lambrides}, Erini and {et al.}},
        title = "{COSMOS-Web: The Overabundance and Physical Nature of ``Little Red Dots''{\textemdash}Implications for Early Galaxy and SMBH Assembly}",
      journal = {Astrophys. J.},
         year = 2025,
        month = sep,
       volume = {991},
       number = {1},
          eid = {37},
        pages = {37},
          doi = {10.3847/1538-4357/ade984},
archivePrefix = {arXiv},
       eprint = {2406.10341},
 primaryClass = {astro-ph.GA},
       adsurl = {https://ui.adsabs.harvard.edu/abs/2025ApJ...991...37A}
}

@ARTICLE{Volonteri2010,
       author = {{Volonteri}, Marta},
        title = "{Formation of supermassive black holes}",
      journal = {Astron. Astrophys. Rev.},
         year = 2010,
        month = jul,
       volume = {18},
       number = {3},
        pages = {279-315},
          doi = {10.1007/s00159-010-0029-x},
archivePrefix = {arXiv},
       eprint = {1003.4404},
 primaryClass = {astro-ph.CO},
       adsurl = {https://ui.adsabs.harvard.edu/abs/2010A&ARv..18..279V}
}

@ARTICLE{Inayoshi2020,
       author = {{Inayoshi}, Kohei and {Visbal}, Eli and {Haiman}, Zolt{\'a}n},
        title = "{The Assembly of the First Massive Black Holes}",
      journal = {Annu. Rev. Astron. Astrophys.},
         year = 2020,
        month = aug,
       volume = {58},
        pages = {27-97},
          doi = {10.1146/annurev-astro-120419-014455},
archivePrefix = {arXiv},
       eprint = {1911.05791},
 primaryClass = {astro-ph.GA},
       adsurl = {https://ui.adsabs.harvard.edu/abs/2020ARA&A..58...27I}
}

@ARTICLE{Fan2023,
       author = {{Fan}, Xiaohui and {Ba{\~n}ados}, Eduardo and {Simcoe}, Robert A.},
        title = "{Quasars and the Intergalactic Medium at Cosmic Dawn}",
      journal = {Annu. Rev. Astron. Astrophys.},
         year = 2023,
        month = aug,
       volume = {61},
        pages = {373-426},
          doi = {10.1146/annurev-astro-052920-102455},
archivePrefix = {arXiv},
       eprint = {2212.06907},
 primaryClass = {astro-ph.GA},
       adsurl = {https://ui.adsabs.harvard.edu/abs/2023ARA&A..61..373F}
}

@ARTICLE{Madau2001,
       author = {{Madau}, Piero and {Rees}, Martin J.},
        title = "{Massive Black Holes as Population III Remnants}",
      journal = {Astrophys. J. Lett.},
         year = 2001,
        month = apr,
       volume = {551},
       number = {1},
        pages = {L27-L30},
          doi = {10.1086/319848},
archivePrefix = {arXiv},
       eprint = {astro-ph/0101223},
 primaryClass = {astro-ph},
       adsurl = {https://ui.adsabs.harvard.edu/abs/2001ApJ...551L..27M}
}

@ARTICLE{Bromm2004,
        author = {{Bromm}, Volker and {Larson}, Richard B.},
        title = "{The First Stars}",
      journal = {Annu. Rev. Astron. Astrophys.},
         year = 2004,
        month = sep,
       volume = {42},
       number = {1},
        pages = {79-118},
          doi = {10.1146/annurev.astro.42.053102.134034},
archivePrefix = {arXiv},
       eprint = {astro-ph/0311019},
 primaryClass = {astro-ph},
       adsurl = {https://ui.adsabs.harvard.edu/abs/2004ARA&A..42...79B}
}

@ARTICLE{BrommLoeb2003,
       author = {{Bromm}, Volker and {Loeb}, Abraham},
        title = "{Formation of the First Supermassive Black Holes}",
      journal = {Astrophys. J.},
         year = 2003,
        month = oct,
       volume = {596},
       number = {1},
        pages = {34-46},
          doi = {10.1086/377529},
archivePrefix = {arXiv},
       eprint = {astro-ph/0212400},
 primaryClass = {astro-ph},
       adsurl = {https://ui.adsabs.harvard.edu/abs/2003ApJ...596...34B}
}

@ARTICLE{Lodato2006,
       author = {{Lodato}, Giuseppe and {Natarajan}, Priyamvada},
        title = "{Supermassive black hole formation during the assembly of pre-galactic discs}",
      journal = {Mon. Not. R. Astron. Soc.},
         year = 2006,
        month = oct,
       volume = {371},
       number = {4},
        pages = {1813-1823},
          doi = {10.1111/j.1365-2966.2006.10801.x},
archivePrefix = {arXiv},
       eprint = {astro-ph/0606159},
 primaryClass = {astro-ph},
       adsurl = {https://ui.adsabs.harvard.edu/abs/2006MNRAS.371.1813L}
}

@ARTICLE{Regan2009,
       author = {{Regan}, John A. and {Haehnelt}, Martin G.},
        title = "{Pathways to massive black holes and compact star clusters in pre-galactic dark matter haloes with virial temperatures >\raisebox{-0.5ex}\textasciitilde10000K}",
      journal = {Mon. Not. R. Astron. Soc.},
         year = 2009,
        month = jun,
       volume = {396},
       number = {1},
        pages = {343-353},
          doi = {10.1111/j.1365-2966.2009.14579.x},
archivePrefix = {arXiv},
       eprint = {0810.2802},
 primaryClass = {astro-ph},
       adsurl = {https://ui.adsabs.harvard.edu/abs/2009MNRAS.396..343R}
}

@ARTICLE{Latif2013,
       author = {{Latif}, M.~A. and {Schleicher}, D.~R.~G. and {Schmidt}, W. and {Niemeyer}, J.},
        title = "{Black hole formation in the early Universe}",
      journal = {Mon. Not. R. Astron. Soc.},
         year = 2013,
        month = aug,
       volume = {433},
       number = {2},
        pages = {1607-1618},
          doi = {10.1093/mnras/stt834},
archivePrefix = {arXiv},
       eprint = {1304.0962},
 primaryClass = {astro-ph.CO},
       adsurl = {https://ui.adsabs.harvard.edu/abs/2013MNRAS.433.1607L}
}

@ARTICLE{Devecchi2009,
       author = {{Devecchi}, B. and {Volonteri}, M.},
        title = "{Formation of the First Nuclear Clusters and Massive Black Holes at High Redshift}",
      journal = {Astrophys. J.},
         year = 2009,
        month = mar,
       volume = {694},
       number = {1},
        pages = {302-313},
          doi = {10.1088/0004-637X/694/1/302},
archivePrefix = {arXiv},
       eprint = {0810.1057},
 primaryClass = {astro-ph},
       adsurl = {https://ui.adsabs.harvard.edu/abs/2009ApJ...694..302D}
}

@ARTICLE{Katz2015,
       author = {{Katz}, Harley and {Sijacki}, Debora and {Haehnelt}, Martin G.},
        title = "{Seeding high-redshift QSOs by collisional runaway in primordial star clusters}",
      journal = {Mon. Not. R. Astron. Soc.},
         year = 2015,
        month = aug,
       volume = {451},
       number = {3},
        pages = {2352-2369},
          doi = {10.1093/mnras/stv1048},
archivePrefix = {arXiv},
       eprint = {1502.03448},
 primaryClass = {astro-ph.GA},
       adsurl = {https://ui.adsabs.harvard.edu/abs/2015MNRAS.451.2352K}
}

@ARTICLE{Zeldovich1967,
       author = {{Zel'dovich}, Ya. B. and {Novikov}, I.~D.},
        title = "{The Hypothesis of Cores Retarded during Expansion and the Hot Cosmological Model}",
      journal = {Soviet Astronomy},
         year = 1967,
        month = feb,
       volume = {10},
        pages = {602},
       adsurl = {https://ui.adsabs.harvard.edu/abs/1967SvA....10..602Z}
}

@ARTICLE{Hawking1971,
       author = {{Hawking}, Stephen},
        title = "{Gravitationally collapsed objects of very low mass}",
      journal = {Mon. Not. R. Astron. Soc.},
         year = 1971,
        month = jan,
       volume = {152},
        pages = {75},
          doi = {10.1093/mnras/152.1.75},
       adsurl = {https://ui.adsabs.harvard.edu/abs/1971MNRAS.152...75H}
}

@ARTICLE{Carr1974,
       author = {{Carr}, B.~J. and {Hawking}, S.~W.},
        title = "{Black holes in the early Universe}",
      journal = {Mon. Not. R. Astron. Soc.},
         year = 1974,
        month = aug,
       volume = {168},
        pages = {399-416},
          doi = {10.1093/mnras/168.2.399},
       adsurl = {https://ui.adsabs.harvard.edu/abs/1974MNRAS.168..399C}
}

@ARTICLE{CarrHawking1974,
       author = {{Carr}, B.~J.},
        title = "{The primordial black hole mass spectrum.}",
      journal = {\apj},
         year = 1975,
        month = oct,
       volume = {201},
        pages = {1-19},
          doi = {10.1086/153853},
       adsurl = {https://ui.adsabs.harvard.edu/abs/1975ApJ...201....1C}
}

@ARTICLE{Khlopov2010,
       author = {{Khlopov}, Maxim Yu.},
        title = "{Primordial black holes}",
      journal = {Research in Astronomy and Astrophysics},
         year = 2010,
        month = jun,
       volume = {10},
       number = {6},
        pages = {495-528},
          doi = {10.1088/1674-4527/10/6/001},
archivePrefix = {arXiv},
       eprint = {0801.0116},
 primaryClass = {astro-ph},
       adsurl = {https://ui.adsabs.harvard.edu/abs/2010RAA....10..495K}
}

@ARTICLE{Abbott2016,
       author = {{Abbott}, B.~P. and {Abbott}, R. and {Abbott}, T.~D. and {et al.}},
        title = "{Observation of Gravitational Waves from a Binary Black Hole Merger}",
      journal = {Phys. Rev. Lett.},
         year = 2016,
        month = feb,
       volume = {116},
       number = {6},
          eid = {061102},
        pages = {061102},
          doi = {10.1103/PhysRevLett.116.061102},
archivePrefix = {arXiv},
       eprint = {1602.03837},
 primaryClass = {gr-qc},
       adsurl = {https://ui.adsabs.harvard.edu/abs/2016PhRvL.116f1102A}
}

@ARTICLE{Bird2016,
       author = {{Bird}, Simeon and {Cholis}, Ilias and {Mu{\~n}oz}, Julian B. and {Ali-Ha{\"\i}moud}, Yacine and {Kamionkowski}, Marc and {Kovetz}, Ely D. and {Raccanelli}, Alvise and {Riess}, Adam G.},
        title = "{Did LIGO Detect Dark Matter?}",
      journal = {Phys. Rev. Lett.},
         year = 2016,
        month = may,
       volume = {116},
       number = {20},
          eid = {201301},
        pages = {201301},
          doi = {10.1103/PhysRevLett.116.201301},
archivePrefix = {arXiv},
       eprint = {1603.00464},
 primaryClass = {astro-ph.CO},
       adsurl = {https://ui.adsabs.harvard.edu/abs/2016PhRvL.116t1301B}
}

@ARTICLE{Sasaki2016,
       author = {{Sasaki}, Misao and {Suyama}, Teruaki and {Tanaka}, Takahiro and {Yokoyama}, Shuichiro},
        title = "{Primordial Black Hole Scenario for the Gravitational-Wave Event GW150914}",
      journal = {Phys. Rev. Lett.},
         year = 2016,
        month = aug,
       volume = {117},
       number = {6},
          eid = {061101},
        pages = {061101},
          doi = {10.1103/PhysRevLett.117.061101},
archivePrefix = {arXiv},
       eprint = {1603.08338},
 primaryClass = {astro-ph.CO},
       adsurl = {https://ui.adsabs.harvard.edu/abs/2016PhRvL.117f1101S}
}

@ARTICLE{Clesse2017,
       author = {{Clesse}, S{\'e}bastien and {Garc{\'\i}a-Bellido}, Juan},
        title = "{The clustering of massive Primordial Black Holes as Dark Matter: Measuring their mass distribution with advanced LIGO}",
      journal = {Phys. Dark Univ.},
         year = 2017,
        month = mar,
       volume = {15},
        pages = {142-147},
          doi = {10.1016/j.dark.2016.10.002},
archivePrefix = {arXiv},
       eprint = {1603.05234},
 primaryClass = {astro-ph.CO},
       adsurl = {https://ui.adsabs.harvard.edu/abs/2017PDU....15..142C}
}

@ARTICLE{Carr2016,
      author = {{Carr}, Bernard and {K{\"u}hnel}, Florian and {Sandstad}, Marit},
        title = "{Primordial black holes as dark matter}",
      journal = {Phys. Rev. D},
         year = 2016,
        month = oct,
       volume = {94},
       number = {8},
          eid = {083504},
        pages = {083504},
          doi = {10.1103/PhysRevD.94.083504},
archivePrefix = {arXiv},
       eprint = {1607.06077},
 primaryClass = {astro-ph.CO},
       adsurl = {https://ui.adsabs.harvard.edu/abs/2016PhRvD..94h3504C}
}

@ARTICLE{CarrKuhnel2020,
       author = {{Carr}, Bernard and {Kohri}, Kazunori and {Sendouda}, Yuuiti and {Yokoyama}, Jun'ichi},
        title = "{Constraints on primordial black holes}",
      journal = {Reports on Progress in Physics},
         year = 2021,
        month = nov,
       volume = {84},
       number = {11},
          eid = {116902},
        pages = {116902},
          doi = {10.1088/1361-6633/ac1e31},
archivePrefix = {arXiv},
       eprint = {2002.12778},
 primaryClass = {astro-ph.CO},
       adsurl = {https://ui.adsabs.harvard.edu/abs/2021RPPh...84k6902C}
}

@ARTICLE{CarrKuhnel2021,
       author = {{Carr}, Bernard and {K{\"u}hnel}, Florian},
        title = "{Primordial Black Holes as Dark Matter: Recent Developments}",
      journal = {Annual Review of Nuclear and Particle Science},
         year = 2020,
        month = oct,
       volume = {70},
        pages = {355-394},
          doi = {10.1146/annurev-nucl-050520-125911},
archivePrefix = {arXiv},
       eprint = {2006.02838},
 primaryClass = {astro-ph.CO},
       adsurl = {https://ui.adsabs.harvard.edu/abs/2020ARNPS..70..355C}
}

@ARTICLE{Green2021,
       author = {{Green}, Anne M. and {Kavanagh}, Bradley J.},
        title = "{Primordial black holes as a dark matter candidate}",
      journal = {Journal of Physics G Nuclear Physics},
         year = 2021,
        month = apr,
       volume = {48},
       number = {4},
          eid = {043001},
        pages = {043001},
          doi = {10.1088/1361-6471/abc534},
archivePrefix = {arXiv},
       eprint = {2007.10722},
 primaryClass = {astro-ph.CO},
       adsurl = {https://ui.adsabs.harvard.edu/abs/2021JPhG...48d3001G}
}

@INCOLLECTION{Escriva2022b,
       author = {{Escriv{\`a}}, Albert and {K{\"u}hnel}, Florian and {Tada}, Yuichiro},
        title = "{Primordial black holes}",
    booktitle = {Black Holes in the Era of Gravitational-Wave Astronomy},
         year = 2024,
       editor = {{Arca Sedda}, Manuel and {Bortolas}, Elisa and {Spera}, Mario},
        pages = {261-377},
          doi = {10.1016/B978-0-32-395636-9.00012-8},
       adsurl = {https://ui.adsabs.harvard.edu/abs/2024bheg.book..261E}
}

@ARTICLE{Chisholm2006,
        author = {{Chisholm}, James R.},
        title = "{Clustering of primordial black holes: Basic results}",
      journal = {Phys. Rev. D},
         year = 2006,
        month = apr,
       volume = {73},
       number = {8},
          eid = {083504},
        pages = {083504},
          doi = {10.1103/PhysRevD.73.083504},
archivePrefix = {arXiv},
       eprint = {astro-ph/0509141},
 primaryClass = {astro-ph},
       adsurl = {https://ui.adsabs.harvard.edu/abs/2006PhRvD..73h3504C}
}

@ARTICLE{Tada2015,
       author = {{Tada}, Yuichiro and {Yokoyama}, Shuichiro},
        title = "{Primordial black holes as biased tracers}",
      journal = {Phys. Rev. D},
         year = 2015,
        month = jun,
       volume = {91},
       number = {12},
          eid = {123534},
        pages = {123534},
          doi = {10.1103/PhysRevD.91.123534},
archivePrefix = {arXiv},
       eprint = {1502.01124},
 primaryClass = {astro-ph.CO},
       adsurl = {https://ui.adsabs.harvard.edu/abs/2015PhRvD..91l3534T}
}

@ARTICLE{Young2015,
       author = {{Young}, Sam and {Byrnes}, Christian T.},
        title = "{Signatures of non-gaussianity in the isocurvature modes of primordial black hole dark matter}",
      journal = {J. Cosmol. Astropart. Phys.},
         year = 2015,
        month = apr,
       volume = {2015},
       number = {4},
        pages = {034-034},
          doi = {10.1088/1475-7516/2015/04/034},
archivePrefix = {arXiv},
       eprint = {1503.01505},
 primaryClass = {astro-ph.CO},
       adsurl = {https://ui.adsabs.harvard.edu/abs/2015JCAP...04..034Y}
}

@ARTICLE{Suyama2019,
       author = {{Suyama}, Teruaki and {Yokoyama}, Shuichiro},
        title = "{A novel formulation of the primordial black hole mass function}",
      journal = {Progress of Theoretical and Experimental Physics},
         year = 2020,
        month = feb,
       volume = {2020},
       number = {2},
          eid = {023E03},
        pages = {023E03},
          doi = {10.1093/ptep/ptaa011},
archivePrefix = {arXiv},
       eprint = {1912.04687},
 primaryClass = {astro-ph.CO},
       adsurl = {https://ui.adsabs.harvard.edu/abs/2020PTEP.2020b3E03S}
}

@ARTICLE{Atal2020,
       author = {{Del Popolo}, Antonino and {Le Delliou}, Morgan},
        title = "{Small Scale Problems of the {\ensuremath{\Lambda}}CDM Model: A Short Review}",
      journal = {Galaxies},
         year = 2017,
        month = feb,
       volume = {5},
       number = {1},
          eid = {17},
        pages = {17},
          doi = {10.3390/galaxies5010017},
archivePrefix = {arXiv},
       eprint = {1606.07790},
 primaryClass = {astro-ph.CO},
       adsurl = {https://ui.adsabs.harvard.edu/abs/2017Galax...5...17D}
}

@ARTICLE{DeLuca2021,
        author = {{De Luca}, V. and {Franciolini}, G. and {Pani}, P. and {Riotto}, A.},
        title = "{Bayesian evidence for both astrophysical and primordial black holes: mapping the GWTC-2 catalog to third-generation detectors}",
      journal = {J. Cosmol. Astropart. Phys.},
         year = 2021,
        month = may,
       volume = {2021},
       number = {5},
          eid = {003},
        pages = {003},
          doi = {10.1088/1475-7516/2021/05/003},
archivePrefix = {arXiv},
       eprint = {2102.03809},
 primaryClass = {astro-ph.CO},
       adsurl = {https://ui.adsabs.harvard.edu/abs/2021JCAP...05..003D}
}

@ARTICLE{Nakamura1997,
       author = {{Nakamura}, Takashi and {Sasaki}, Misao and {Tanaka}, Takahiro and {Thorne}, Kip S.},
        title = "{Gravitational Waves from Coalescing Black Hole MACHO Binaries}",
      journal = {Astrophys. J. Lett.},
         year = 1997,
        month = oct,
       volume = {487},
       number = {2},
        pages = {L139-L142},
          doi = {10.1086/310886},
archivePrefix = {arXiv},
       eprint = {astro-ph/9708060},
 primaryClass = {astro-ph},
       adsurl = {https://ui.adsabs.harvard.edu/abs/1997ApJ...487L.139N}
}

@ARTICLE{Ioka1998,
       author = {{Ioka}, Kunihito and {Chiba}, Takeshi and {Tanaka}, Takahiro and {Nakamura}, Takashi},
        title = "{Black hole binary formation in the expanding universe: Three body problem approximation}",
      journal = {Phys. Rev. D},
         year = 1998,
        month = sep,
       volume = {58},
       number = {6},
          eid = {063003},
        pages = {063003},
          doi = {10.1103/PhysRevD.58.063003},
archivePrefix = {arXiv},
       eprint = {astro-ph/9807018},
 primaryClass = {astro-ph},
       adsurl = {https://ui.adsabs.harvard.edu/abs/1998PhRvD..58f3003I}
}

@ARTICLE{AliHaimoud2017,
       author = {{Ali-Ha{\"\i}moud}, Yacine and {Kovetz}, Ely D. and {Kamionkowski}, Marc},
        title = "{Merger rate of primordial black-hole binaries}",
      journal = {Phys. Rev. D},
         year = 2017,
        month = dec,
       volume = {96},
       number = {12},
          eid = {123523},
        pages = {123523},
          doi = {10.1103/PhysRevD.96.123523},
archivePrefix = {arXiv},
       eprint = {1709.06576},
 primaryClass = {astro-ph.CO},
       adsurl = {https://ui.adsabs.harvard.edu/abs/2017PhRvD..96l3523A}
}

@ARTICLE{Raidal2017,
       author = {{Raidal}, Martti and {Vaskonen}, Ville and {Veerm{\"a}e}, Hardi},
        title = "{Gravitational waves from primordial black hole mergers}",
      journal = {J. Cosmol. Astropart. Phys.},
         year = 2017,
        month = sep,
       volume = {2017},
       number = {9},
          eid = {037},
        pages = {037},
          doi = {10.1088/1475-7516/2017/09/037},
archivePrefix = {arXiv},
       eprint = {1707.01480},
 primaryClass = {astro-ph.CO},
       adsurl = {https://ui.adsabs.harvard.edu/abs/2017JCAP...09..037R}
}

@ARTICLE{Ballesteros2018,
        author = {{Ballesteros}, Guillermo and {Serpico}, Pasquale D. and {Taoso}, Marco},
        title = "{On the merger rate of primordial black holes: effects of nearest neighbours distribution and clustering}",
      journal = {J. Cosmol. Astropart. Phys.},
         year = 2018,
        month = oct,
       volume = {2018},
       number = {10},
          eid = {043},
        pages = {043},
          doi = {10.1088/1475-7516/2018/10/043},
archivePrefix = {arXiv},
       eprint = {1807.02084},
 primaryClass = {astro-ph.CO},
       adsurl = {https://ui.adsabs.harvard.edu/abs/2018JCAP...10..043B}
}

@ARTICLE{Vaskonen2020,
       author = {{Vaskonen}, Ville and {Veerm{\"a}e}, Hardi},
        title = "{Lower bound on the primordial black hole merger rate}",
      journal = {Phys. Rev. D},
         year = 2020,
        month = feb,
       volume = {101},
       number = {4},
          eid = {043015},
        pages = {043015},
          doi = {10.1103/PhysRevD.101.043015},
archivePrefix = {arXiv},
       eprint = {1908.09752},
 primaryClass = {astro-ph.CO},
       adsurl = {https://ui.adsabs.harvard.edu/abs/2020PhRvD.101d3015V}
}

@ARTICLE{Jedamzik2020,
       author = {{Jedamzik}, Karsten},
        title = "{Primordial black hole dark matter and the LIGO/Virgo observations}",
      journal = {J. Cosmol. Astropart. Phys.},
         year = 2020,
        month = sep,
       volume = {2020},
       number = {9},
          eid = {022},
        pages = {022},
          doi = {10.1088/1475-7516/2020/09/022},
archivePrefix = {arXiv},
       eprint = {2006.11172},
 primaryClass = {astro-ph.CO},
       adsurl = {https://ui.adsabs.harvard.edu/abs/2020JCAP...09..022J}
}

@ARTICLE{Hutsi2021,
        author = {{H{\"u}tsi}, Gert and {Raidal}, Martti and {Vaskonen}, Ville and {Veerm{\"a}e}, Hardi},
        title = "{Two populations of LIGO-Virgo black holes}",
      journal = {J. Cosmol. Astropart. Phys.},
         year = 2021,
        month = mar,
       volume = {2021},
       number = {3},
          eid = {068},
        pages = {068},
          doi = {10.1088/1475-7516/2021/03/068},
archivePrefix = {arXiv},
       eprint = {2012.02786},
 primaryClass = {astro-ph.CO},
       adsurl = {https://ui.adsabs.harvard.edu/abs/2021JCAP...03..068H}
}

@ARTICLE{Franciolini2022,
       author = {{Franciolini}, Gabriele and {Baibhav}, Vishal and {De Luca}, Valerio and {Ng}, Ken K.~Y. and {Wong}, Kaze W.~K. and {Berti}, Emanuele and {Pani}, Paolo and {Riotto}, Antonio and {Vitale}, Salvatore},
        title = "{Searching for a subpopulation of primordial black holes in LIGO-Virgo gravitational-wave data}",
      journal = {Phys. Rev. D},
         year = 2022,
        month = apr,
       volume = {105},
       number = {8},
          eid = {083526},
        pages = {083526},
          doi = {10.1103/PhysRevD.105.083526},
archivePrefix = {arXiv},
       eprint = {2105.03349},
 primaryClass = {gr-qc},
       adsurl = {https://ui.adsabs.harvard.edu/abs/2022PhRvD.105h3526F}
}

@ARTICLE{Fakhry2021ellip,
       author = {{Fakhry}, Saeed and {Firouzjaee}, Javad T. and {Farhoudi}, Mehrdad},
        title = "{Primordial black hole merger rate in ellipsoidal-collapse dark matter halo models}",
      journal = {Phys. Rev. D},
         year = 2021,
        month = jun,
       volume = {103},
       number = {12},
          eid = {123014},
        pages = {123014},
          doi = {10.1103/PhysRevD.103.123014},
archivePrefix = {arXiv},
       eprint = {2012.03211},
 primaryClass = {astro-ph.CO},
       adsurl = {https://ui.adsabs.harvard.edu/abs/2021PhRvD.103l3014F}
}

@ARTICLE{Fakhry2021sidm,
       author = {{Fakhry}, Saeed and {Naseri}, Mahdi and {Firouzjaee}, Javad T. and {Farhoudi}, Mehrdad},
        title = "{Primordial black hole merger rate in self-interacting dark matter halo models}",
      journal = {Phys. Rev. D},
         year = 2022,
        month = feb,
       volume = {105},
       number = {4},
          eid = {043525},
        pages = {043525},
          doi = {10.1103/PhysRevD.105.043525},
archivePrefix = {arXiv},
       eprint = {2106.06265},
 primaryClass = {astro-ph.CO},
       adsurl = {https://ui.adsabs.harvard.edu/abs/2022PhRvD.105d3525F}
}

@ARTICLE{Fakhry2022,
        author = {{Fakhry}, Saeed and {Salehnia}, Zahra and {Shirmohammadi}, Azin and {Firouzjaee}, Javad T.},
        title = "{The Merger Rate of Primordial Black Hole-Neutron Star Binaries in Ellipsoidal-collapse Dark Matter Halo Models}",
      journal = {Astrophys. J.},
         year = 2022,
        month = dec,
       volume = {941},
       number = {1},
          eid = {36},
        pages = {36},
          doi = {10.3847/1538-4357/aca523},
archivePrefix = {arXiv},
       eprint = {2209.08909},
 primaryClass = {astro-ph.CO},
       adsurl = {https://ui.adsabs.harvard.edu/abs/2022ApJ...941...36F}
}

@ARTICLE{Fakhry2023voids,
       author = {{Fakhry}, Saeed and {Tabasi}, Seyed Sajad and {Firouzjaee}, Javad T.},
        title = "{On the merger rate of primordial black holes in cosmic voids}",
      journal = {Phys. Dark Univ.},
         year = 2023,
        month = aug,
       volume = {41},
          eid = {101244},
        pages = {101244},
          doi = {10.1016/j.dark.2023.101244},
archivePrefix = {arXiv},
       eprint = {2210.13558},
 primaryClass = {astro-ph.CO},
       adsurl = {https://ui.adsabs.harvard.edu/abs/2023PDU....4101244F}
}

@ARTICLE{Fakhry2023spikes,
       author = {{Fakhry}, Saeed and {Salehnia}, Zahra and {Shirmohammadi}, Azin and {Yengejeh}, Mina Ghodsi and {Firouzjaee}, Javad T.},
        title = "{Compact Binary Merger Rate in Dark-matter Spikes}",
      journal = {Astrophys. J.},
         year = 2023,
        month = apr,
       volume = {947},
       number = {2},
          eid = {46},
        pages = {46},
          doi = {10.3847/1538-4357/acc1dd},
archivePrefix = {arXiv},
       eprint = {2301.02349},
 primaryClass = {astro-ph.CO},
       adsurl = {https://ui.adsabs.harvard.edu/abs/2023ApJ...947...46F}
}

@ARTICLE{HuSawicki2007,
       author = {{Hu}, Wayne and {Sawicki}, Ignacy},
        title = "{Models of f(R) cosmic acceleration that evade solar system tests}",
      journal = {Phys. Rev. D},
         year = 2007,
        month = sep,
       volume = {76},
       number = {6},
          eid = {064004},
        pages = {064004},
          doi = {10.1103/PhysRevD.76.064004},
archivePrefix = {arXiv},
       eprint = {0705.1158},
 primaryClass = {astro-ph},
       adsurl = {https://ui.adsabs.harvard.edu/abs/2007PhRvD..76f4004H}
}

@ARTICLE{Sotiriou2010,
       author = {{Sotiriou}, Thomas P. and {Faraoni}, Valerio},
        title = "{f(R) theories of gravity}",
      journal = {Reviews of Modern Physics},
         year = 2010,
        month = jan,
       volume = {82},
       number = {1},
        pages = {451-497},
          doi = {10.1103/RevModPhys.82.451},
archivePrefix = {arXiv},
       eprint = {0805.1726},
 primaryClass = {gr-qc},
       adsurl = {https://ui.adsabs.harvard.edu/abs/2010RvMP...82..451S}
}

@ARTICLE{DeFelice2010,
      author = {{De Felice}, Antonio and {Tsujikawa}, Shinji},
        title = "{f( R) Theories}",
      journal = {Living Reviews in Relativity},
         year = 2010,
        month = dec,
       volume = {13},
       number = {1},
          eid = {3},
        pages = {3},
          doi = {10.12942/lrr-2010-3},
archivePrefix = {arXiv},
       eprint = {1002.4928},
 primaryClass = {gr-qc},
       adsurl = {https://ui.adsabs.harvard.edu/abs/2010LRR....13....3D}
}

@ARTICLE{Nojiri2011,
       author = {{Nojiri}, Shin'Ichi and {Odintsov}, Sergei D.},
        title = "{Unified cosmic history in modified gravity: From F(R) theory to Lorentz non-invariant models}",
      journal = {Phys. Rep.},
         year = 2011,
        month = aug,
       volume = {505},
       number = {2},
        pages = {59-144},
          doi = {10.1016/j.physrep.2011.04.001},
archivePrefix = {arXiv},
       eprint = {1011.0544},
 primaryClass = {gr-qc},
       adsurl = {https://ui.adsabs.harvard.edu/abs/2011PhR...505...59N}
}

@ARTICLE{Khoury2004,
       author = {{Khoury}, Justin and {Weltman}, Amanda},
        title = "{Chameleon cosmology}",
      journal = {Phys. Rev. D},
         year = 2004,
        month = feb,
       volume = {69},
       number = {4},
          eid = {044026},
        pages = {044026},
          doi = {10.1103/PhysRevD.69.044026},
archivePrefix = {arXiv},
       eprint = {astro-ph/0309411},
 primaryClass = {astro-ph},
       adsurl = {https://ui.adsabs.harvard.edu/abs/2004PhRvD..69d4026K}
}

@ARTICLE{Brax2008,
       author = {{Brax}, Philippe and {van de Bruck}, Carsten and {Davis}, Anne-Christine and {Shaw}, Douglas J.},
        title = "{f(R) gravity and chameleon theories}",
      journal = {Phys. Rev. D},
         year = 2008,
        month = nov,
       volume = {78},
       number = {10},
          eid = {104021},
        pages = {104021},
          doi = {10.1103/PhysRevD.78.104021},
archivePrefix = {arXiv},
       eprint = {0806.3415},
 primaryClass = {astro-ph},
       adsurl = {https://ui.adsabs.harvard.edu/abs/2008PhRvD..78j4021B}
}

@ARTICLE{Lombriser2014,
       author = {{Lombriser}, Lucas},
        title = "{Constraining chameleon models with cosmology}",
      journal = {Annalen der Physik},
         year = 2014,
        month = aug,
       volume = {264},
       number = {7-8},
        pages = {259-282},
          doi = {10.1002/andp.201400058},
archivePrefix = {arXiv},
       eprint = {1403.4268},
 primaryClass = {astro-ph.CO},
       adsurl = {https://ui.adsabs.harvard.edu/abs/2014AnP...526..259L}
}

@ARTICLE{Burrage2018,
       author = {{Burrage}, Clare and {Sakstein}, Jeremy},
        title = "{Tests of chameleon gravity}",
      journal = {Living Rev. Relativ.},
         year = 2018,
        month = dec,
       volume = {21},
       number = {1},
          eid = {1},
        pages = {1},
          doi = {10.1007/s41114-018-0011-x},
archivePrefix = {arXiv},
       eprint = {1709.09071},
 primaryClass = {astro-ph.CO},
       adsurl = {https://ui.adsabs.harvard.edu/abs/2018LRR....21....1B}
}

@ARTICLE{Fakhry2024fR,
       author = {{Fakhry}, Saeed},
        title = "{Primordial Black Hole Merger Rate in f(R) Gravity}",
      journal = {Astrophys. J.},
         year = 2024,
        month = jan,
       volume = {961},
       number = {1},
          eid = {8},
        pages = {8},
          doi = {10.3847/1538-4357/ad0e66},
archivePrefix = {arXiv},
       eprint = {2308.11049},
 primaryClass = {gr-qc},
       adsurl = {https://ui.adsabs.harvard.edu/abs/2024ApJ...961....8F}
}

@ARTICLE{Fakhry2024pbhns,
       author = {{Fakhry}, Saeed and {Shiravand}, Maryam and {Farhang}, Marzieh},
        title = "{Primordial Black Hole─Neutron Star Merger Rate in Modified Gravity}",
      journal = {Astrophys. J.},
         year = 2024,
        month = may,
       volume = {966},
       number = {2},
          eid = {235},
        pages = {235},
          doi = {10.3847/1538-4357/ad3a66},
archivePrefix = {arXiv},
       eprint = {2401.15171},
 primaryClass = {gr-qc},
       adsurl = {https://ui.adsabs.harvard.edu/abs/2024ApJ...966..235F}
}

@ARTICLE{Fakhry:2024kjj,
        author = {{Fakhry}, Saeed and {Gholamhoseinian}, Sara and {Farhang}, Marzieh},
        title = "{Compact Binary Merger Rate with Modified Gravity in Dark Matter Spikes}",
      journal = {Astrophys. J.},
         year = 2024,
        month = dec,
       volume = {976},
       number = {2},
          eid = {248},
        pages = {248},
          doi = {10.3847/1538-4357/ad8917},
archivePrefix = {arXiv},
       eprint = {2408.11995},
 primaryClass = {gr-qc},
       adsurl = {https://ui.adsabs.harvard.edu/abs/2024ApJ...976..248F}
}

@ARTICLE{Liu2019,
       author = {{Liu}, Lang and {Guo}, Zong-Kuan and {Cai}, Rong-Gen},
        title = "{Effects of the merger history on the merger rate density of primordial black hole binaries}",
      journal = {Eur. Phys. J. C},
         year = 2019,
        month = aug,
       volume = {79},
       number = {8},
          eid = {717},
        pages = {717},
          doi = {10.1140/epjc/s10052-019-7227-0},
archivePrefix = {arXiv},
       eprint = {1901.07672},
 primaryClass = {astro-ph.CO},
       adsurl = {https://ui.adsabs.harvard.edu/abs/2019EPJC...79..717L}
}

@ARTICLE{Wu2020,
       author = {{Wu}, You},
        title = "{Merger history of primordial black-hole binaries}",
      journal = {Phys. Rev. D},
         year = 2020,
        month = apr,
       volume = {101},
       number = {8},
          eid = {083008},
        pages = {083008},
          doi = {10.1103/PhysRevD.101.083008},
archivePrefix = {arXiv},
       eprint = {2001.03833},
 primaryClass = {astro-ph.CO},
       adsurl = {https://ui.adsabs.harvard.edu/abs/2020PhRvD.101h3008W}
}

@ARTICLE{Liu2023,
        author = {{Liu}, Lang and {You}, Zhi-Qiang and {Wu}, You and {Chen}, Zu-Cheng},
        title = "{Constraining the merger history of primordial-black-hole binaries from GWTC-3}",
      journal = {Phys. Rev. D},
         year = 2023,
        month = mar,
       volume = {107},
       number = {6},
          eid = {063035},
        pages = {063035},
          doi = {10.1103/PhysRevD.107.063035},
archivePrefix = {arXiv},
       eprint = {2210.16094},
 primaryClass = {astro-ph.CO},
       adsurl = {https://ui.adsabs.harvard.edu/abs/2023PhRvD.107f3035L}
}

@ARTICLE{Smoluchowski1916,
        author = {{Smoluchowski}, M.~V.},
        title = "{Drei Vortrage uber Diffusion, Brownsche Bewegung und Koagulation von Kolloidteilchen}",
      journal = {Zeitschrift fur Physik},
         year = 1916,
        month = jan,
       volume = {17},
        pages = {557-585},
       adsurl = {https://ui.adsabs.harvard.edu/abs/1916ZPhy...17..557S}
}

@ARTICLE{Silk1993,
       author = {{Silk}, Joseph and {Szalay}, Alexander S.},
        title = "{Primeval Galaxies and Cold Dark Matter}",
      journal = {Astrophys. J. Lett.},
         year = 1987,
        month = dec,
       volume = {323},
        pages = {L107},
          doi = {10.1086/185067},
       adsurl = {https://ui.adsabs.harvard.edu/abs/1987ApJ...323L.107S}
}

@ARTICLE{SigurdssonHernquist1993,
       author = {{Sigurdsson}, Steinn and {Hernquist}, Lars},
        title = "{Primordial black holes in globular clusters}",
      journal = {Nature},
         year = 1993,
        month = jul,
       volume = {364},
       number = {6436},
        pages = {423-425},
          doi = {10.1038/364423a0},
       adsurl = {https://ui.adsabs.harvard.edu/abs/1993Natur.364..423S}
}

@ARTICLE{NunoSilesGarciaBellido2025,
       author = {{Nu{\~n}o Siles}, Jos{\'e} Francisco and {Garc{\'\i}a-Bellido}, Juan},
        title = "{Primordial black hole clusters, phenomenology \& implications}",
      journal = {Phys. Dark Univ.},
         year = 2025,
        month = feb,
       volume = {47},
          eid = {101789},
        pages = {101789},
          doi = {10.1016/j.dark.2024.101789},
archivePrefix = {arXiv},
       eprint = {2405.06391},
 primaryClass = {astro-ph.CO},
       adsurl = {https://ui.adsabs.harvard.edu/abs/2025PDU....4701789N}
}

@ARTICLE{Planck2018,
       author = {{Planck Collaboration} and {Aghanim}, N. and {Akrami}, Y. and {et al.}},
        title = "{Planck 2018 results. VI. Cosmological parameters}",
      journal = {Astron. Astrophys.},
         year = 2020,
        month = sep,
       volume = {641},
          eid = {A6},
        pages = {A6},
          doi = {10.1051/0004-6361/201833910},
archivePrefix = {arXiv},
       eprint = {1807.06209},
 primaryClass = {astro-ph.CO},
       adsurl = {https://ui.adsabs.harvard.edu/abs/2020A&A...641A...6P}
}

@ARTICLE{Sotiriou2012,
       author = {{Sotiriou}, Thomas P. and {Faraoni}, Valerio},
        title = "{Black Holes in Scalar-Tensor Gravity}",
      journal = {Phys. Rev. Lett.},
         year = 2012,
        month = feb,
       volume = {108},
       number = {8},
          eid = {081103},
        pages = {081103},
          doi = {10.1103/PhysRevLett.108.081103},
archivePrefix = {arXiv},
       eprint = {1109.6324},
 primaryClass = {gr-qc},
       adsurl = {https://ui.adsabs.harvard.edu/abs/2012PhRvL.108h1103S}
}

@ARTICLE{Campanelli2007,
       author = {{Campanelli}, Manuela and {Lousto}, Carlos O. and {Zlochower}, Yosef and {Merritt}, David},
        title = "{Maximum Gravitational Recoil}",
      journal = {Phys. Rev. Lett.},
         year = 2007,
        month = jun,
       volume = {98},
       number = {23},
          eid = {231102},
        pages = {231102},
          doi = {10.1103/PhysRevLett.98.231102},
archivePrefix = {arXiv},
       eprint = {gr-qc/0702133},
 primaryClass = {gr-qc},
       adsurl = {https://ui.adsabs.harvard.edu/abs/2007PhRvL..98w1102C}
}

@ARTICLE{Gonzalez2007,
       author = {{Gonz{\'a}lez}, Jos{\'e} A. and {Hannam}, Mark and {Sperhake}, Ulrich and {Br{\"u}gmann}, Bernd and {Husa}, Sascha},
        title = "{Supermassive Recoil Velocities for Binary Black-Hole Mergers with Antialigned Spins}",
      journal = {Phys. Rev. Lett.},
         year = 2007,
        month = jun,
       volume = {98},
       number = {23},
          eid = {231101},
        pages = {231101},
          doi = {10.1103/PhysRevLett.98.231101},
archivePrefix = {arXiv},
       eprint = {gr-qc/0702052},
 primaryClass = {gr-qc},
       adsurl = {https://ui.adsabs.harvard.edu/abs/2007PhRvL..98w1101G}
}

@ARTICLE{Baker2006,
       author = {{Baker}, John G. and {Centrella}, Joan and {Choi}, Dae-Il and {Koppitz}, Michael and {van Meter}, James R. and {Miller}, M. Coleman},
        title = "{Getting a Kick Out of Numerical Relativity}",
      journal = {Astrophys. J. Lett.},
         year = 2006,
        month = dec,
       volume = {653},
       number = {2},
        pages = {L93-L96},
          doi = {10.1086/510448},
archivePrefix = {arXiv},
       eprint = {astro-ph/0603204},
 primaryClass = {astro-ph},
       adsurl = {https://ui.adsabs.harvard.edu/abs/2006ApJ...653L..93B}
}

@ARTICLE{Bardeen1972,
       author = {{Bardeen}, James M. and {Press}, William H. and {Teukolsky}, Saul A.},
        title = "{Rotating Black Holes: Locally Nonrotating Frames, Energy Extraction, and Scalar Synchrotron Radiation}",
      journal = {Astrophys. J.},
         year = 1972,
        month = dec,
       volume = {178},
        pages = {347-370},
          doi = {10.1086/151796},
       adsurl = {https://ui.adsabs.harvard.edu/abs/1972ApJ...178..347B}
}

@ARTICLE{Bardeen1970,
        author = {{Bardeen}, James M.},
        title = "{Kerr Metric Black Holes}",
      journal = {Nature},
         year = 1970,
        month = apr,
       volume = {226},
       number = {5240},
        pages = {64-65},
          doi = {10.1038/226064a0},
       adsurl = {https://ui.adsabs.harvard.edu/abs/1970Natur.226...64B}
}

@ARTICLE{Thorne1974,
       author = {{Thorne}, Kip S.},
        title = "{Disk-Accretion onto a Black Hole. II. Evolution of the Hole}",
      journal = {Astrophys. J.},
         year = 1974,
        month = jul,
       volume = {191},
        pages = {507-520},
          doi = {10.1086/152991},
       adsurl = {https://ui.adsabs.harvard.edu/abs/1974ApJ...191..507T}
}

@ARTICLE{BardeenPetterson1975,
       author = {{Bardeen}, James M. and {Petterson}, Jacobus A.},
        title = "{The Lense-Thirring Effect and Accretion Disks around Kerr Black Holes}",
      journal = {Astrophys. J. Lett.},
         year = 1975,
        month = jan,
       volume = {195},
        pages = {L65},
          doi = {10.1086/181711},
       adsurl = {https://ui.adsabs.harvard.edu/abs/1975ApJ...195L..65B}
}

@ARTICLE{KingPringle2006,
       author = {{King}, A.~R. and {Pringle}, J.~E.},
        title = "{Growing supermassive black holes by chaotic accretion}",
      journal = {Mon. Not. R. Astron. Soc.},
         year = 2006,
        month = nov,
       volume = {373},
       number = {1},
        pages = {L90-L92},
          doi = {10.1111/j.1745-3933.2006.00249.x},
archivePrefix = {arXiv},
       eprint = {astro-ph/0609598},
 primaryClass = {astro-ph},
       adsurl = {https://ui.adsabs.harvard.edu/abs/2006MNRAS.373L..90K}
}

@ARTICLE{King2008,
       author = {{King}, A.~R. and {Pringle}, J.~E. and {Hofmann}, J.~A.},
        title = "{The evolution of black hole mass and spin in active galactic nuclei}",
      journal = {Mon. Not. R. Astron. Soc.},
         year = 2008,
        month = apr,
       volume = {385},
       number = {3},
        pages = {1621-1627},
          doi = {10.1111/j.1365-2966.2008.12943.x},
archivePrefix = {arXiv},
       eprint = {0801.1564},
 primaryClass = {astro-ph},
       adsurl = {https://ui.adsabs.harvard.edu/abs/2008MNRAS.385.1621K}
}

@ARTICLE{Barausse2009,
       author = {{Barausse}, Enrico and {Rezzolla}, Luciano},
        title = "{Predicting the Direction of the Final Spin from the Coalescence of Two Black Holes}",
      journal = {Astrophys. J. Lett.},
         year = 2009,
        month = oct,
       volume = {704},
       number = {1},
        pages = {L40-L44},
          doi = {10.1088/0004-637X/704/1/L40},
archivePrefix = {arXiv},
       eprint = {0904.2577},
 primaryClass = {gr-qc},
       adsurl = {https://ui.adsabs.harvard.edu/abs/2009ApJ...704L..40B}
}

@ARTICLE{Fishbach2017,
       author = {{Fishbach}, Maya and {Holz}, Daniel E. and {Farr}, Ben},
        title = "{Are LIGO's Black Holes Made from Smaller Black Holes?}",
      journal = {Astrophys. J. Lett.},
         year = 2017,
        month = may,
       volume = {840},
       number = {2},
          eid = {L24},
        pages = {L24},
          doi = {10.3847/2041-8213/aa7045},
archivePrefix = {arXiv},
       eprint = {1703.06869},
 primaryClass = {astro-ph.HE},
       adsurl = {https://ui.adsabs.harvard.edu/abs/2017ApJ...840L..24F}
}

@ARTICLE{GerosaBerti2017,
       author = {{Gerosa}, Davide and {Berti}, Emanuele},
        title = "{Are merging black holes born from stellar collapse or previous mergers?}",
      journal = {Phys. Rev. D},
         year = 2017,
        month = jun,
       volume = {95},
       number = {12},
          eid = {124046},
        pages = {124046},
          doi = {10.1103/PhysRevD.95.124046},
archivePrefix = {arXiv},
       eprint = {1703.06223},
 primaryClass = {gr-qc},
       adsurl = {https://ui.adsabs.harvard.edu/abs/2017PhRvD..95l4046G}
}

@ARTICLE{Taylor2025,
       author = {{Taylor}, Anthony J. and {Finkelstein}, Steven L. and {Kocevski}, Dale D. and {et al.}},
        title = "{Broad-line AGNs at 3.5 < z < 6: The Black Hole Mass Function and a Connection with Little Red Dots}",
      journal = {Astrophys. J.},
         year = 2025,
        month = jun,
       volume = {986},
       number = {2},
          eid = {165},
        pages = {165},
          doi = {10.3847/1538-4357/add15b},
archivePrefix = {arXiv},
       eprint = {2409.06772},
 primaryClass = {astro-ph.GA},
       adsurl = {https://ui.adsabs.harvard.edu/abs/2025ApJ...986..165T}
}

@ARTICLE{Jones2025,
       author = {{Jones}, Brenda L. and {Kocevski}, Dale D. and {Pacucci}, Fabio and {et al.}},
        title = "{The $M_{\rm BH}-M_{*}$ Relationship at $3<z<7$: Big Black Holes in Little Red Dots}",
      journal = {arXiv e-prints},
         year = 2025,
        month = oct,
          eid = {arXiv:2510.07376},
        pages = {arXiv:2510.07376},
          doi = {10.48550/arXiv.2510.07376},
archivePrefix = {arXiv},
       eprint = {2510.07376},
 primaryClass = {astro-ph.GA},
       adsurl = {https://ui.adsabs.harvard.edu/abs/2025arXiv251007376J}
}

@ARTICLE{NANOGrav:2023hde,
       author = {{Agazie}, Gabriella and {Anumarlapudi}, Akash and {Archibald}, Anne M. and {et al.}},
        title = "{The NANOGrav 15 yr Data Set: Evidence for a Gravitational-wave Background}",
      journal = {Astrophys. J. Lett.},
         year = 2023,
        month = jul,
       volume = {951},
       number = {1},
          eid = {L8},
        pages = {L8},
          doi = {10.3847/2041-8213/acdac6},
archivePrefix = {arXiv},
       eprint = {2306.16213},
 primaryClass = {astro-ph.HE},
       adsurl = {https://ui.adsabs.harvard.edu/abs/2023ApJ...951L...8A}
}

@ARTICLE{Janssen:2014dka,
       author = {{Janssen}, G. and {Hobbs}, G. and {McLaughlin}, M. and {Bassa}, C. and {Deller}, A. and {Kramer}, M. and {Lee}, K. and {Mingarelli}, C. and {Rosado}, P. and {Sanidas}, S. and {Sesana}, A. and {Shao}, L. and {Stairs}, I. and {Stappers}, B. and {Verbiest}, J.~P.~W.},
        title = "{Gravitational Wave Astronomy with the SKA}",
    journal = {PoS(AASKA14)},
         year = 2015,
        month = apr,
          eid = {37},
        pages = {37},
          doi = {10.22323/1.215.0037},
archivePrefix = {arXiv},
       eprint = {1501.00127},
 primaryClass = {astro-ph.IM},
       adsurl = {https://ui.adsabs.harvard.edu/abs/2015aska.confE..37J}
}

@ARTICLE{LISA:2017cee,
       author = {{Amaro-Seoane}, Pau and {Audley}, Heather and {Babak}, Stanislav and {et al.}},
        title = "{Laser Interferometer Space Antenna}",
      journal = {arXiv e-prints},
         year = 2017,
        month = feb,
          eid = {arXiv:1702.00786},
        pages = {arXiv:1702.00786},
          doi = {10.48550/arXiv.1702.00786},
archivePrefix = {arXiv},
       eprint = {1702.00786},
 primaryClass = {astro-ph.IM},
       adsurl = {https://ui.adsabs.harvard.edu/abs/2017arXiv170200786A}
}

@ARTICLE{KAGRA:2021kki,
       author = {{Abbott}, R. and {Abbott}, T.~D. and {Abraham}, S. and {et al.}},
        title = "{Upper limits on the isotropic gravitational-wave background from Advanced LIGO and Advanced Virgo's third observing run}",
      journal = {Phys. Rev. D},
         year = 2021,
        month = jul,
       volume = {104},
       number = {2},
          eid = {022004},
        pages = {022004},
          doi = {10.1103/PhysRevD.104.022004},
archivePrefix = {arXiv},
       eprint = {2101.12130},
 primaryClass = {gr-qc},
       adsurl = {https://ui.adsabs.harvard.edu/abs/2021PhRvD.104b2004A}
}

@ARTICLE{Kawamura:2021ggy,
       author = {{Kawamura}, Seiji and {Ando}, Masaki and {Seto}, Naoki and {et al.}},
        title = "{Current status of space gravitational wave antenna DECIGO and B-DECIGO}",
      journal = {Prog. Theor. Exp. Phys.},
         year = 2021,
        month = may,
       volume = {2021},
       number = {5},
          eid = {05A105},
        pages = {05A105},
          doi = {10.1093/ptep/ptab019},
archivePrefix = {arXiv},
       eprint = {2006.13545},
 primaryClass = {gr-qc},
       adsurl = {https://ui.adsabs.harvard.edu/abs/2021PTEP.2021eA105K}
}

@ARTICLE{Punturo:2010zz,
       author = {{Punturo}, M. and {Abernathy}, M. and {Acernese}, F. and {et al.}},
        title = "{The Einstein Telescope: a third-generation gravitational wave observatory}",
      journal = {Class. Quantum Grav.},
         year = 2010,
        month = oct,
       volume = {27},
       number = {19},
          eid = {194002},
        pages = {194002},
          doi = {10.1088/0264-9381/27/19/194002},
       adsurl = {https://ui.adsabs.harvard.edu/abs/2010CQGra..27s4002P}
}

@ARTICLE{AllenRomano1999,
       author = {{Allen}, Bruce and {Romano}, Joseph D.},
        title = "{Detecting a stochastic background of gravitational radiation: Signal processing strategies and sensitivities}",
      journal = {Phys. Rev. D},
         year = 1999,
        month = may,
       volume = {59},
       number = {10},
          eid = {102001},
        pages = {102001},
          doi = {10.1103/PhysRevD.59.102001},
archivePrefix = {arXiv},
       eprint = {gr-qc/9710117},
 primaryClass = {gr-qc},
       adsurl = {https://ui.adsabs.harvard.edu/abs/1999PhRvD..59j2001A}
}

@ARTICLE{BelliniSawicki2012,
       author = {{Bellini}, Emilio and {Sawicki}, Ignacy},
        title = "{Maximal freedom at minimum cost: linear large-scale structure in general modifications of gravity}",
      journal = {J. Cosmol. Astropart. Phys.},
         year = 2014,
        month = jul,
       volume = {2014},
       number = {7},
          eid = {050},
        pages = {050},
          doi = {10.1088/1475-7516/2014/07/050},
archivePrefix = {arXiv},
       eprint = {1404.3713},
 primaryClass = {astro-ph.CO},
       adsurl = {https://ui.adsabs.harvard.edu/abs/2014JCAP...07..050B}
}

@ARTICLE{Belgacem2018,
       author = {{Belgacem}, Enis and {Dirian}, Yves and {Foffa}, Stefano and {Maggiore}, Michele},
        title = "{Gravitational-wave luminosity distance in modified gravity theories}",
      journal = {Phys. Rev. D},
         year = 2018,
        month = may,
       volume = {97},
       number = {10},
          eid = {104066},
        pages = {104066},
          doi = {10.1103/PhysRevD.97.104066},
archivePrefix = {arXiv},
       eprint = {1712.08108},
 primaryClass = {astro-ph.CO},
       adsurl = {https://ui.adsabs.harvard.edu/abs/2018PhRvD..97j4066B}
}

@ARTICLE{Nishizawa2018,
       author = {{Nishizawa}, Atsushi},
        title = "{Generalized framework for testing gravity with gravitational-wave propagation. I. Formulation}",
      journal = {Phys. Rev. D},
         year = 2018,
        month = may,
       volume = {97},
       number = {10},
          eid = {104037},
        pages = {104037},
          doi = {10.1103/PhysRevD.97.104037},
archivePrefix = {arXiv},
       eprint = {1710.04825},
 primaryClass = {gr-qc},
       adsurl = {https://ui.adsabs.harvard.edu/abs/2018PhRvD..97j4037N}
}

@ARTICLE{Baraldini2023,
       author = {{Babak}, Stanislav and {Caprini}, Chiara and {Figueroa}, Daniel G. and {Karnesis}, Nikolaos and {Marcoccia}, Paolo and {Nardini}, Germano and {Pieroni}, Mauro and {Ricciardone}, Angelo and {Sesana}, Alberto and {Torrado}, Jes{\'u}s},
        title = "{Stochastic gravitational wave background from stellar origin binary black holes in LISA}",
      journal = {J. Cosmol. Astropart. Phys.},
         year = 2023,
        month = aug,
       volume = {2023},
       number = {8},
          eid = {034},
        pages = {034},
          doi = {10.1088/1475-7516/2023/08/034},
archivePrefix = {arXiv},
       eprint = {2304.06368},
 primaryClass = {astro-ph.CO},
       adsurl = {https://ui.adsabs.harvard.edu/abs/2023JCAP...08..034B}
}

@BOOK{Maggiore2007,
       author = {{Maggiore}, Michele},
        title = "{Gravitational Waves: Volume 1: Theory and Experiments}",
         year = 2007,
          doi = {10.1093/acprof:oso/9780198570745.001.0001},
       adsurl = {https://ui.adsabs.harvard.edu/abs/2007gwte.book.....M},
       publisher = {Oxford University Press}
}

@ARTICLE{2025PhRvD.112l3503F,
       author = {{Fakhry}, Saeed and {Salmani}, Reyhaneh Vojoudi and {Firouzjaee}, Javad T.},
        title = "{High-redshift galaxies from JWST observations in more realistic dark matter halo models}",
      journal = {Phys. Rev. D},
         year = 2025,
        month = dec,
       volume = {112},
       number = {12},
          eid = {123503},
        pages = {123503},
          doi = {10.1103/9cmb-kf3x},
archivePrefix = {arXiv},
       eprint = {2507.23742},
 primaryClass = {astro-ph.GA},
       adsurl = {https://ui.adsabs.harvard.edu/abs/2025PhRvD.112l3503F}
}

@ARTICLE{2026ApJ...998..178F,
       author = {{Fakhry}, Saeed and {Shiravand}, Maryam and {Del Popolo}, Antonino},
        title = "{Matching JWST Ultraviolet Luminosity Functions with Refined {\ensuremath{\Lambda}}CDM Halo Models}",
      journal = {Astrophys. J.},
         year = 2026,
        month = feb,
       volume = {998},
       number = {1},
          eid = {178},
        pages = {178},
          doi = {10.3847/1538-4357/ae371b},
archivePrefix = {arXiv},
       eprint = {2510.04709},
 primaryClass = {astro-ph.GA},
       adsurl = {https://ui.adsabs.harvard.edu/abs/2026ApJ...998..178F}
}

@ARTICLE{2017PhR...692....1N,
       author = {{Nojiri}, S. and {Odintsov}, S.~D. and {Oikonomou}, V.~K.},
        title = "{Modified gravity theories on a nutshell: Inflation, bounce and late-time evolution}",
      journal = {Phys. Rep.},
         year = 2017,
        month = jun,
       volume = {692},
        pages = {1-104},
          doi = {10.1016/j.physrep.2017.06.001},
archivePrefix = {arXiv},
       eprint = {1705.11098},
 primaryClass = {gr-qc},
       adsurl = {https://ui.adsabs.harvard.edu/abs/2017PhR...692....1N}
}

@ARTICLE{2026PhRvL.136w1402D,
       author = {{De Luca}, Valerio and {Del Grosso}, Loris and {Franciolini}, Gabriele and {Kritos}, Konstantinos and {Berti}, Emanuele and {D'Orazio}, Daniel J. and {Silk}, Joseph},
        title = "{Primordial-Black-Hole-Based Pathways to Little Red Dots}",
      journal = {Phys. rev. Lett.},
         year = 2026,
        month = jun,
       volume = {136},
       number = {23},
          eid = {231402},
        pages = {231402},
          doi = {10.1103/6y1w-87pd},
archivePrefix = {arXiv},
       eprint = {2512.19666},
 primaryClass = {astro-ph.CO},
       adsurl = {https://ui.adsabs.harvard.edu/abs/2026PhRvL.136w1402D}
}

@ARTICLE{2025JCAP...04..040Z,
       author = {{Ziparo}, F. and {Gallerani}, S. and {Ferrara}, A.},
        title = "{Primordial black holes as supermassive black hole seeds}",
      journal = {J. Cosmol. Astropart. Phys.},
         year = 2025,
        month = apr,
       volume = {2025},
       number = {4},
          eid = {040},
        pages = {040},
          doi = {10.1088/1475-7516/2025/04/040},
archivePrefix = {arXiv},
       eprint = {2411.03448},
 primaryClass = {astro-ph.CO},
       adsurl = {https://ui.adsabs.harvard.edu/abs/2025JCAP...04..040Z}
}

@ARTICLE{2026arXiv260628296A,
      author = {{Allegrini}, S. and {Iovino}, A.~J. and {Perna}, G. and {Veerm{\"a}e}, H.},
        title = "{Memoirs of the curvaton: non-perturbative non-Gaussianity and supermassive primordial black holes}",
      journal = {arXiv e-prints},
         year = 2026,
        month = jun,
          eid = {arXiv:2606.28296},
        pages = {arXiv:2606.28296},
          doi = {10.48550/arXiv.2606.28296},
archivePrefix = {arXiv},
       eprint = {2606.28296},
 primaryClass = {astro-ph.CO},
       adsurl = {https://ui.adsabs.harvard.edu/abs/2026arXiv260628296A}
}

@ARTICLE{2024Univ...10..297F,
       author = {{Fischer}, Hauke and {K{\"a}ding}, Christian and {Pitschmann}, Mario},
        title = "{Screened Scalar Fields in the Laboratory and the Solar System}",
      journal = {Universe},
         year = 2024,
        month = jul,
       volume = {10},
       number = {7},
          eid = {297},
        pages = {297},
          doi = {10.3390/universe10070297},
archivePrefix = {arXiv},
       eprint = {2405.14638},
 primaryClass = {gr-qc},
       adsurl = {https://ui.adsabs.harvard.edu/abs/2024Univ...10..297F}
}

@ARTICLE{Hawking1972,
        author = "Hawking, S. W.",
    title = "{Black holes in the Brans-Dicke theory of gravitation}",
    doi = "10.1007/BF01877518",
    journal = "Commun. Math. Phys.",
    volume = "25",
    pages = "167--171",
    year = "1972"
}

@ARTICLE{HuiNicolisStubbs2009,
       author = {{Hui}, Lam and {Nicolis}, Alberto and {Stubbs}, Christopher W.},
        title = "{Equivalence principle implications of modified gravity models}",
      journal = {Phys. Rev. D},
         year = 2009,
       volume = {80},
       number = {10},
          eid = {104002},
        pages = {104002},
          doi = {10.1103/PhysRevD.80.104002},
archivePrefix = {arXiv},
       eprint = {0905.2966},
 primaryClass = {astro-ph.CO}
}

@ARTICLE{Jacobson1999,
       author = {{Jacobson}, Ted},
        title = "{Primordial Black Hole Evolution in Tensor-Scalar Cosmology}",
      journal = {\prl},
         year = 1999,
        month = oct,
       volume = {83},
       number = {14},
        pages = {2699-2702},
          doi = {10.1103/PhysRevLett.83.2699},
archivePrefix = {arXiv},
       eprint = {astro-ph/9905303},
 primaryClass = {astro-ph},
       adsurl = {https://ui.adsabs.harvard.edu/abs/1999PhRvL..83.2699J}
}

@ARTICLE{HorbatschBurgess2012,
       author = {{Horbatsch}, M.~W. and {Burgess}, C.~P.},
        title = "{Cosmic black-hole hair growth and quasar OJ287}",
      journal = {J. Cosmol. Astropart. Phys.},
         year = 2012,
        month = may,
       volume = {2012},
       number = {5},
          eid = {010},
        pages = {010},
          doi = {10.1088/1475-7516/2012/05/010},
archivePrefix = {arXiv},
       eprint = {1111.4009},
 primaryClass = {gr-qc},
       adsurl = {https://ui.adsabs.harvard.edu/abs/2012JCAP...05..010H}
}

@ARTICLE{Berti2013,
       author = {{Berti}, Emanuele and {Cardoso}, Vitor and {Gualtieri}, Leonardo and {Horbatsch}, Michael and {Sperhake}, Ulrich},
        title = "{Numerical simulations of single and binary black holes in scalar-tensor theories: Circumventing the no-hair theorem}",
      journal = {Phys. Rev. D},
         year = 2013,
        month = jun,
       volume = {87},
       number = {12},
          eid = {124020},
        pages = {124020},
          doi = {10.1103/PhysRevD.87.124020},
archivePrefix = {arXiv},
       eprint = {1304.2836},
 primaryClass = {gr-qc},
       adsurl = {https://ui.adsabs.harvard.edu/abs/2013PhRvD..87l4020B}
}

@ARTICLE{Cataneo2015,
       author = {{Cataneo}, Matteo and {Rapetti}, David and {Schmidt}, Fabian and {Mantz}, Adam B. and {Allen}, Steven W. and {Applegate}, Douglas E. and {Kelly}, Patrick L. and {von der Linden}, Anja and {Morris}, R. Glenn},
        title = "{New constraints on f(R) gravity from clusters of galaxies}",
      journal = {Phys. Rev. D},
         year = 2015,
       volume = {92},
       number = {4},
          eid = {044009},
        pages = {044009},
          doi = {10.1103/PhysRevD.92.044009},
archivePrefix = {arXiv},
       eprint = {1412.0133},
 primaryClass = {astro-ph.CO}
}

@ARTICLE{DesmondFerreira2020,
       author = {{Desmond}, Harry and {Ferreira}, Pedro G.},
        title = "{Galaxy morphology rules out astrophysically relevant Hu-Sawicki f(R) gravity}",
      journal = {Phys. Rev. D},
         year = 2020,
       volume = {102},
       number = {10},
          eid = {104060},
        pages = {104060},
          doi = {10.1103/PhysRevD.102.104060},
archivePrefix = {arXiv},
       eprint = {2009.08743},
 primaryClass = {astro-ph.CO}
}

@ARTICLE{BryanNorman1998,
       author = {{Bryan}, Greg L. and {Norman}, Michael L.},
        title = "{Statistical Properties of X-Ray Clusters: Analytic and Numerical Comparisons}",
      journal = {Astrophys. J.},
         year = 1998,
       volume = {495},
       number = {1},
        pages = {80-99},
          doi = {10.1086/305262},
archivePrefix = {arXiv},
       eprint = {astro-ph/9710107}
}

@ARTICLE{King2005,
       author = {{King}, A.~R. and {Lubow}, S.~H. and {Ogilvie}, G.~I. and {Pringle}, J.~E.},
        title = "{Aligning spinning black holes and accretion discs}",
      journal = {Mon. Not. R. Astron. Soc.},
         year = 2005,
       volume = {363},
       number = {1},
        pages = {49-56},
          doi = {10.1111/j.1365-2966.2005.09378.x},
archivePrefix = {arXiv},
       eprint = {astro-ph/0507098}
}

@ARTICLE{HughesBlandford2003,
       author = {{Hughes}, Scott A. and {Blandford}, Roger D.},
        title = "{Black Hole Mass and Spin Coevolution by Mergers}",
      journal = {Astrophys. J. Lett.},
         year = 2003,
       volume = {585},
       number = {2},
        pages = {L101-L104},
          doi = {10.1086/375495},
archivePrefix = {arXiv},
       eprint = {astro-ph/0208484}
}

@ARTICLE{BertiVolonteri2008,
       author = {{Berti}, Emanuele and {Volonteri}, Marta},
        title = "{Cosmological Black Hole Spin Evolution by Mergers and Accretion}",
      journal = {Astrophys. J.},
         year = 2008,
       volume = {684},
       number = {2},
        pages = {822-828},
          doi = {10.1086/590379},
archivePrefix = {arXiv},
       eprint = {0802.0025}
}

@ARTICLE{DeLuca2019spin,
       author = {{De Luca}, V. and {Desjacques}, V. and {Franciolini}, G. and {Malhotra}, A. and {Riotto}, A.},
        title = "{The initial spin probability distribution of primordial black holes}",
      journal = {J. Cosmol. Astropart. Phys.},
         year = 2019,
       volume = {2019},
       number = {05},
          eid = {018},
        pages = {018},
          doi = {10.1088/1475-7516/2019/05/018},
archivePrefix = {arXiv},
       eprint = {1903.01179}
}

@ARTICLE{GonzalezTotalRecoil2007,
       author = {{Gonz{\'a}lez}, Jos{\'e} A. and {Sperhake}, Ulrich and {Br{\"u}gmann}, Bernd and {Hannam}, Mark and {Husa}, Sascha},
        title = "{Total recoil: the maximum kick from nonspinning black-hole binary inspiral}",
      journal = {Phys. Rev. Lett.},
         year = 2007,
       volume = {98},
       number = {9},
          eid = {091101},
        pages = {091101},
          doi = {10.1103/PhysRevLett.98.091101},
archivePrefix = {arXiv},
       eprint = {gr-qc/0610154}
}

@ARTICLE{BarausseMorozovaRezzolla2012,
       author = {{Barausse}, Enrico and {Morozova}, Viktoriya and {Rezzolla}, Luciano},
        title = "{On the Mass Radiated by Coalescing Black Hole Binaries}",
      journal = {Astrophys. J.},
         year = 2012,
       volume = {758},
       number = {1},
          eid = {63},
        pages = {63},
          doi = {10.1088/0004-637X/758/1/63},
archivePrefix = {arXiv},
       eprint = {1206.3803}
}

@ARTICLE{Dotti2013,
       author = {{Dotti}, M. and {Colpi}, M. and {Pallini}, S. and {Perego}, A. and {Volonteri}, M.},
        title = "{On the Orientation and Magnitude of the Black Hole Spin in Galactic Nuclei}",
      journal = {Astrophys. J.},
         year = 2013,
       volume = {762},
       number = {2},
          eid = {68},
        pages = {68},
          doi = {10.1088/0004-637X/762/2/68},
archivePrefix = {arXiv},
       eprint = {1211.4871}
}

@ARTICLE{ShethTormen1999,
       author = {{Sheth}, Ravi K. and {Tormen}, Giuseppe},
        title = "{Large-scale bias and the peak background split}",
      journal = {Mon. Not. R. Astron. Soc.},
         year = 1999,
       volume = {308},
       number = {1},
        pages = {119-126},
          doi = {10.1046/j.1365-8711.1999.02692.x},
archivePrefix = {arXiv},
       eprint = {astro-ph/9901122}
}

@ARTICLE{WangLiu2023,
    author = {{Wang}, Deng and {Liu}, Yizhou},
        title = "{JWST high redshift galaxy observations have a strong tension with Planck CMB measurements}",
      journal = {arXiv e-prints},
         year = 2022,
        month = dec,
          eid = {arXiv:2301.00347},
        pages = {arXiv:2301.00347},
          doi = {10.48550/arXiv.2301.00347},
archivePrefix = {arXiv},
       eprint = {2301.00347},
 primaryClass = {astro-ph.CO},
       adsurl = {https://ui.adsabs.harvard.edu/abs/2023arXiv230100347W}
}

@ARTICLE{NakamuraSuto1997,
       author = {{Nakamura}, Takashi T. and {Suto}, Yasushi},
        title = "{Strong Gravitational Lensing and Velocity Function as Tools to Probe Cosmological Parameters: Current Constraints and Future Predictions}",
      journal = {Prog. Theor. Phys.},
         year = 1997,
       volume = {97},
       number = {1},
        pages = {49-81},
          doi = {10.1143/PTP.97.49},
archivePrefix = {arXiv},
       eprint = {astro-ph/9612074}
}

@ARTICLE{Schmidt2009,
       author = {{Schmidt}, Fabian and {Vikhlinin}, Alexey and {Hu}, Wayne},
        title = "{Cluster constraints on f(R) gravity}",
      journal = {Phys. Rev. D},
         year = 2009,
       volume = {80},
       number = {8},
          eid = {083505},
        pages = {083505},
          doi = {10.1103/PhysRevD.80.083505},
archivePrefix = {arXiv},
       eprint = {0908.2457}
}

@ARTICLE{Hagstotz2019,
       author = {{Hagstotz}, Steffen and {Costanzi}, Matteo and {Baldi}, Marco and {Weller}, Jochen},
        title = "{Joint halo-mass function for modified gravity and massive neutrinos - I. Simulations and cosmological forecasts}",
      journal = {Mon. Not. R. Astron. Soc.},
         year = 2019,
       volume = {486},
       number = {3},
        pages = {3927-3941},
          doi = {10.1093/mnras/stz1051},
archivePrefix = {arXiv},
       eprint = {1806.07400}
}

@ARTICLE{Kopp2013,
       author = {{Kopp}, Michael and {Appleby}, Stephen A. and {Achitouv}, Ixandra and {Weller}, Jochen},
        title = "{Spherical collapse and halo mass function in f(R) theories}",
      journal = {Phys. Rev. D},
         year = 2013,
       volume = {88},
       number = {8},
          eid = {084015},
        pages = {084015},
          doi = {10.1103/PhysRevD.88.084015},
archivePrefix = {arXiv},
       eprint = {1306.3233}
}

@ARTICLE{Ma2021,
       author = {{Ma}, Linhao and {Hopkins}, Philip F. and {Ma}, Xiangcheng and {Angl{\'e}s-Alc{\'a}zar}, Daniel and {Faucher-Gigu{\`e}re}, Claude-Andr{\'e} and {Kelley}, Luke Zoltan},
        title = "{Seeds don't sink: even massive black hole `seeds' cannot migrate to galaxy centres efficiently}",
      journal = {Mon. Not. R. Astron. Soc.},
         year = 2021,
       volume = {508},
       number = {2},
        pages = {1973-1985},
          doi = {10.1093/mnras/stab2713},
archivePrefix = {arXiv},
       eprint = {2101.02727}
}

@ARTICLE{Ma2026lowz,
       author = {{Ma}, Yilun and {Greene}, Jenny E. and {Setton}, David J. and {Goulding}, Andy D. and {Annunziatella}, Marianna and {Fan}, Xiaohui and {Kokorev}, Vasily and {Labbe}, Ivo and {Li}, Jiaxuan and {Lin}, Xiaojing and {Marchesini}, Danilo and {Matthee}, Jorryt and {Robbins}, Luke and {Sajina}, Anna and {Sawicki}, Marcin and {Telford}, O. Grace},
        title = "{Counting Little Red Dots at z < 4 with Ground-based Surveys and Spectroscopic Follow-up}",
      journal = {\apj},
         year = 2026,
        month = mar,
       volume = {1000},
       number = {1},
          eid = {59},
        pages = {59},
          doi = {10.3847/1538-4357/ae4596},
archivePrefix = {arXiv},
       eprint = {2504.08032},
 primaryClass = {astro-ph.GA},
       adsurl = {https://ui.adsabs.harvard.edu/abs/2026ApJ..1000...59M}
}

@ARTICLE{Kapoor2026,
       author = {{Kapoor}, Shrriya and {Matthee}, Jorryt and {Torralba}, Alberto and {Kramarenko}, Ivan G. and {Bordoloi}, Rongmon and {Greene}, Jenny E. and {Iani}, Edoardo and {Kashino}, Daichi and {Liu}, Zhaoran and {Mackenzie}, Ruari and {Mascia}, Sara and {Naidu}, Rohan P. and {Simcoe}, Rob},
        title = "{Little Red Dots at z\raisebox{-0.5ex}\textasciitilde2 in EIGER reveal a gentle decline with respect to their peak number density at z\raisebox{-0.5ex}\textasciitilde5}",
      journal = {arXiv e-prints},
         year = 2026,
        month = jun,
          eid = {arXiv:2607.00084},
        pages = {arXiv:2607.00084},
          doi = {10.48550/arXiv.2607.00084},
archivePrefix = {arXiv},
       eprint = {2607.00084},
 primaryClass = {astro-ph.GA},
       adsurl = {https://ui.adsabs.harvard.edu/abs/2026arXiv260700084K}
}

@ARTICLE{Kocevski2025,
       author = {{Kocevski}, Dale D. and {et al.}},
        title = "{The Rise of Faint, Red Active Galactic Nuclei at z > 4: A Sample of Little Red Dots in the JWST Extragalactic Legacy Fields}",
      journal = {\apj},
         year = 2025,
        month = jun,
       volume = {986},
       number = {2},
          eid = {126},
        pages = {126},
          doi = {10.3847/1538-4357/adbc7d},
archivePrefix = {arXiv},
       eprint = {2404.03576},
 primaryClass = {astro-ph.GA},
       adsurl = {https://ui.adsabs.harvard.edu/abs/2025ApJ...986..126K}
}

\appendix

\section{\label{sec:app_sgwb}Stochastic Gravitational-Wave Background}


Hierarchical coalescences of PBH binaries across cosmic time generate an unresolvable SGWB, with dimensionless spectral energy density~\citep{AllenRomano1999}:
\begin{equation}
\begin{aligned}
\Omega_{\rm GW}(f)
&=
\dfrac{f}{\rho_{\rm crit,0} c^2}\times\\
&\qquad\int_{z_{\rm min}}^{z_{\rm max}}
\dfrac{\mathcal{R}_{\rm PBH}(z, f_{R0})}
{(1+z)\,H(z)}\,
\dfrac{\dd E_{\rm GW}}{\dd f_s}(f_s)
\mathcal{P}_{f(R)}(z)\,\dd z,
\label{eq:OmegaGW}
\end{aligned}
\end{equation}
where $\rho_{\rm crit,0} = 3 H_0^2 / (8\pi G)$ is the present critical density, $f$ the observer-frame frequency, $f_s = (1+z)f$ the source-frame frequency, $\mathcal{R}_{\rm PBH}(z, f_{R0})$ is the intrinsic merger-rate density, $\mathcal{P}_{f(R)}(z)$ denotes the propagation factor, and the integral spans the active merger epoch $z \in [6.0, 11.5]$.

For quasi-circular post-Newtonian inspirals, the angle-averaged energy spectrum per coalescence is~\citep{Maggiore2007}
\begin{equation}
\dfrac{\dd E_{\rm GW}}{\dd f_s}(f_s)
=
\dfrac{(\pi G)^{2/3}}{3}\,\mathcal{M}_c^{5/3}\,f_s^{-1/3},
\quad \text{for } f_s < f_{\rm ISCO},
\label{eq:dEdfs}
\end{equation}
where $\mathcal{M}_c \equiv (m_1 m_2)^{3/5}/(m_1+m_2)^{1/5}$ is the chirp mass. The spectrum terminates at the source-frame ISCO frequency
\begin{equation}
f_{\rm ISCO} = \dfrac{c^3}{6^{3/2}\pi G M_{\rm tot}}.
\label{eq:f_isco_gw}
\end{equation}
This relation gives $f_{\rm ISCO} \approx 73.3\,{\rm Hz}$ for $M_{\rm tot}=60\,M_\odot$, corresponding to an observed cut-off frequency of $f_{\rm ISCO}^{\rm obs}\approx7.7\,{\rm Hz}$ at the representative redshift $z=8.5$, taken as the midpoint of the merger epoch.

The intrinsic merger-rate density $\mathcal{R}_{\rm PBH}(z)$ comes directly from the coagulation calculation of Sec.~\ref{ssec:smolu}, evaluated with the GR kernel of Eq.~\eqref{eq:Kij}, for field-field mergers of equal-mass binaries of $30\,\Msun$; the captures of field PBHs by the central seed are not included. The rate is anchored to $\mathcal{R}(z=6)=20\;{\rm Gpc}^{-3}\,{\rm yr}^{-1}$, comparable to dynamical PBH-merger rate estimates for dense clusters~\citep{Hutsi2021}, adopted as an illustrative normalization; the spectrum scales linearly with it. Since black holes carry no scalar charge, the merger rate is the same in GR and in all Hu-Sawicki models, and the gravity model can enter only through the propagation of the waves.

In metric $f(R)$ gravity, tensor perturbations propagate as $\ddot{h}_{ij} + (3 + \alpha_M) H \dot{h}_{ij} + (k^2/a^2) h_{ij} = 0$, with Planck-mass running rate $\alpha_M \equiv \dd\ln F / \dd\ln a$ and $F \equiv \dd f/\dd R = 1 + f_R$~\citep{BelliniSawicki2012}. This introduces a ratio between electromagnetic and GW luminosity distances, the propagation factor~\citep{Belgacem2018,Nishizawa2018}
\begin{equation}
\mathcal{P}_{f(R)}(z)
\equiv
\left[\dfrac{d_L^{\rm EM}(z)}{d_L^{\rm GW}(z)}\right]^2
=
\dfrac{F(z=0)}{F(z)}
=
\dfrac{1 + f_{R0}}{1 + f_R(z)}.
\label{eq:PropFactor}
\end{equation}
For observationally viable Hu-Sawicki models with $|f_{R0}| \le 10^{-4}$, $|f_R(z)| \ll |f_{R0}| \ll 1$ at high redshift, so $\mathcal{P}_{f(R)}(z)$ deviates from unity by less than $0.01\%$ across $z \in [6.0, 11.5]$. Neither propagation nor the merger kernel therefore produces an appreciable modified-gravity signature in $\Omega_{\rm GW}(f)$.


The amplitude of this background scales linearly with the adopted merger-rate normalization, $\mathcal{R}(z=6)=20\,{\rm Gpc^{-3}\,yr^{-1}}$, which is illustrative and is not fixed by the calibration of Sec.~\ref{ssec:bhmf_results}. The spectral slope, set by the quadrupole inspiral, and the cutoff frequency, set by the innermost stable orbit of $30+30\,\Msun$ binaries at the redshifts of the mergers, do not depend on this normalization. A further approximation concerns the field population. The coagulation of Sec.~\ref{ssec:smolu} is solved independently of the seed, so the PBHs captured by the seed are not removed from the field distribution (Appendix~\ref{sec:app_numerics}). After the seed consumes its cluster, field-field mergers therefore continue in the calculation, although the field would by then be depleted. This approximation overestimates the field-field merger rate at redshifts below the exhaustion redshift of each cluster; its quantitative impact on the background has not been computed and would require the self-consistent treatment of Appendix~\ref{sec:app_numerics} to be extended to the merger rate.

\begin{figure}[!ht]
\centering
\includegraphics[width=\linewidth]{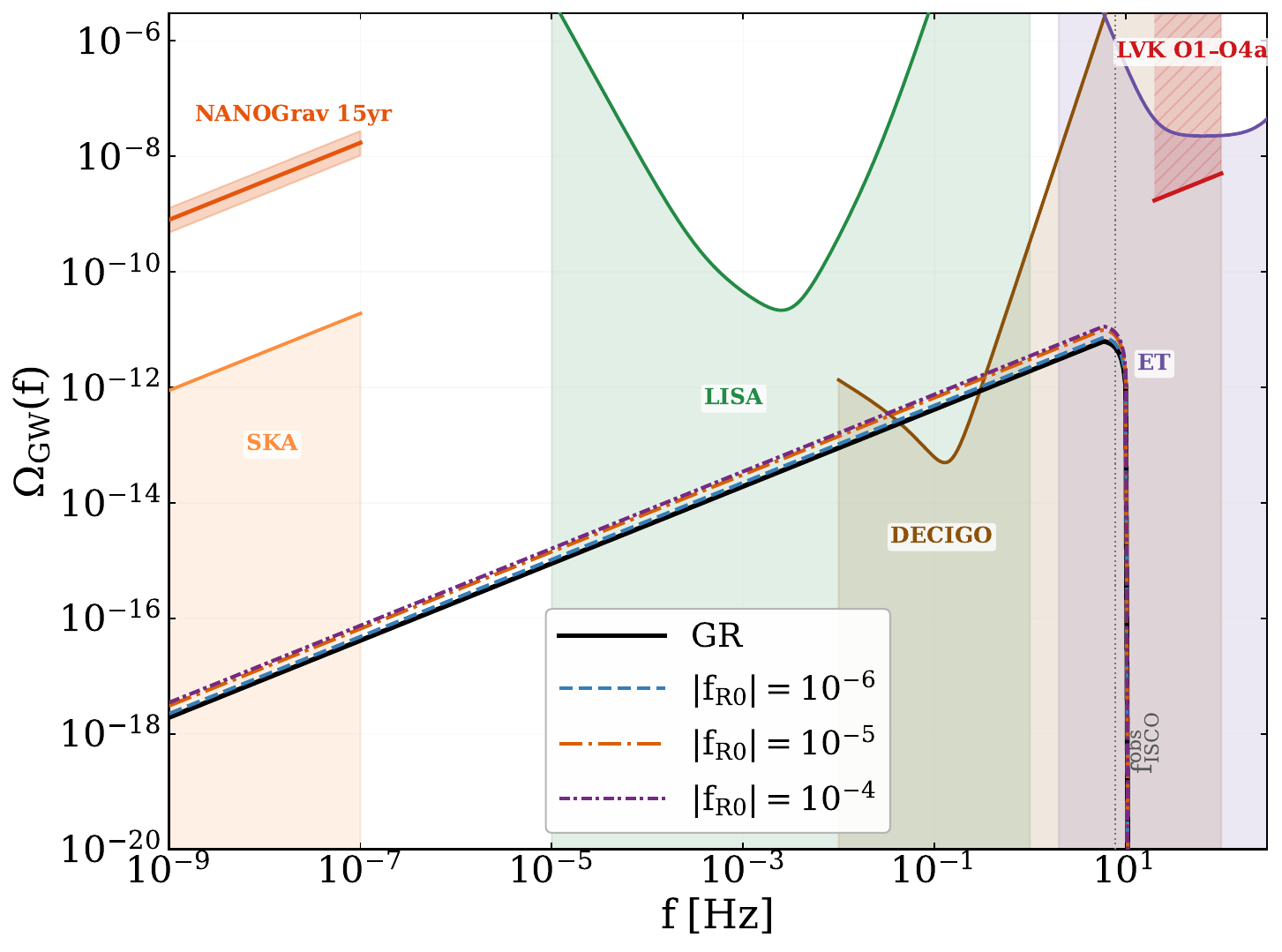}
\caption{SGWB spectrum $\Omega_{\rm GW}(f)$ from PBH mergers in dense clusters, Eq.~\eqref{eq:OmegaGW}, for GR and Hu-Sawicki models with $|f_{R0}|=10^{-6}$, $10^{-5}$ and $10^{-4}$, computed with the GR capture kernel for equal-mass $30\,\Msun$ binaries and normalized to $\mathcal{R}(z=6)=20\,{\rm Gpc^{-3}\,yr^{-1}}$. The four curves coincide. The vertical dotted line marks the observer-frame ISCO frequency at $z=8.5$. Detector curves show the LVK O1-O4a upper limit, the NANOGrav 15-year signal, and the projected sensitivities of SKA, LISA, DECIGO and ET.}
\label{fig:sgwb}
\end{figure}

In Fig.~\ref{fig:sgwb}, we have evaluated the SGWB energy density spectrum $\Omega_{\rm GW}(f)$ across frequencies $f \in [10^{-9}, 10^2]\,\mathrm{Hz}$ using Eq.~\eqref{eq:OmegaGW}. The underlying PBH merger rates are obtained from the Smoluchowski kinetic equation with the GR kernel over redshifts $z \in [6.0, 11.5]$, normalized to $\mathcal{R}(z=6) = 20\,\mathrm{Gpc}^{-3}\,\mathrm{yr}^{-1}$. The spectrum exhibits the characteristic $f^{2/3}$ power-law slope across lower frequencies, corresponding to the quadrupole inspiral regime of equal-mass $30\,\mathrm{M}_\odot$ binaries ($M_{\rm tot} = 60\,\mathrm{M}_\odot$, chirp mass $\mathcal{M}_c \approx 26.1\,\mathrm{M}_\odot$), and peaks at $\Omega_{\rm GW}\simeq6\times10^{-12}$ near $6$~Hz. It sharply terminates at the observer-frame ISCO frequency, Eq.~\eqref{eq:f_isco_gw}, evaluated for $M_{\rm tot} = 60\,\mathrm{M}_\odot$ at a mean redshift of $\bar{z} = 8.5$, indicated by the vertical dotted line. Because this cutoff occurs below $20\,\mathrm{Hz}$, the signal terminates before entering the LVK O1-O4a sensitivity band ($f \gtrsim 20\,\mathrm{Hz}$).

The four curves for GR and the Hu-Sawicki models coincide. Since black holes carry no scalar charge, the capture kernel, and hence the merger rate, is the same in all models, and the propagation factor of Eq.~\eqref{eq:PropFactor} deviates from unity by less than $0.01\%$ for $z\leq11.5$. The SGWB from PBH mergers in clusters is therefore insensitive to viable $f(R)$ modifications and does not discriminate them from GR.

Comparing this prediction with observational limits and sensitivity curves in Fig.~\ref{fig:sgwb}, the signal lies several orders of magnitude below the North American Nanohertz Observatory for Gravitational Waves (NANOGrav 15-year) signal~\cite{NANOGrav:2023hde} and the projected sensitivities of the Square Kilometre Array (SKA)~\cite{Janssen:2014dka} and the Laser Interferometer Space Antenna (LISA)~\cite{LISA:2017cee}, and below the LVK limits~\cite{KAGRA:2021kki}. It rises above the projected noise curve of the Decihertz Interferometer Gravitational wave Observatory (DECIGO)~\cite{Kawamura:2021ggy} for $f\simeq0.05$-$0.3$~Hz, by up to an order of magnitude, while remaining well below the Einstein Telescope (ET)~\cite{Punturo:2010zz} curve. For the adopted illustrative rate, and subject to the field-depletion approximation discussed above, deci-hertz observations with DECIGO could therefore probe the background from field-field PBH mergers in dense clusters, but not the modified-gravity parameter.

\section{\label{sec:app_clustering}PBH Spatial Clustering}

\subsection{\label{ssec:xi}Framework}

We characterize the clustering of PBHs at $z=6$ using the scale-dependent growth of the host matter and of the PBH perturbations, Eqs.~\eqref{eq:growth_m} and~\eqref{eq:growth_P}. The PBH power spectrum adopted below is phenomenological: its shape and its amplitude normalization, $A_0$ and $\eta_0$, are prescribed rather than derived from the seed-growth model, so the correlation function is not a parameter-free prediction of the scenario. It serves to quantify the relative change between GR and the Hu-Sawicki models.

This scale dependence alters the coupling between large-scale density fluctuations and small-scale clustering modes. In the squeezed-limit bispectrum, the mode-coupling amplitude of the PBH distribution is modified as~\citep{Baraldini2023}
\begin{equation}
\eta(k_L, k_s) = K_{\rm GR}(k_s)\,\dfrac{g_{\rm P}(k_s)}{g_{\rm P}(k_L)},
\label{eq:eta_mg}
\end{equation}
where $K_{\rm GR}(k_s) \equiv \frac{68}{21} - \frac{1}{3}\bigl[n_{\rm eff}(k_s) - 3\bigr]$ is the standard GR squeezed-kernel amplitude, with $n_{\rm eff}(k) \equiv \dd\ln P(k)/\dd\ln k$ the effective power-law index.

The resulting PBH spatial distribution follows the phenomenological cluster power spectrum
\begin{equation}
\begin{aligned}
P_{\rm PBH}(k)=A(k_L, k_s)\,
\left(\dfrac{k}{k_0}\right)^{-2}\times \qquad\qquad\qquad \\
\exp\!\left[-\left(\dfrac{k}{k_s}\right)^2\right] \exp\!\left[-\left(\dfrac{k_L}{k}\right)^2\right],
\label{eq:Pk}
\end{aligned}
\end{equation}
where $k_0$ is a pivot scale, $k_l,k_s$ are large- and small-scale cutoffs. Also,
\begin{equation}
A(k_L, k_s)
=
A_0\,\left[\dfrac{\eta(k_L, k_s)}{\eta_0}\right]^2 g_{\rm P}^2(k_s),
\label{eq:A_amp}
\end{equation}
with $A_0 = 0.90$, $\eta_0 = 14$, evaluated at $k_L = 0.05\;{\rm Mpc}^{-1}$ and $k_s = 50\;{\rm Mpc}^{-1}$. The factor $g_{\rm P}^2(k_s)$ carries the modified growth of the PBH perturbations; it replaces the coupling $\mu^2(k_s)$ of non-relativistic matter, which PBHs do not feel.

Eventually, the isotropic real-space two-point correlation function follows from the spherical Bessel transform
\begin{equation}
\xi_{\rm PBH}(r)
=
\dfrac{1}{2\pi^2}\int_0^\infty k^2\,P_{\rm PBH}(k)\,j_0(kr)\,\dd k,
\label{eq:xi}
\end{equation}
where $j_0(x) \equiv \sin(x)/x$. Equation~\eqref{eq:xi} therefore translates the scale-dependent clustering encoded in $P_{\rm PBH}(k)$ into the real-space correlation function $\xi_{\rm PBH}(r)$, providing a direct connection to spatial clustering observables of high-redshift black holes and their host galaxies.

\subsection{\label{ssec:screening_res}Linear Clustering Diagnostics}

\begin{figure*}[!ht]
\centering
\includegraphics[width=\textwidth]{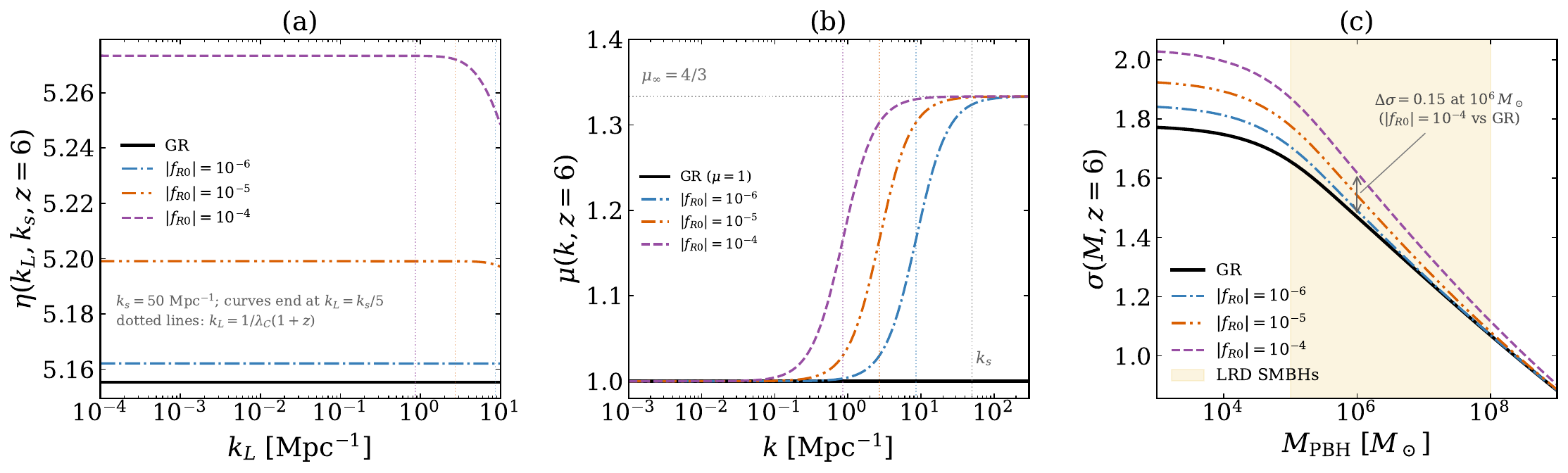}
\caption{Linear clustering diagnostics at $z=6$ for GR and Hu-Sawicki models with $|f_{R0}| = 10^{-6}$, $10^{-5}$, and $10^{-4}$, computed with the background field $|f_R(z)|$.
\textbf{(a)}~Squeezed-limit mode-coupling parameter $\eta(k_L, k_s)$ of Eq.~\eqref{eq:eta_mg} as a function of the long-mode wavenumber $k_L$, at $k_s = 50\,\mathrm{Mpc}^{-1}$, using the growth response of the PBHs; the curves extend to the squeezed-limit boundary $k_L=k_s/5$, and dotted lines mark the inverse comoving Compton wavelengths.
\textbf{(b)}~Scale-dependent coupling $\mu(k,z=6)$ of non-relativistic matter, Eq.~\eqref{eq:mu_hs}, from the GR limit ($\mu = 1$) to the scalar-tensor limit ($\mu_\infty = 4/3$, horizontal dotted line); vertical dotted lines mark the inverse comoving Compton wavelengths and $k_s$.
\textbf{(c)}~Linear rms mass variance $\sigma(M,z=6)$ from the scale-dependent growth of Eq.~\eqref{eq:growth_m}, with host screening. The arrow marks $\Delta\sigma$ at $10^6\,\Msun$ for $|f_{R0}|=10^{-4}$. The shaded gold region highlights the LRD SMBH mass interval $10^5\text{-}10^8\,\Msun$.}
\label{fig:screening}
\end{figure*}

In Fig.~\ref{fig:screening}, we have evaluated the linear clustering quantities described in Sec.~\ref{ssec:xi} at $z=6$, which govern the amplitude of the spatial correlation function $\xi_{\rm PBH}$. At this redshift, the comoving Compton wavelength is $\lambda_C(1+z)\simeq0.12$, $0.37$ and $1.2$~Mpc for $|f_{R0}|=10^{-6}$, $10^{-5}$ and $10^{-4}$, so that the modification operates only at $k\gtrsim8.6$, $2.7$ and $0.86\,{\rm Mpc}^{-1}$, respectively.

In Fig.~\ref{fig:screening}(b), we have shown the scale-dependent coupling $\mu(k,z=6)$ of Eq.~\eqref{eq:mu_hs}. Each Hu-Sawicki model interpolates from the GR limit ($\mu = 1$) at large scales to the scalar-tensor limit ($\mu_\infty = 4/3$) at small scales, with the transition at $k\simeq[\lambda_C(1+z)]^{-1}$ (vertical dotted lines). The short-mode scale $k_s = 50\,{\rm Mpc}^{-1}$ lies on the saturated plateau for all three models. This coupling acts on the particle dark matter, and only for a short time: since $|f_R(z)|$ grows toward low redshift, the scale-dependent enhancement of the matter growth accumulates mainly at late times.

In Fig.~\ref{fig:screening}(c), we have mapped the modified growth into the linear mass variance $\sigma(M,z=6)$, computed from the scale-dependent growth of Eq.~\eqref{eq:growth_m} with host screening. At $M=10^6\,\Msun$, $\sigma$ increases from $1.47$ in GR to $1.49$, $1.54$ and $1.62$ for $|f_{R0}|=10^{-6}$, $10^{-5}$ and $10^{-4}$, i.e., by $1.5\%$, $5\%$ and $10\%$ ($\Delta\sigma=0.15$ for $|f_{R0}|=10^{-4}$). The enhancement is largest at low masses, where the hosts are least screened and the modes lie furthest below the Compton wavelength, and vanishes toward $10^{9}\,\Msun$. The shaded gold region indicates the mass range of SMBHs observed in LRDs.

In Fig.~\ref{fig:screening}(a), we have presented the squeezed-limit coupling $\eta(k_L, k_s)$ of Eq.~\eqref{eq:eta_mg} as a function of the long-mode wavenumber $k_L$. Since the PBHs respond to the enhanced matter perturbations only through their gravitational potential, their growth response is much weaker than that of the particle dark matter: at $k_s$, $g_{\rm P}\simeq1.001$, $1.009$ and $1.023$, compared with $g_m\simeq1.05$, $1.10$ and $1.17$ before screening. For long modes outside the Compton wavelength, $g_{\rm P}(k_L)=1$, and $\eta$ reaches a plateau of $5.162$, $5.199$ and $5.273$, above the GR value of $5.155$ by $0.13\%$, $0.85\%$ and $2.3\%$; it decreases only when $k_L$ approaches the Compton scale.

\subsection{\label{ssec:xi_res}PBH Correlation Function}

\begin{figure}[!ht]
\centering
\includegraphics[width=\columnwidth]{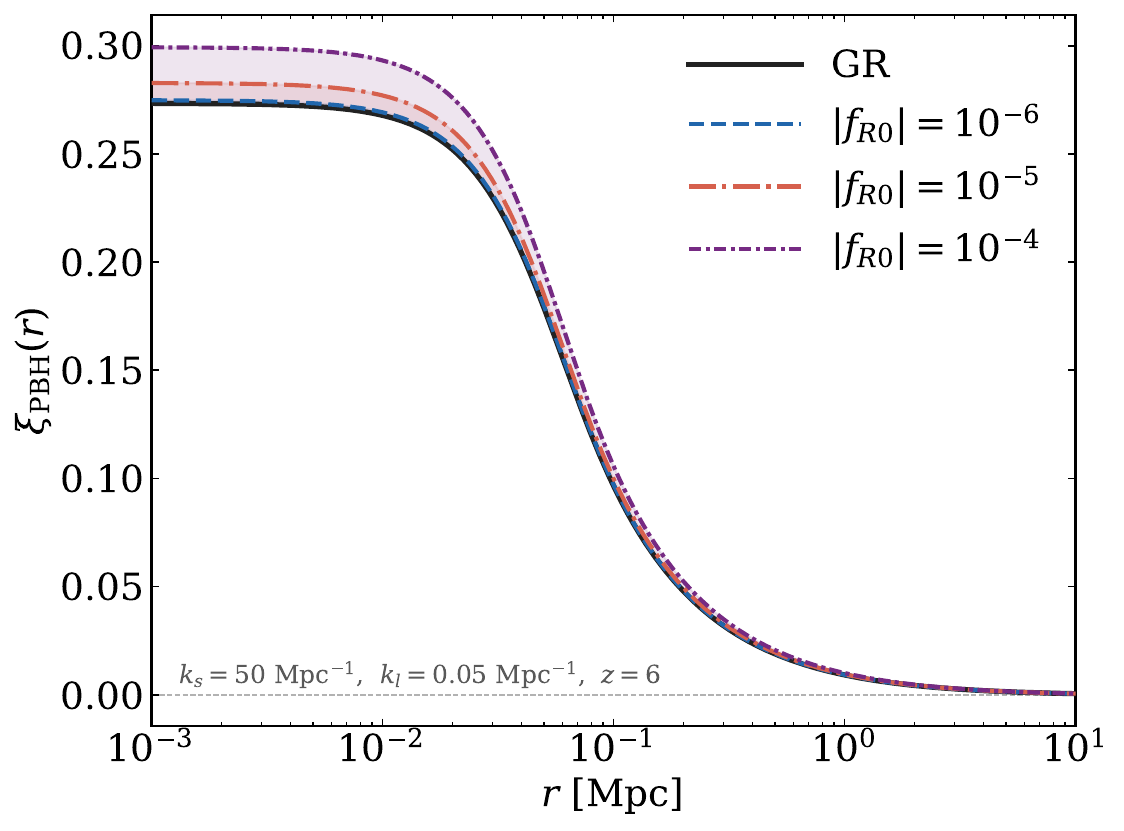}
\caption{Real-space PBH correlation function $\xi_{\rm PBH}(r)$ at $z = 6$ from Eq.~\eqref{eq:xi} for GR and Hu-Sawicki models with $|f_{R0}| = 10^{-6}$, $10^{-5}$, and $10^{-4}$, evaluated at $k_s = 50\,\mathrm{Mpc}^{-1}$ and $k_L = 0.05\,\mathrm{Mpc}^{-1}$. The amplitude includes the growth response of the PBHs, which carry no scalar charge; the enhancement over GR is $0.5\%$, $3.4\%$ and $9.5\%$, respectively.}
\label{fig:xi}
\end{figure}

In Fig.~\ref{fig:xi}, we have evaluated the real-space PBH correlation function $\xi_{\rm PBH}(r)$ at $z = 6$, derived from Eq.~\eqref{eq:xi} with the power spectrum of Eq.~\eqref{eq:Pk} anchored at $k_s = 50\,{\rm Mpc}^{-1}$ and $k_L = 0.05\,{\rm Mpc}^{-1}$. All four profiles share the same radial shape, with a plateau at $r \lesssim 0.01\,{\rm Mpc}$, a steep decline around $r \sim 0.1\,{\rm Mpc}$, and a large-scale cutoff at $r \gtrsim 1\,{\rm Mpc}$ set by $k_l$. They differ only through the amplitude $A \propto \eta^2 g_{\rm P}^2(k_s)$ of Eq.~\eqref{eq:A_amp}.

GR gives $\eta = 5.155$, $g_{\rm P}=1$ and $A = 0.1220$. The Hu-Sawicki models give $A=0.1227$, $0.1262$ and $0.1336$ for $|f_{R0}|=10^{-6}$, $10^{-5}$ and $10^{-4}$, so that the correlation function is enhanced by a scale-independent factor of $1.005$, $1.034$ and $1.095$, respectively. The amplitudes follow the expected hierarchy, $\xi_{\rm GR}<\xi_{10^{-6}}<\xi_{10^{-5}}<\xi_{10^{-4}}$, and converge to GR as $|f_{R0}|\to0$. The enhancement is set by the weak background field at $z=6$ and because the PBHs, which carry no scalar charge, are affected only indirectly through the growth of the surrounding matter. Physically, $\xi_{\rm PBH}(r)$ characterizes the spatial clustering of the PBH population across $0.001$-$1\,{\rm Mpc}$ scales, independently of the internal cluster dynamics ($R_{\rm cl} \approx 0.062\,{\rm pc}$) that govern coagulation.


\section{\label{sec:app_spin}Remnant-Spin Evolution}

\subsection{\label{ssec:spin_framework}Framework}

The spin vector of a black hole records its mass-assembly history. In dense cluster environments, seed PBHs grow through gas accretion and hierarchical binary coalescences, each imparting characteristic torques whose imprint on the spin magnitude and orientation reflects the dominant channel. We first present the relativistic framework common to both channels, then describe how each regime drives spin evolution, and finally couple this framework to the cluster kinetics of Sec.~\ref{sec:assembly}.

An isolated, stationary vacuum black hole is uniquely described by the Kerr metric, whose spin state is characterized by the dimensionless spin vector
\begin{equation}
\bm{\chi} \equiv \dfrac{c\,\mathbf{S}}{G M^2},
\label{eq:chi_def}
\end{equation}
where $\mathbf{S}$ is the spin angular-momentum vector, $\chi\equiv|\bm{\chi}|$ is bounded by $0\le\chi<1$, and $\hat{\mathbf{S}}\equiv\mathbf{S}/|\mathbf{S}|$ is the unit orientation vector in the cluster frame.

The radiative efficiency and angular-momentum transfer of accreted matter can be set by the location of the innermost stable circular orbit (ISCO). In units of the gravitational radius $R_g \equiv G M / c^2$, the equatorial ISCO radius is~\citep{Bardeen1972}
\begin{equation}
r_{\rm ISCO}(\chi)
=
3 + Z_2 \mp \sqrt{(3-Z_1)(3+Z_1+2Z_2)},
\label{eq:r_isco}
\end{equation}
where the lower (upper) sign corresponds to prograde (retrograde) orbits, and
\begin{align}
Z_1 &\equiv 1 + \left(1-\chi^2\right)^{1/3} \left[(1+\chi)^{1/3} + (1-\chi)^{1/3}\right], \label{eq:Z1}\\
Z_2 &\equiv \sqrt{3\chi^2 + Z_1^2}. \label{eq:Z2}
\end{align}
The specific energy and angular momentum per unit rest mass carried across the horizon at the ISCO follow as
\begin{align}
E_{\rm ISCO}(\chi)
&=
\dfrac{r_{\rm ISCO}^{3/2} - 2 r_{\rm ISCO}^{1/2} \pm \chi}
     {r_{\rm ISCO}^{3/4}\sqrt{r_{\rm ISCO}^{3/2} - 3 r_{\rm ISCO}^{1/2} \pm 2\chi}},
\label{eq:E_isco}\\~\nonumber\\
L_{\rm ISCO}(\chi)
&=
\pm \dfrac{G M}{c}\,
\dfrac{r_{\rm ISCO}^2 \mp 2\chi\sqrt{r_{\rm ISCO}} + \chi^2}
     {r_{\rm ISCO}^{3/4}\sqrt{r_{\rm ISCO}^{3/2} - 3 r_{\rm ISCO}^{1/2} \pm 2\chi}}.
\label{eq:L_isco}
\end{align}
Based on this, one can obtain the Novikov-Thorne thin-disk accretion efficiency as $\eta(\chi)=1-E_{\rm ISCO}(\chi)$ \footnote{For a Schwarzschild black hole, $r_{\rm ISCO}=6\,R_g$ and $\eta\approx0.057$. For the extremal prograde limit, $r_{\rm ISCO}\to R_g$ and $\eta\to0.423$, while, for the extremal retrograde limit, $r_{\rm ISCO}\to9\,R_g$ and $\eta\approx0.038$. These limits bracket the radiative efficiencies of spinning black holes; the reference value $\epsilon=0.10$ of Sec.~\ref{ssec:accretion_model} corresponds to an intermediate spin.}.

The spin evolution of the central seed proceeds through the two growth channels of Sec.~\ref{ssec:accretion_model}: gas accretion and the capture of field PBHs. We compute both along the coupled growth track of Eq.~\eqref{eq:full_growth}, rather than prescribing their outcome.

\paragraph*{Accretion.} When gas is supplied with a coherent angular momentum $\mathbf{J}_{\rm disk}$, frame dragging exerts a Lense-Thirring torque on the misaligned accretion flow,
\begin{equation}
\bm{\tau}_{\rm LT}
=
\dfrac{2G}{c^2 r^3}\left(\mathbf{S}\times\mathbf{J}_{\rm disk}\right),
\label{eq:tau_LT}
\end{equation}
which warps the inner disk and aligns $\hat{\mathbf{S}}$ with $\hat{\mathbf{J}}_{\rm disk}$ through the Bardeen-Petterson effect~\citep{BardeenPetterson1975}, on a timescale much shorter than the mass-doubling time. Accretion from the ISCO then changes the spin according to the Bardeen equation~\citep{Bardeen1970},
\begin{equation}
\dfrac{\dd\chi}{\dd\ln M}
=
\dfrac{c\,L_{\rm ISCO}(\chi)}{G M\,E_{\rm ISCO}(\chi)} - 2\chi,
\label{eq:bardeen_ode}
\end{equation}
where $\chi$ is signed relative to the disk axis, so that retrograde accretion ($\chi<0$) spins the black hole down. Prograde accretion drives the spin toward Thorne's limit~\citep{Thorne1974}, $\chi_{\rm max}=0.998$, at which the spin-up torque is balanced by the capture of disk photons; we cap the spin at this value. A non-spinning black hole reaches $\chi_{\rm max}$ after growing by a factor $\simeq\sqrt{6}$ in mass. We consider two feeding geometries. In \emph{coherent} accretion the disk axis is fixed, and since the disk angular momentum greatly exceeds that of the hole, the spin is aligned with the disk. In \emph{chaotic} accretion, gas arrives in episodes with randomly oriented angular momenta~\citep{KingPringle2006,King2008}. We follow~\citet{King2008} and set the mass of each episode by the self-gravity of the accretion disk: a Shakura-Sunyaev disk becomes self-gravitating outside the radius $R_{\rm sg}$, so that an episode carries at most $M_{\rm sg}\simeq(H/R)M$, which is $\sim2\%$ of the black-hole mass at $M\sim10^{3}\,\Msun$ and $\sim0.4\%$ at $M\sim10^{8}\,\Msun$ for $\lambda_{\rm Edd}=0.5$. The angular momentum delivered by an episode is $J_{\rm d}=M_{\rm sg}(GMR)^{1/2}$, evaluated at the warp radius $R_{\rm warp}$ or at $R_{\rm sg}$ if smaller, and that of the hole is $J_{\rm h}=\chi GM^2/c$~\citep{King2008}. A disk misaligned by an angle $\theta$ counter-aligns with the hole if~\citep{King2005}
\begin{equation}
\cos\theta<-\dfrac{J_{\rm d}}{2J_{\rm h}},
\label{eq:counter}
\end{equation}
and co-aligns otherwise; during the episode the spin direction moves along the total angular momentum $\mathbf{J}_{\rm h}+\mathbf{J}_{\rm d}$. We adopt the disk parameters of~\citet{King2008}, $\alpha_1=\alpha_2=0.03$. Since retrograde episodes spin the hole down more efficiently than prograde episodes spin it up, the spin settles at a low, slowly mass-dependent mean value, with no free episode size. Between these limits, the fuelling of galactic nuclei is expected to retain some net orientation. Following~\citet{Dotti2013}, we parameterize this by the fraction $F$ of episodes whose angular momentum points into the hemisphere opposite to a fixed axis, with isotropic orientations within each hemisphere: $F=0.5$ recovers chaotic accretion, and $F=0$ is the most anisotropic case, still far from coherent.

\paragraph*{Mergers.} Each capture of a field PBH is a merger of the seed with a much lighter, non-spinning companion; PBHs formed from the collapse of density perturbations are born with negligible spin~\citep{DeLuca2019spin}. For a merger of masses $m_1\geq m_2$ with mass ratio $q\equiv m_2/m_1$ and spins $\bm{\chi}_1$, $\bm{\chi}_2$, we use the final-spin prescription of~\citet{Barausse2009}, which is calibrated to numerical relativity and reproduces the test-particle limit,
\begin{equation}
\bm{\chi}_f
=
\dfrac{\bm{\chi}_1 + \bm{\chi}_2\,q^2 + \ell\,q\,\hat{\mathbf{L}}}{(1+q)^2},
\label{eq:chi_f}
\end{equation}
where $\hat{\mathbf{L}}$ is the orbital angular-momentum direction and
\begin{align}
\ell
={}&
\dfrac{s_4}{(1+q^2)^2}\left(\chi_1^2+\chi_2^2q^4+2\chi_1\chi_2q^2\cos\alpha\right)
+2\sqrt{3}+t_2\nu
\nonumber\\
&+t_3\nu^2+\dfrac{s_5\nu+t_0+2}{1+q^2}\left(\chi_1\cos\beta+\chi_2q^2\cos\gamma\right),
\label{eq:lrem}
\end{align}
with $\nu\equiv q/(1+q)^2$, $\cos\alpha=\hat{\bm{\chi}}_1\cdot\hat{\bm{\chi}}_2$, $\cos\beta=\hat{\bm{\chi}}_1\cdot\hat{\mathbf{L}}$, $\cos\gamma=\hat{\bm{\chi}}_2\cdot\hat{\mathbf{L}}$, and $s_4=-0.1229$, $s_5=0.4537$, $t_0=-2.8904$, $t_2=-3.5171$, $t_3=2.5763$. For two non-spinning equal-mass black holes, Eq.~\eqref{eq:chi_f} gives $\chi_f=0.686$, the value approached by hierarchical sequences of comparable-mass mergers~\citep{Fishbach2017,GerosaBerti2017}. The central seed, however, grows by captures at mass ratios $q\ll1$. In this regime, each capture deposits a small orbital angular momentum in a random direction, and repeated isotropic minor mergers drive the spin down rather than toward $0.686$~\citep{HughesBlandford2003,BertiVolonteri2008}. We draw the orbital axes isotropically and the captured masses from the field distribution weighted by the kernel factor $m^{2/7}$. Each capture radiates the energy given by the fit of~\citet{BarausseMorozovaRezzolla2012}, which reduces to $[1-E_{\rm ISCO}(\chi)]\,m$ in the test-particle limit ($5.7\%$ of the captured mass for a non-spinning hole), and imparts a GW recoil, which we compute from the standard mass-ratio and spin terms~\citep{GonzalezTotalRecoil2007,Campanelli2007} and compare with the central escape speed of the cluster, $v_{\rm esc}=\sqrt{3}\,v_{\rm vir}\simeq820\,{\rm km\,s^{-1}}$.

\paragraph*{Implementation.} We follow four channels: pure mergers ($\lambda_{\rm Edd}=0$), and mergers combined with coherent, anisotropic ($F=0$) or chaotic ($F=0.5$) accretion. For the accreting channels we adopt $\lambda_{\rm Edd}=0.5$ and $\epsilon=0.10$, which bring the seed into the LRD mass range by $z\simeq5\text{-}6.5$ (Sec.~\ref{ssec:growth_res}). Along each track, the captures (about $9\times10^{4}$ for the accreting channels) and the accretion between them are applied in sequence, and the spin vector is evolved with Eqs.~\eqref{eq:bardeen_ode}-\eqref{eq:lrem}. In every channel, the capture times of each realization are drawn from a Poisson process with the rate of the coupled growth track. The seed starts non-spinning at $z_{\rm ini}$. We keep $\epsilon=0.10$ in the growth equation for all channels. A spin-dependent efficiency, $\epsilon=1-E_{\rm ISCO}(\chi)$, would slow the growth of coherently accreting seeds ($\epsilon\to0.32$ at $\chi=0.998$) and accelerate that of chaotically accreting ones; this coupling is beyond the scope of this work. We evolve $4000$ Monte Carlo realizations per channel; curves show the mean spin, and we quote medians and $16\text{-}84\%$ intervals of the distributions, with orientations shown relative to the disk axis of the coherent channel. To illustrate the intrinsic behavior of each mechanism, we also evolve each channel in isolation, as a function of mass growth and for two initial spins.


\subsection{\label{ssec:spin_res}Results}

\begin{figure*}[!ht]
\centering
\includegraphics[width=\textwidth]{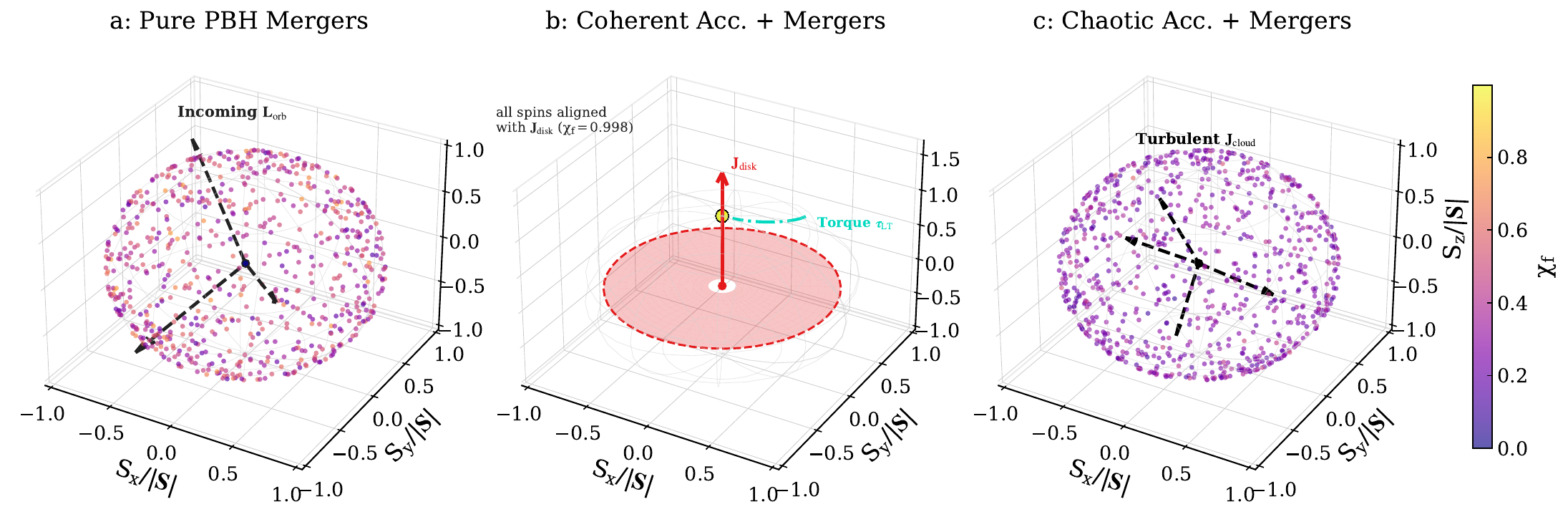}
\caption{Spin orientations $\hat{\mathbf{S}}=\mathbf{S}/|\mathbf{S}|$ of the central seed at $z=6$, colored by the spin magnitude $\chi_f$, for $800$ randomly selected Monte Carlo realizations per channel along the GR growth tracks of Fig.~\ref{fig:growth}. \textbf{(a)}~Pure PBH mergers ($\lambda_{\rm Edd}=0$): captures with isotropic orbital angular momenta $\mathbf{L}_{\rm orb}$ give isotropic orientations. \textbf{(b)}~Coherent accretion and mergers ($\lambda_{\rm Edd}=0.5$): the Lense-Thirring torque aligns the spin with $\mathbf{J}_{\rm disk}$, and all realizations have $\chi_f=0.998$ along the disk axis. \textbf{(c)}~Chaotic accretion and mergers ($\lambda_{\rm Edd}=0.5$): randomly oriented, self-gravity-limited episodes~\citep{King2008} give isotropic orientations and moderate spins. These distributions hold for all $|f_{R0}|$ considered, since the scalar field acts on the medium surrounding the seed.}
\label{fig:3d_spin_orientations}
\end{figure*}

\begin{figure*}[!ht]
\centering
\includegraphics[width=\textwidth]{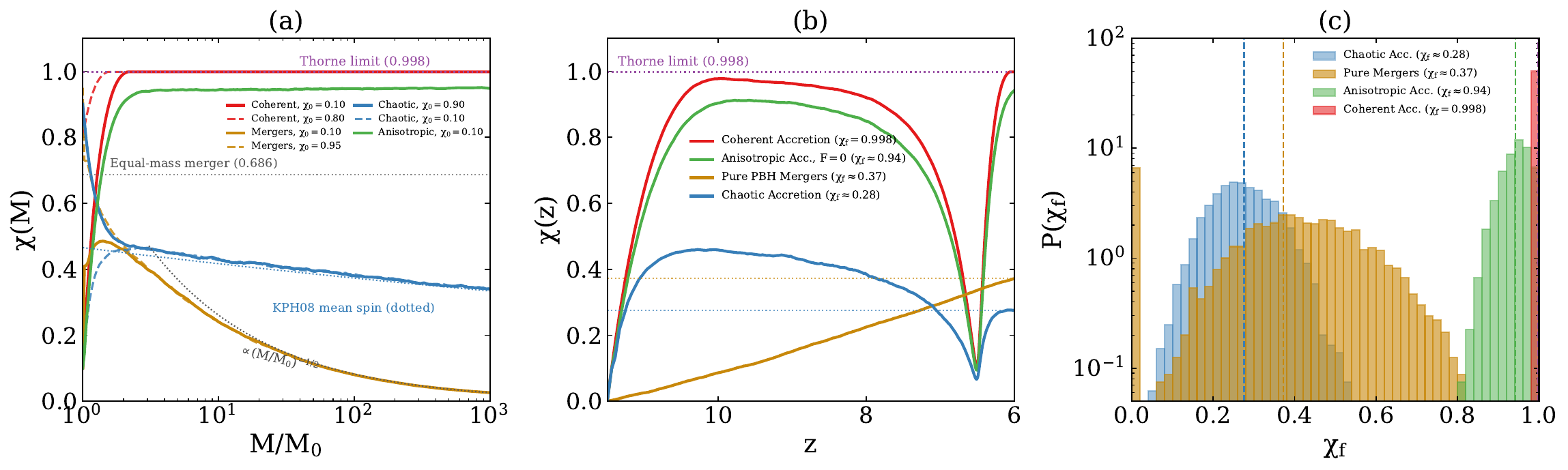}
\caption{Spin evolution of the central seed, computed from the Bardeen equation [Eq.~\eqref{eq:bardeen_ode}], capped at the Thorne limit, with self-gravity-limited accretion episodes and the alignment criterion of Eq.~\eqref{eq:counter}~\citep{King2008} for non-coherent feeding, and from the final-spin prescription of Eq.~\eqref{eq:chi_f} for each capture. \textbf{(a)}~Each channel in isolation, as a function of the mass growth $M/M_0$ from $M_0\simeq170\,\Msun$, for two initial spins $\chi_0$ (solid and dashed; means of $4000$ realizations for the stochastic channels); green: anisotropic accretion with $F=0$. Dotted lines mark the Thorne limit and the equal-mass merger value $0.686$; the blue dotted curve is the mean spin of~\citet{King2008}, and the grey dotted curve the random-walk prediction of Eq.~\eqref{eq:chi_rw} for successive minor mergers. \textbf{(b)}~Mean spin along the coupled growth tracks of Fig.~\ref{fig:growth}, with $\lambda_{\rm Edd}=0$ for pure mergers and $\lambda_{\rm Edd}=0.5$ for the accreting channels. \textbf{(c)}~Distribution of spin magnitudes at $z=6$ from $4000$ realizations per channel; dashed lines mark the means. These distributions hold for all $|f_{R0}|$ considered, since the scalar field acts on the medium surrounding the seed.}
\label{fig:spin_trajectories_and_distributions}
\end{figure*}

In Figs.~\ref{fig:3d_spin_orientations} and~\ref{fig:spin_trajectories_and_distributions}, we have shown the spin orientations and magnitudes of the central seed for the channels of Sec.~\ref{ssec:spin_framework}. All results are computed along the growth tracks of Fig.~\ref{fig:growth}, which hold for every $|f_{R0}|$ considered.

Figure~\ref{fig:spin_trajectories_and_distributions}(a) shows each channel in isolation. Coherent accretion drives the spin to the Thorne limit after a mass growth of $\simeq\sqrt{6}$, independently of the initial spin. Anisotropic accretion with $F=0$ drives it to $\chi\simeq0.95$ on a similar mass scale: even though the episode axes are randomly distributed within a hemisphere, the persistent net orientation spins the hole up, in agreement with~\citet{Dotti2013}. Chaotic accretion relaxes from either initial spin to a low mean value within a mass growth of a factor of $\sim2$, after which the mean spin decreases slowly with mass, from $\simeq0.45$ at $M\sim10^{3}\,\Msun$ to $\simeq0.34$ at $M\sim2\times10^{5}\,\Msun$. This reproduces the mean spin found by~\citet{King2008} for self-gravity-limited episodes, $\bar\chi\simeq0.246\,(M/10^{8}\,\Msun)^{-0.048}$ (dotted curve), to within a few per cent, without any adjustable episode size.

Successive captures of field PBHs do not approach the equal-mass value $0.686$. The first captures, at $q\simeq0.2$, bring the spin to $\chi\simeq0.5$ regardless of its initial value, because a single companion of this mass ratio deposits orbital angular momentum comparable to the spin of the hole. Further captures at decreasing mass ratio deposit orbital angular momenta in random directions, which add as a random walk while the mass grows linearly. For $N\simeq M/\bar m$ captures of mean mass $\bar m$, the accumulated angular momentum is $J_{\rm rms}\simeq\ell\,(G/c)\sqrt{\bar m/3}\,M^{3/2}$, so that the mean spin decays as
\begin{equation}
\langle\chi\rangle\simeq\sqrt{\dfrac{8}{3\pi}}\;2\sqrt{3}\left(\dfrac{\bar m}{3M}\right)^{1/2}\propto\left(\dfrac{M}{M_0}\right)^{-1/2},
\label{eq:chi_rw}
\end{equation}
where $\sqrt{8/3\pi}$ converts the rms to the mean of a three-dimensional random vector. The Monte Carlo, which includes the energy radiated in each capture, follows Eq.~\eqref{eq:chi_rw} closely for $M/M_0\gtrsim30$, reaching $\chi\simeq0.24$ after a tenfold and $\simeq0.03$ after a thousandfold mass growth. Minor mergers therefore have no finite attractor: they drive the spin toward zero~\citep{HughesBlandford2003}.

Along the coupled growth tracks [Figs.~\ref{fig:spin_trajectories_and_distributions}(b) and~(c)], the pure-merger channel captures on average only $\simeq4$ field PBHs by $z=5$, at mass ratios $q\simeq0.1\text{-}0.2$. Its spin reaches a broad distribution with mean $\chi\simeq0.37$, median $\simeq0.39$ and $16\text{-}84\%$ range $0.18\text{-}0.56$ at $z=6$, with isotropic orientations [Fig.~\ref{fig:3d_spin_orientations}(a)].

In the coherent channel, accretion spins the seed up to the Thorne limit before $z=10$. Early captures, at $q\sim0.1$, reduce the spin by $\sim10\%$ each, so that the mean spin declines slowly from $z\simeq10$. During the runaway phase, the seed captures $\sim3\times10^{6}\,\Msun$ of PBHs with isotropic orbits within $\simeq50$~Myr, while gas accretion supplies only $\sim10\%$ of its growth. These minor mergers spin the seed down efficiently, and the mean spin falls to $\chi\simeq0.10$ at $z\simeq6.5$, when $M_{\rm seed}\simeq2\times10^{6}\,\Msun$. Once the reservoir is exhausted, accretion restores $\chi=0.998$ after a further mass growth of $\simeq\sqrt{6}$, by \textcolor{blue}{$z\simeq6.1$}. At $z=6$, all realizations have $\chi=0.998$ and are aligned with the disk axis [Fig.~\ref{fig:3d_spin_orientations}(b)]; at $z=6.5$, however, the mean spin is only \textcolor{blue}{$\simeq0.11$}. The anisotropic channel follows the same pattern at slightly lower spin: $\chi\simeq0.91$ before the runaway, a minimum of $\simeq0.09$ during it, and mean $\chi\simeq0.94$ (median $0.95$, $16\text{-}84\%$ range $0.91\text{-}0.98$) at $z=6$, with spins aligned to within $\langle\cos\theta\rangle\simeq0.95$ of the preferred axis.

In the chaotic channel, the self-gravity-limited episodes keep the spin moderate throughout. Its mean is $\simeq0.45$ at $z\simeq10$, declines slowly as the seed grows, drops to $\simeq0.07$ during the runaway, and recovers to the King et al. value afterwards. At $z=6$, the spin has mean \textcolor{blue}{$\chi\simeq0.28$} (median $0.27$) and $16\text{-}84\%$ range $0.20\text{-}0.36$, with isotropic orientations [Fig.~\ref{fig:3d_spin_orientations}(c)].

GW recoil does not alter these conclusions. The kicks from captures at $q\lesssim0.2$ exceed the central escape speed of the cluster in only $\lesssim0.4\%$ of the realizations, and those at the small mass ratios of the runaway are negligible, so that the seed is retained, consistent with the assumption of Sec.~\ref{ssec:cluster_dynamics}.

The channels thus remain distinguishable at $z=6$ [Fig.~\ref{fig:spin_trajectories_and_distributions}(c)]. Coherent and anisotropic accretion give high, aligned spins ($\chi\simeq0.95\text{-}0.998$); chaotic accretion gives moderate, isotropic spins ($\chi\simeq0.2\text{-}0.4$); and the merger channel gives intermediate, isotropic spins with a broad distribution, below the equal-mass value. The distributions of the merger and chaotic channels overlap, and every accreting channel passes through low spins during the runaway. Spin measurements alone therefore identify the growth channel only when combined with the epoch of observation and the inferred mass.

\begin{table*}[!ht]
\caption{Growth of the central seed with the reservoir approximation of Eq.~\eqref{eq:full_growth} and with explicit depletion of the field population.}
\label{tab:depletion}
\begin{ruledtabular}
\begin{tabular}{cccc}
$\lambda_{\rm Edd}$ & $z_{\rm exh}$ (reservoir / depleted) & $M_{\rm seed}(z=6)\,[\Msun]$ & $M_{\rm seed}(z=5)\,[\Msun]$\\
\hline
0.3 & 5.55 / 5.55 & $2.03\times10^{4}$ / $2.03\times10^{4}$ & $8.38\times10^{6}$ / $8.37\times10^{6}$\\
0.5 & 6.46 / 6.46 & $8.59\times10^{6}$ / $8.58\times10^{6}$ & $9.60\times10^{7}$ / $9.59\times10^{7}$\\
0.8 & 7.32 / 7.31 & $1.17\times10^{8}$ / $1.17\times10^{8}$ & $5.58\times10^{9}$ / $5.58\times10^{9}$\\
\end{tabular}
\end{ruledtabular}
\end{table*}

\section{\label{sec:app_numerics}Reservoir Normalization and Field Depletion}

The initial PBH distribution is log-normal in $\ln m$ with median $m_c=30\,\Msun$ and width $\sigma_{\ln m}=0.40$, normalized to $N_{\rm cl}=10^{5}$ PBHs. The corresponding cluster mass is $M_{\rm cl}=N_{\rm cl}\langle m\rangle=N_{\rm cl}\,m_c\,e^{\sigma_{\ln m}^{2}/2}=3.25\times10^{6}\,\Msun$, which sets both the reservoir in Eq.~\eqref{eq:full_growth} and the virial velocity of the cluster. The mass grid of the coagulation solver extends from $0.1\,m_c$, so that the discretized distribution resolves the low-mass tail of the log-normal and reproduces this cluster mass; the solver conserves the total PBH mass to better than $10^{-5}$.

In Eq.~\eqref{eq:full_growth}, the field population entering the capture rate of the seed is taken from the coagulation of Sec.~\ref{ssec:smolu}, solved independently of the seed, and the depletion of the field by the seed is represented by the reservoir factor $(1-M_{\rm cap}/M_{\rm cl})$. To test this approximation, we solved the coagulation equation and the growth of the seed simultaneously, removing each captured PBH from the field through an explicit sink term, $\dot n_i\supset-K(M,m_i)\,n_i/V_{\rm cl}$, without a reservoir factor. Table~\ref{tab:depletion} compares the two treatments. The exhaustion redshifts agree to within $0.01$ and the seed masses at $z=6$ and $z=5$ agree to within $0.2\%$, because the depletion of the field reduces the capture rate by the same fraction as the reservoir factor. The reservoir approximation is therefore accurate for the growth of the seed. It does not, however, deplete the field population itself: in the self-consistent solution the field is fully consumed by the seed by $z=5$ for all three Eddington ratios, whereas in the independent coagulation the field persists. This affects only quantities derived from field-field mergers, in particular the SGWB of Appendix~\ref{sec:app_sgwb}.

\end{document}